\documentclass[aip,jcp,amsmath,amssymb,reprint,floatfix,notitlepage]{revtex4-1}

\usepackage{graphicx}
\usepackage{dcolumn}
\usepackage{bm}
\usepackage{placeins} 
\usepackage{tabularx}
\usepackage[version=3]{mhchem} 
\usepackage{array}
\usepackage{natmove}
\usepackage{mciteplus}
\usepackage[caption=false]{subfig}
\usepackage{graphicx}
\usepackage{xr}
\usepackage{xcolor}
\usepackage[normalem]{ ulem }
\usepackage{soul}
\usepackage{hyperref}
\hypersetup{%
  colorlinks=true, 
  breaklinks=true, 
  urlcolor=blue, 
  linkcolor=black, 
  citecolor=black, 
}

\makeatletter
\providecommand*{\input@path}{}
\g@addto@macro\input@path{{./Figures/}}
\makeatother

\graphicspath{{.}{Figures/}} 

\usepackage[justification=justified]{caption}
\definecolor{light-gray}{gray}{0.9}
\begin{document}
\title{The \ce{MgCO3}--\ce{CaCO3}--\ce{Li2CO3}--\ce{Na2CO3}--\ce{K2CO3} Carbonate Melts: Thermodynamics and Transport Properties by Atomistic Simulations}

\author{Elsa Desmaele}\email{elsa.desmaele@gmail.com}
\affiliation{Sorbonne Universit\'e, CNRS, Laboratoire de Physique Th\'eorique de la Mati\`{e}re Condens\'ee, LPTMC, F75005, Paris, France}
\author{Nicolas Sator}
\affiliation{Sorbonne Universit\'e, CNRS, Laboratoire de Physique Th\'eorique de la Mati\`{e}re Condens\'ee, LPTMC, F75005, Paris, France}
\author{Rodolphe Vuilleumier}
\affiliation{PASTEUR, D\'epartement de chimie, \'Ecole normale sup\'erieure, PSL University, Sorbonne Universit\'e, CNRS, 75005 Paris, France}
\author{Bertrand Guillot}\email{guillot@lptmc.jussieu.fr}
\affiliation{Sorbonne Universit\'e, CNRS, Laboratoire de Physique Th\'eorique de la Mati\`{e}re Condens\'ee, LPTMC, F75005, Paris, France}


\begin{abstract}
Atomistic simulations provide a meaningful way to determine the physico-chemical properties of liquids in a consistent theoretical framework. This approach takes on a particular usefulness for the study of molten  carbonates, in a context where thermodynamic and transport data are crucially needed over a large domain of temperatures and pressures (to ascertain the role of these melts in geochemical processes) but are very scarce in the literature, especially for the calco-magnesian compositions prevailing in the Earth's mantle. \\
Following our work on \ce{Li2CO3}--\ce{Na2CO3}--\ce{K2CO3} melts,\cite{moi2018,moi} we extend our force field to incorporate Ca and Mg components. The empirical interaction potentials are benchmarked on the density data available in the experimental literature (for the crystals and the \ce{K2Ca(CO3)2} melt) and on the liquid structure issued from ab initio molecular dynamics simulations. Molecular dynamics simulations are then performed to study the thermodynamics, the microscopic structure, the diffusion coefficients, the electrical conductivity and the viscosity of molten Ca, Mg-bearing carbonates up to 2073~K and 15~GPa. Additionally, the equation of state of a Na--Ca--K mixture representative of the lavas emitted at Ol Doinyo Lengai (Tanzania) is evaluated.\\
The overall agreement between the MD results and the existing experimental data is very satisfying and provides evidence for the ability of the force field to accurately model any \ce{MgCO3}--\ce{CaCO3}--\ce{Li2CO3}--\ce{Na2CO3}--\ce{K2CO3} melt over a large $T-P$ range. Moreover it is the first report of a force field allowing to study the transport properties of molten magnesite (\ce{MgCO3}) and molten dolomite (\ce{CaMg(CO3)2}).
\end{abstract}

\maketitle

\section{Introduction}
\label{sec1} 
Whether it is for the applied or the fundamental fields in which carbonate melts are implied (electrochemistry, geochemistry), having a reliable model for their alkali, alkaline-earth mixtures (and their end-members) is essential. For instance the longevity of molten carbonate fuel cells, usually based on alkali eutectic mixtures, is known to be improved by the addition of alkaline-earth cations,\cite{Lair2012} in a small enough amount that doesn't compromise the efficiency of the cell.\cite{Chery2015,Chery2015rev,Cassir2012,Cassir2013,Lair2012} From the viewpoint of geosciences, carbonate melts, although they constitute a very minor phase of the Earth's mantle (the most part is silicate), are thought to have important implications in the lowering of the melting temperature of silicate rocks, the deep carbon cycle or the high conductivity anomalies observed in the 70--200~km depth region.\cite{Dasgupta2006,Gaillard2008,Dasgupta2010,Dasgupta2013,Hammouda2015}
On Earth, evidence of a past volcanic activity induced by carbonatitic lava (\ce{SiO2} $<20$~wt\%) is given by petrology.\cite{Mitchell2005,Woolley2008} Nowadays, Ol Doinyo Lengai in Tanzania is the only active volcano to produce these remarkable (low temperature, low viscosity) carbonatitic lavas. 
The carbonatitic melt is mainly composed of a \ce{Na2CO3}--\ce{K2CO3}--\ce{CaCO3} mixture (in proportions 55:9:36 mol\% according to \citet{Keller2012}), called natrocarbonatite, whereas the majority of other 
inventoried carbonatites are of calco-magnesian composition.\cite{Woolley2005} In addition, in the Earth's mantle the most abundant carbonate compositions are calcite, \ce{CaCO3}, predominating at shallow depth and magnesite, \ce{MgCO3}, progressively taking over as depth increases. As a consequence their mixture and particularly dolomite, \ce{CaMg(CO3)2}, are of major interest. For pressure of a few kbars and beyond, these Ca,Mg-carbonates form stable liquid phases over a wide temperature domain.\cite{Spivak2012,Solopova2013,Solopova2015} But at atmospheric pressure they break down into \ce{CO2} + oxide (CaO, MgO) at temperatures below their melting point (decarbonation occurs for $P$ below $\sim$1~GPa for calcite and $\sim$3~GPa for magnesite and dolomite).\cite{Irving1975,Suito2001} Consequently only a few data are available in the literature for calco-magnesian compositions (most of which are alkali-bearing mixtures stable at low pressure),\cite{Gaillard2008,Kojima2009,Lair2012} and very few at high pressures.\cite{Dobson1996,Kono2014,Sifre2015} Thus for molten \ce{MgCO3} there is simply no data to our knowledge.\cite{Hurt2019} \\
In this context where experimental data are sparse in terms of thermodynamic conditions, chemical composition and physical properties, a noticeable advantage of molecular dynamics simulations (classical and ab initio) is that a variety of properties (structure, equation of state and transport coefficients) can be computed in the same theoretical framework. However, one has to be aware that collective quantities such as the ionic conductivity and the viscosity are calculated from slowly converging time correlation functions, that necessarily require (to be accurate) a rather large number of atoms and long time runs. For this reason, ab initio molecular dynamics simulations (AIMD), deriving from electronic density calculations, are either restricted to the study of thermodynamics and structure properties\cite{Zhang2015,Li2017} or only provide crude estimates of transport coefficients, especially when it is question to establish their evolution with pressure and temperature.\cite{Vuilleumier2014,Du2018} As for classical molecular dynamics simulations (MD), based on a more empirical approach, very few studies have provided estimates of the viscosity and electrical conductivity of carbonate melts, namely \ce{CaCO3},\cite{Vuilleumier2014} the \ce{Li2CO3}-\ce{K2CO3} eutectic mixture (62:38 mol\%)\cite{Corradini2016} and more recently the \ce{Li2CO3}--\ce{Na2CO3}--\ce{K2CO3} mixtures and end-members.\cite{moi2018} Obviously, the scarcity of thermodynamic data\cite{Liu2003,Hudspeth2018,Hurt2019} does not help for developing empirical potentials. As a consequence, very few force fields (FF) are available for Ca-Mg and most of them were developed for crystals \cite{Yuen1978,Dove1992,Pavese1992,Pavese1996,Fisler2000,Archer2003,Raiteri2010} based on the Born model of solids\cite{Born1954}. A first step towards an accurate description of the \ce{CaCO3} melt was made by Genge \emph{et al.}\cite{Genge1995} by adapting the FF of \citet{Dove1992} (for the crystal). The resulting FF reproduces quite well the liquid structure issued from a AIMD calculation published long after\cite{Vuilleumier2014}, but the calculated pressure is overestimated by $\sim +15$~kbar, compared to the recently published equation of state of \citet{Zhang2015} based on AIMD calculations. Moreover the melting temperature is underestimated. Both these features point toward a too weakly cohesive melt, that may result from the absence of dispersion interaction in the FF.\cite{moi2018} In another attempt \citet{Hurt2018} adapted the model of \citet{Archer2003} and proposed a FF fitted on crystalline properties for carbonates in the \ce{CaCO3}--\ce{SrCO3}--\ce{BaCO3} system.\\ 
For \ce{MgCO3}, in absence of experimental data, \citet{Hurt2018}  only list in their study some FF parameters, but no explanation on the way they were derived and no result is given for this composition. In the present study, we have developed an empirical force field to describe the thermodynamics and transport properties of \ce{MgCO3}--\ce{CaCO3} melts. \\
In a previous study devoted to the \ce{Li2CO3}--\ce{Na2CO3}--\ce{K2CO3} system, we have demonstrated the ability of molecular dynamics (MD) simulations to reproduce experimental data for alkali carbonate melts.\cite{moi2018} Here we extend the latter study to incorporate alkaline-earth components (Mg, Ca). In our study of alkali carbonates we evidenced the need to take into account in the FF the dispersion interactions, which were not included in the FF previously developed by our team for \ce{CaCO3}.\cite{Vuilleumier2014} Thus for the sake of consistency the interactions between the carbonate anions and the alkali or alkaline-earth cations are treated on the same footing in the present FF. In section~\ref{smetho} the details of the simulations are presented. In section~\ref{sthemo} the thermodynamic properties (equation of state) and the liquid structure are reported, whereas in section~\ref{strans} we evaluate the transport coefficients (diffusion coefficients, electrical conductivity and viscosity). Furthermore, the heuristicity of the Nernst-Einstein equation (relating the electrical conductivity and diffusion coefficients) and of the Stokes-Einstein equation (relating the viscosity and diffusion coefficients) is evaluated and commented.

\section{Method}
\label{smetho}
\subsection{Computational details}
\subsubsection{Ab initio molecular dynamics (AIMD)}
Two AIMD simulations were run for this study, one for molten magnesite (\ce{MgCO3}) and one for molten dolomite (\ce{CaMg(CO3)2}). They were based on the density functional theory (DFT) within the Born-Oppenheimer approximation using the freely available QUICKSTEP/CP2K software \cite{VandeVondele2005a} that applies a hybrid Gaussian/plane-wave method \cite{Lippert1997}. For the study of \ce{CaMg(CO3)2} we used for the valence electrons of carbon and oxygen a triple- zeta valence plus polarization basis set optimized for molecules (TZV2P)\cite{VandeVondele2005b}. Otherwise we used a double-zeta plus polarization basis set (DZVP) \cite{VdVHutter2007}. Core electrons were treated by the Goedecker-Teter-Hutter (GTH) norm conserving pseudo-potentials \cite{Goedecker1996,Hartwigsen1998,Krack2005}. The cutoff for the electronic density was set to 700 Ry. Exchange and correlation interactions were accounted for by the gradient corrected BLYP functional \cite{Becke1988,Lee1988} using a semi-empirical D3 dispersion correction scheme with a cutoff $\geq\sqrt{3} L$, where $L$ is the length of the simulation box.\cite{Grimme2010} All DFT calculations were run in the $NVT$ ensemble with the temperature set constant by a Nos\'e-Hoover thermostat. \cite{Nose1984a,Nose1984b} \\
For \ce{MgCO3} and for \ce{CaMg(CO3)2} the simulations were run with 200 atoms for 20 ps and with 640 atoms for 12~ps at state points (1873~K, 2.49~g/cm$^{3}$) and (1773~K, 2.25~g/cm$^{3}$), leading to an average pressure of 4.5 $\pm$ 1.5 and 1.8 $\pm$ 1.0 GPa, respectively. For information, congruent melting is known to occur at 2.7~GPa and 1850~K for magnesite\cite{Irving1975} and at 2.7~GPa and 1653~K for dolomite.\cite{Irving1975} As for molten calcite (\ce{CaCO3}), the AIMD  simulations of \citet{Vuilleumier2014}, based on the same DFT approach, were used as benchmark. 

\subsection{Classical molecular dynamics (MD)}
Classical MD simulations were carried out using the DL\_POLY~2 software \cite{Smith1996}, with a timestep of 1~fs. Density calculations were performed in the $NPT$ ensemble with a Nos\'e-Hoover thermostat for a simulation time of $t\simeq$ 0.9~ns (including a 0.5~ns equilibration run) allowing to reach an accuracy on the density value of $\Delta n/n \sim \pm1\%$. To evaluate  the transport coefficients, simulations were performed in the $NVE$ ensemble with an equilibration run of 0.5~ns, followed by a production run of 10 to 30~ns. Structural data were extracted from the same simulations. All simulations had a system size $N\simeq$ 2000 atoms, except for specific calculations that used the number of atoms as the AIMD simulations in order to compare at best the pair distribution functions obtained by both approaches.

\subsection{Force Field}
\begin{table}
\begin{tabular}{|l|l|r|l|r|r|}
\hline 
$i$  & $j$  & $A_{ij}$ (kJ/mol)  & $\rho_{ij}$ (\AA)  & $C_{ij}$ (\AA$^{6}$/mol)  & $q_{i}$ (e) \tabularnewline
\hline 
Mg  & O  & 243 000  & 0.24335  & 1 439  & +1.64202 \tabularnewline \hline 
Ca  & O  & 200 000  & 0.2935  & 5 000  & +1.64202 \tabularnewline \hline 
Li  & O  & 300 000  & 0.2228  & 1 210  & +0.82101\tabularnewline \hline 
Na  & O  & 1 100 000  & 0.2228  & 3 000  & +0.82101 \tabularnewline \hline 
K  & O  & 900 000  & 0.2570  & 7 000  & +0.82101 \tabularnewline \hline 
O  & O  & 500 000  & 0.252525  & 2 300  & $-$0.89429 \tabularnewline 
\hline 
\end{tabular}\caption{Intermolecular Buckingham parameters and partial charges. Note that
for electroneutrality considerations $q_{{\rm {C}}}=+1.04085$~e.
Intramolecular Born repulsion parameters between oxygen atoms of a same carbonate ion are
$A_{{\rm {OO}}}^{intra-{\rm {CO}_{3}}}=2611707.2$~kJ/mol and $\rho_{{\rm {OO}}}^{intra-{\rm {CO}_{3}}}=0.22$~\AA.}
\label{tffsh} 
\end{table}
%
\begin{figure}
\includegraphics{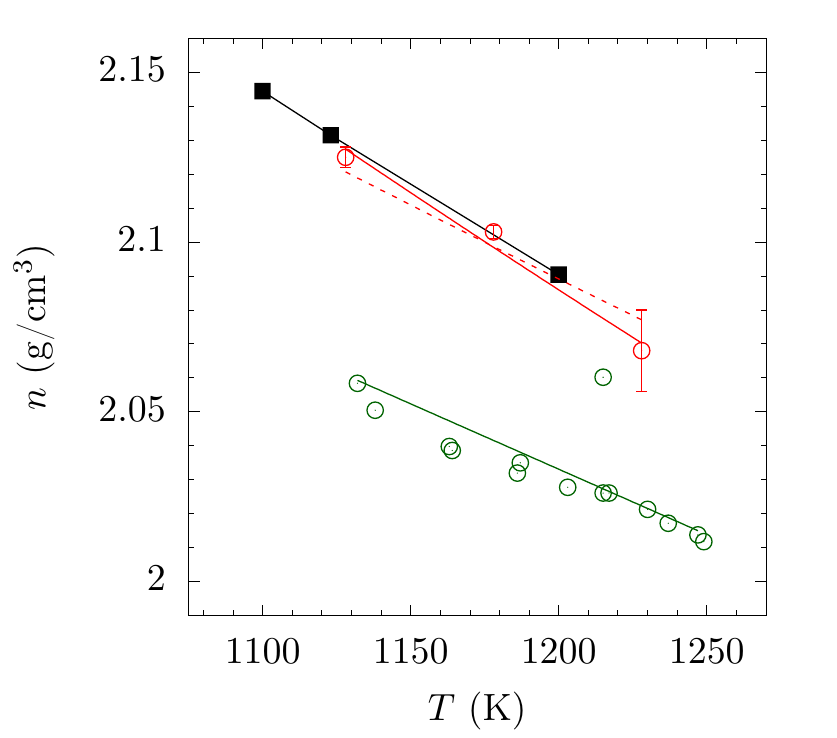}
\caption{Density of the molten \ce{K2Ca(CO3)2} mixture from MD simulations (squares) and from experimental measurements (circles) of \citet{Liu2003} (red) and \citet{Dobson1996}(green). The lines correspond to a linear fit. The dotted line is the extrapolation of  \citet{Liu2003} using an ideal mixing rule.}
\label{falphakc} 
\end{figure}
\begin{figure}
\includegraphics{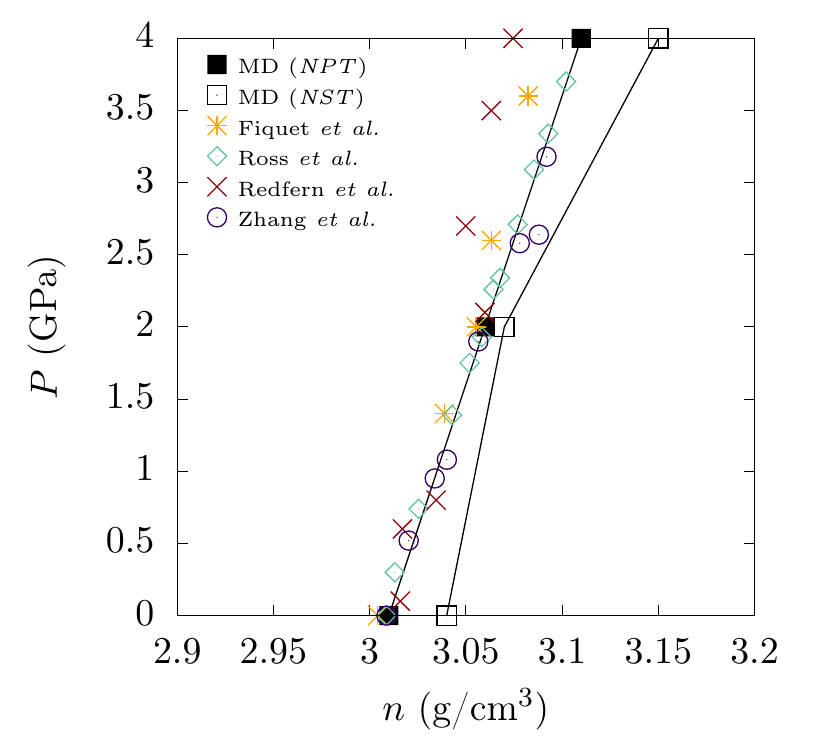}
\caption{Density-pressure diagram for crystalline magnesite \ce{MgCO3} at 300~K from the experimental literature (colored symbols)\cite{Redfern1993,Fiquet1994,Zhang1997,Ross1997} and obtained from MD calculations in the $NPT$ ensemble (filled black squares and black line as a guide to the eye) and in the $NST$ ensemble (empty black squares and black line as a guide to the eye).}
\label{fmagcr} 
\end{figure}

\begin{table*}
\begin{tabular}{|l|c|c|c|c|c|}
\hline 
 & \ce{MgCO3}  & \ce{CaCO3}  & \ce{CaMg(CO3)2} & \ce{K2Ca(CO3)2}  & Natro\tabularnewline \hline 
$n_{MD}^{0}$ (g/cm$^{3}$)  & 2.49 & 2.44 \scriptsize{(2.47$^{\dagger}$, 2.44$^\ddagger$)}  & 2.49 &2.14  & 2.11 \\ \hline
$n_{MD,\,mix}^{0}$(g/cm$^{3}$)  & -- & -- & 2.46 & 2.12 & 2.11 \\ \hline
$n_{Exp,\,mix}^{0}$ (g/cm$^{3}$) & 2.45 & 2.49 & 2.47 & 2.13  & 2.13\tabularnewline \hline
$K_{MD}^{0}$ (GPa)  & 15.0 & 17.6 \scriptsize{(15.7$^{\dagger}$, 15.1$^\ddagger$)} & 12.8 & 9.1 & 11.4\tabularnewline \hline
$K_{MD,\,mix}^{0}$ (GPa)  & -- &  -- & -- & 8.6 & 10.4\\ \hline
$K_{Exp,\,mix}^{0}$ (GPa) & -- & 18.7 & -- & 8.9 & 10.4\tabularnewline\hline 
\end{tabular}
\centering
\caption{Density and bulk modulus of \ce{MgCO3}, \ce{CaCO3} and \ce{CaCO3}-bearing
mixtures at $T=1100$~K calculated either by MD simulations ($n_{MD}^{0}$,
$K_{MD}^{0}$) or by using ideal mixing rules ($n_{MD,\,mix}^{0}$,
$K_{MD,\,mix}^{0}$). Extrapolation from experiments $n_{Exp,\,mix}^{0}$,
$K_{Exp,\,mix}^{0}$ are given for comparison.\cite{Liu2003,OLeary2015} Note that for \ce{MgCO3} an additional assumption relative to the cationic mass is needed to estimate $n_{Exp,\,mix}^{0}$ (see \citet{Hurt2019}). In parenthesis are given the MD results from the models of \citet{Vuilleumier2014} ($^{\dagger}$) and \citet{Hurt2018} ($^{\ddagger}$).}
\label{tid} 
\end{table*}
The force field is composed of interionic pair potentials as presented in a previous paper\cite{moi2018}. The intramolecular part (interactions within a carbonate molecule) contains an oxygen-oxygen term: $V_{\text{OO}}^{intra-{\rm {CO}_{3}}}({r}_{\text{OO}})={A_{\text{OO}}^{intra-{\rm {CO}_{3}}}}\exp(-{r}_{\text{OO}}/{\rho_{\text{OO}}^{intra-{\rm {CO}_{3}}}})$, and a carbon-oxygen term: $V_{\text{CO}}^{intra-{\rm {CO}_{3}}}(r_{\text{CO}})={k_{\text{CO}}}(r_{\text{CO}}-{r_{0,\text{CO}}})^{2}/2+{q_{\text{O}}q_{\text{C}}}/{4\pi\epsilon_{0}r_{\text{CO}}}$, with a force constant ${k_{\text{CO}}}=6118.17$~kJ/mol and an harmonic equilibrium distance $r_{0,\text{CO}}=$1.30 \AA\, adjusted so as to obtain a mean C--O distance of 1.29~\AA, as found by X-ray diffraction measurements and AIMD simulations \cite{Antao2010,Vuilleumier2014,moi2018}. Two ions $i$ and $j$, with $i,j$= Li, Na, K, Ca, Mg, O and C (with O and C not belonging to a same carbonate group) interact through the following intermolecular potential: $V_{ij}^{inter}({r}_{ij})={A_{ij}}\exp(-{r}_{ij}/{\rho_{ij}})-{C_{ij}}/{{r}_{ij}^{6}}+{q_{i}q_{j}}/{4\pi\epsilon_{0}{r}_{ij}} $. Table~\ref{tffsh} recaps the previously published parameters for the Li--Na--K melts\cite{moi2018}, and provides the parameters for Mg and Ca. The latter parameters were adjusted so as to reproduce at best (i) the density measurements of crystalline \ce{MgCO3}, \ce{CaCO3} and \ce{CaMg(CO3)2} at 300~K and 1~bar, the density and the compressibility of the \ce{K2CO3}--\ce{CaCO3} melt at 1~bar and 1100--1200~K, and (ii) the microscopic structure, in the form of atomic pair distribution functions (PDFs), issued from AIMD simulations of the \ce{CaCO3}, \ce{MgCO3} and \ce{CaMg(CO3)2} melts.\\  
At variance with alkali carbonates, alkaline-earth carbonates do not form stable melts under atmospheric pressure \cite{Irving1975,Suito2001,Buob2006,Spivak2012,Shat2015} and no reliable measurement of their density under high pressure has been published yet. However molten calcium carbonate is stable at 1~bar in mixtures with alkalis. In particular, the density of the equimolar \ce{K2CO3}--\ce{CaCO3} mixture has been measured by \citet{Liu2003} using an advanced double-bob Archimedean method overcoming artifacts due to surface tension and to the mass of the bob, and which reproduced with a great accuracy the density of many liquid standards.\cite{Janz1988} As shown on Fig.~\ref{falphakc} the agreement between the density measurements of \citet{Liu2003} on the K--Ca melt and our MD calculation is excellent. Note that Dobson \emph{et al.}\cite{Dobson1996} also measured the density of this mixture but the values they report are by $\sim$~5\% lower than the ones from Liu and Lange.\cite{Liu2003} Additionally, \citet{Liu2003} have suggested that the density of molten carbonates virtually has an ideal behavior regarding composition \cite{Liu2003,moi2018}:  
\begin{equation}
n_{mix}=\cfrac{\sum_{i}x_{i}{M}_{i}}{\sum_{i}x_{i}\bar{V}_{i}}\enspace,
\label{emelid}
\end{equation}
where $x_{i}$ is the molar fraction of species $i$ of molar mass ${M}_{i}$ and molar volume $\bar{V}_{i}$. Although this approximation is fairly good for purely alkali mixtures, it seems less accurate for mixtures containing both alkali and alkaline-earth cations (see Fig.~\ref{falphakc} and next section). Still this mixing rule allows to extrapolate the density of an hypothetical \ce{CaCO3} liquid at 1~bar.\cite{Liu2003} In a MD simulation such a liquid is metastable, meaning that it is possible to evaluate its density straightforwardly, at variance with the experiments. Satisfactorily the density calculated by MD (e.g. 2.44~g/cm$^{3}$ at 1100~K) is fairly compatible with the extrapolation proposed by \citet{Liu2003} (2.49~g/cm$^{3}$ at 1100~K). \\
For \ce{MgCO3} no measurement of density, not even in a mixture, has been published yet. But based on the density of various carbonates (in the \ce{Li2CO3}--\ce{Na2CO3}--\ce{K2CO3}--\ce{Rb2CO3}--\ce{Cs2CO3}-\ce{CaCO3}--\ce{SrCO3}--\ce{BaCO3} system) and considering how this property evolve as a function of the cationic radius, \citet{Hurt2019} suggested that an hypothetical \ce{MgCO3} melt at 1~bar and 1100~K would have a density of 2.45 g/cm$^3$. This guess is within 2\% of the density we obtained by simulating the metastable liquid (2.49~g/cm$^3$). So by introducing in Eq.~(\ref{emelid}) the density of molten calcite (from \citet{Liu2003}) and that of magnesite (from \citet{Hurt2019}) at the chosen reference state of 1~bar and 1100~K gives a density of 2.47~g/cm$^3$ for the \ce{CaMg(CO3)2} melt, in close match with the value of 2.49~g/cm$^3$ as obtained by MD (Table~\ref{tid}).\\
As the densities of molten magnesite and calcite proposed by \citet{Hurt2019} are merely an approximate estimation, we chose to test the accuracy of our FF by calculating the density of the crystalline phases that are well constrained even under high $T-P$. \cite{Markgraf1985,Fiquet1994,Zhang1997,Ross1997,Redfern1993} We chose to focus on the calcite structure (rhombohedral) as it is a common polymorph to \ce{MgCO3}, \ce{CaCO3} and \ce{CaMg(CO3)2}.\cite{Hazen20132}. For magnesite at 300~K up to 4~GPa (Fig.~\ref{fmagcr}), the calculated density is in excellent agreement with the experimental values, well within the scattered data of the experimental literature. Moreover MD simulations reproduce very well the compressibility: $K_{T}$ = 125~GPa as compared to 117~GPa according to Ross.\cite{Ross1997} We believe this should lead to a reliable evaluation of densities for the liquid phase. We also calculated the density of calcite (\ce{CaCO3}) at room conditions and obtained 2.63~g/cm$^{3}$, against 2.71 and 2.72~g/cm$^{3}$ according to Redfern \emph{et al.}\cite{Redfern1993} and Fiquet \emph{et al.}\cite{Fiquet1994}, respectively, that is within $\sim$3\%. For comparison, at the same conditions AIMD yielded 2.67~g/cm$^{3}$.\cite{Vuilleumier2014} For dolomite, the agreement between MD ($n=2.81$~g/cm$^3$) and the X-ray diffraction data ($n=2.84$~g/cm$^3$)\cite{Fiquet1994} is even better with a deviation of $-1$\%. The densities were also calculated by anisotropic relaxation of the crystal structure (MD simulations in the $NST$ ensemble). The accuracy of the MD-calculated values deteriorates slightly for magnesite ($n=3.04$~g/cm$^3$ instead of 3.01, see Fig.~\ref{fmagcr}) but improves for dolomite ($n=2.84$~g/cm$^3$) and remains unchanged for calcite.  Tables~S1 and S2 in the supplementary material detail the lattice parameters and the calculated density values up to 4~GPa.

\section{Thermodynamics}
\label{sthemo}
\subsection{Equation of State}\label{seos} 
\begin{figure*}
\resizebox*{!}{0.35\textheight}{\subfloat[\label{feosm}]{\includegraphics{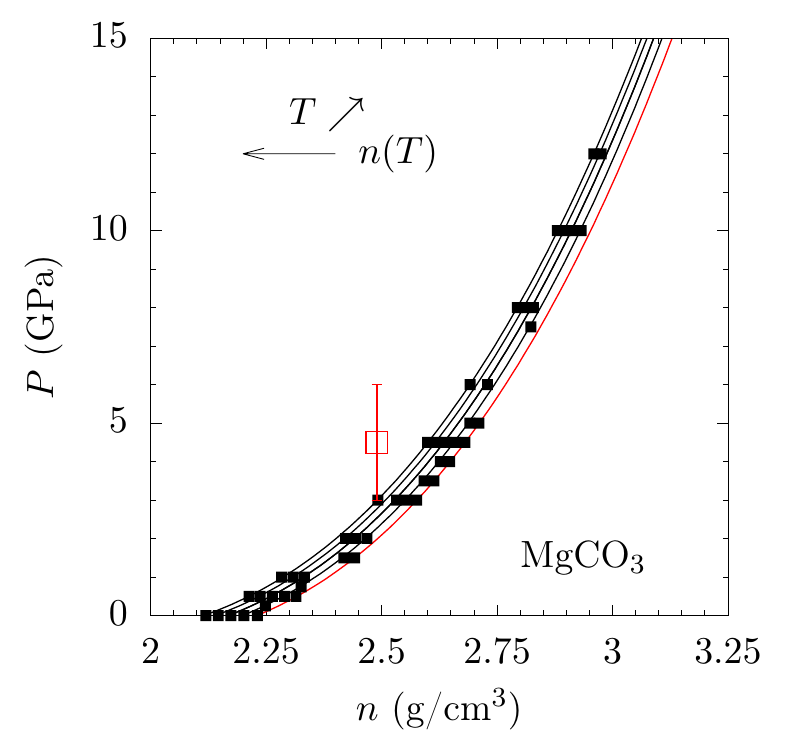}}}
\resizebox*{!}{0.35\textheight}{\subfloat[\label{feosc}]{\includegraphics{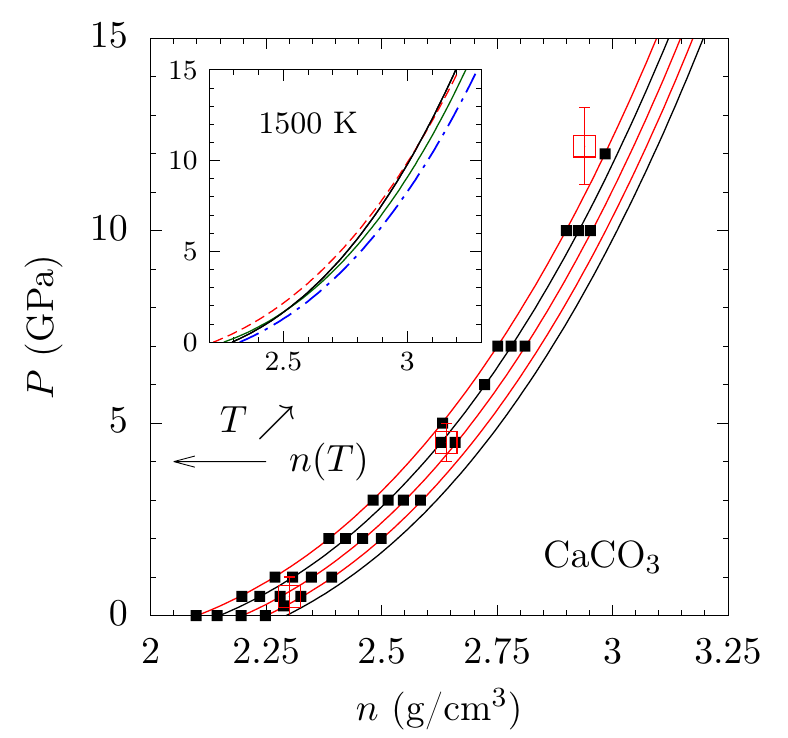}}}\\
\resizebox*{!}{0.35\textheight}{\subfloat[\label{feosmc}]{\includegraphics{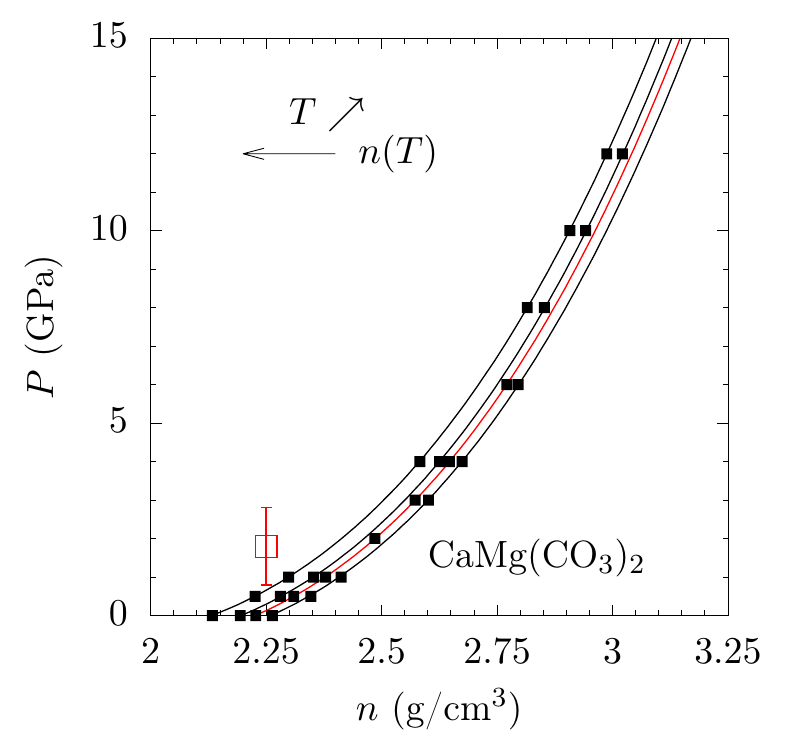}}}
\resizebox*{!}{0.35\textheight}{\subfloat[\label{feosn}]{\includegraphics{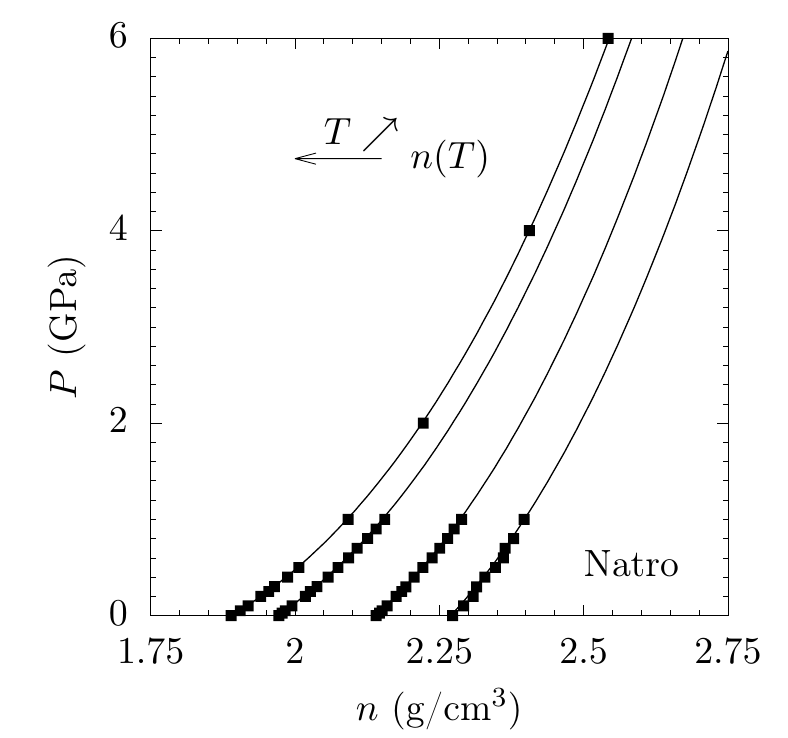}}}
\caption{MD (black) and AIMD (red) calculated density-pressure points. Compression isotherms (black or red curves) are interpolated from the MD points by the Birch-Murnaghan equation of state (BMEoS). The temperature of the MD isotherms are: \{1873~K, 1973~K, 2073~K, 2173~K, 2273~K\} for $\ce{MgCO3}$ (a), \{1500~K, 1623~K, 1773~K, 1923~K, 2073~K\} for $\ce{CaCO3}$ (b), \{1653~K, 1773~K, 1873~K, 2073~K\} for $\ce{CaMg(CO3)2}$ (c) and \{823~K, 1073~K, 1423~K, 1600~K\} for the natrocarbonatite (Natro) Na$_{1.1}$K$_{0.18}$Ca$_{0.36}$CO$_3$ (d). Red squares are AIMD points: 1873~K for \ce{MgCO3}, 1623~K, 1773~K and 2073~K for \ce{CaCO3} and 1773~K for \ce{CaMg(CO3)2}. The compression curved at the corresponding temperatures are plotted in red to facilitate the comparison between MD and AIMD results.\\
For \ce{CaCO3} a BMEoS isotherm issued from the experimental thermodynamics data and fusion curve analysis at 1500~K is also plotted in the inset (dashed dotted blue curve)\cite{Liu2003,OLeary2009,OLeary2015} along with the isotherms calculated from AIMD by~\citet{Zhang2015} (dashed red curve) and from MD by \citet{Vuilleumier2014} (green) at the same temperature (the present study is the black curve).}
\label{feosall}
\end{figure*}

Although there exist no measured data of the density of calco-magnesian carbonate melts, an estimation can be made by assuming that carbonate
liquids mix linearly with respect to carbonate components (ideal mixing assumption). As discussed above, this allowed \citet{Liu2003} to estimate the density of molten \ce{CaCO3} at 1-bar. Now, to infer the high pressure properties of these liquids, constraining the 1-bar compressibility is also needed. Then the density of molten carbonates can be accurately modeled by the third-order Birch Murnaghan equation of state (BMEoS) \cite{Birch1947}: 
\begin{equation}
\begin{split}P= & \cfrac{3}{2}K^{0}(T)\Bigg(\Big(\cfrac{n}{n_{T}^{0}}\Big)^{7/3}-\Big(\cfrac{n}{n_{T}^{0}}\Big)^{5/3}\Bigg)\\
&\times\left(1-\cfrac{3}{4}(4-K^{\prime0})\Bigg(\Big(\cfrac{n}{n_{T}^{0}}\Big)^{2/3}-1\Bigg)\right)\enspace,\end{split}
\label{eBMEOS}
\end{equation}
where $n_{T}^{0}$ is the atmospheric density at temperature $T$, $K_{T}^{0}$ the bulk modulus (inverse of the compressibility) and $K^{\prime0}$ its pressure derivative at 1~bar (note that it is a constant under the thermodynamic conditions of this study). After fitting our MD data, we propose an equation of state for \ce{MgCO3}, \ce{CaMg(CO3)2} and for a natrocarbonatite (Na$_{1.10}$K$_{0.18}$Ca$_{0.36}$CO$_3$) modeling the carbonatitic lavas from Ol Doinyo Lengai (see Table~\ref{teo}).  
O'Leary \emph{et al.}\cite{OLeary2015} measured the compressibility of the alkali end-members and several mixtures of the system \ce{CaCO3}--\ce{Na2CO3}--\ce{K2CO3}--\ce{Li2CO3}, including mixtures containing various ratios of \ce{CaCO3}. Then they extrapolated the 1~bar compressibility of molten \ce{CaCO3}, that is a metastable liquid at this pressure, by using an ideal mixing rule: 
\begin{equation}
K_{mix}=\cfrac{\sum_{i}x_{i}\bar{V}_{i}}{\sum_{i}x_{i}\bar{V}_{i}/K_{i}}\enspace .\label{emelidbeta}
\end{equation}
The comparison of our MD results with their study is good as shown in Table~\ref{tid}. In particular our results are in a better agreement than when using the FFs developed by \citet{Vuilleumier2014} and by Hurt and Wolf\cite{Hurt2018}. Then O'Leary \emph{et al.}\cite{OLeary2009} used fusion curve analysis to constrain $K^{\prime0}=7\pm1$, still for \ce{CaCO3}. Note that this value is closer to the one produced using our FF ($K^{\prime0}=8.2$, as in \cite{Vuilleumier2014}) than using the one of \citet{Hurt2018} ($K^{\prime0}=10.3$). Inserting this value, as well as the 1-bar density from \citet{Liu2003} and the 1-bar compressibility from O'Leary \emph{et al.}\cite{OLeary2009} in Eq.~(\ref{eBMEOS}) we built an experimental compression curve for \ce{CaCO3} at 1500~K (see the inset in Fig.~\ref{feosc}). This estimation is
very close to the one issued from our MD simulations.\\ 
Based on AIMD simulations, \citet{Zhang2015} proposed an equation of state for \ce{CaCO3}. However, because of the LDA approximation they used to compute exchange-correlation energies, an approximation which tends to overestimate the density (at a given $P)$, the authors have applied a rescaling method. The compression curve given by their equation of state at 1500~K is plotted in the inset of Fig.~\ref{feosc} for \ce{CaCO3}. It fits our MD compression curve quite well although there remains a slight deviation at low pressure. Moreover, the compressibility provided by our model is slightly lower than the one from \citet{Zhang2015}. However that may be, at 1500~K, we get $K_{1500K}^{0}=13.7$~GPa, just like O'Leary \emph{et al.}\cite{OLeary2015}, whereas \citet{Zhang2015} give a value of 11.9~GPa. As for the AIMD calculations we performed (this study and \citet{Vuilleumier2014}), they are based on the GGA approximation and include a (semi-empirical) correction for dispersion forces. The introduction of dispersion interactions enhances interionic cohesion in the liquid, but maybe not sufficiently, a feature which could account for the slight remaining difference between MD and AIMD ($P_{MD} < P_{AIMD}$ in general, see Fig.~\ref{feosall}).\\
If the number density, $\rho$, is considered (rather than mass density $n$), it goes as follow at 3~GPa and 1873~K: $\rho_{\ce{MgCO3}}=$ 40.1~mol/L $>$ $\rho_{\ce{CaCO3}}=$ 25.2~mol/L, which is consistent with the observation made for alkali melts\cite{moi2018}, that the number density is negatively correlated to the cation size. However, the bulk modulus is the same in magnesite and calcite ($K_{\ce{MgCO}_{3}}^{0}=K_{\ce{CaCO}_{3}}^{0}=11.0$~GPa), at variance with alkali melts where $K^{0}$ increases with increasing cation radius. This suggests that the compressibility of calco-magnesian carbonate melts is not similar to that of a hard sphere system and instead is dominated by the coulombic repulsion between divalent cations. If we consider the compressibility at 1~bar and 1100~K of the metastable \ce{MgCO3} and \ce{CaCO3} melts ($K_{\ce{MgCO}_{3}}^{0}=14.0$~GPa and $K_{\ce{CaCO}}^{0}=17.6$~GPa), it is greater by a factor~$\sim$ 2 than that of alkali carbonates at the same $T-P$ conditions. However, the compressibilities calculated in corresponding states (i.e. near the melting point, 1823~K and 3~GPa for \ce{MgCO3}, and 1623~K and 1~GPa for \ce{CaCO3}) are very similar ($K_{1823\,\ce{K}}^{3\,\ce{GPa}}=7.5$~GPa for \ce{MgCO3} and  $K_{1623\,\text{K}}^{1\,\text{GPa}}=6.4$~GPa for \ce{CaCO3}) for both alkali and alkaline-earth carbonates.\\
Knowing the equations of state for \ce{MgCO3}, \ce{CaCO3}, \ce{Na2CO3}, \ce{K2CO3} and \ce{Li2CO3} and some mixtures (dolomite and natrocarbonatite) it is worth checking how these latter melts behave regarding ideality (Table~\ref{tid}). For that we consider, for instance, the density of molten dolomite at 1~bar and at  3~GPa, let's say at 1873~K. The comparison between the raw MD data and the ideal mixing rules (Eqs.~(\ref{emelid}) and (\ref{emelidbeta})) using the data of Table~\ref{teo} gives: $n_{T}^{0}=n_{T}^{0,\,mix}=2.19$~g/cm$^{3}$ and $K_{T}^{0}=K_{T}^{0,\,mix}=11.0$~GPa,  and $n_{T}^{3\,\text{GPa}}=n_{T}^{3\,\text{GPa},\,mix}=2.55$~g/cm$^{3}$. Proceeding the same way for the natrocarbonatite at 1~bar and 1073~K we get $n_{T}^{0}=2.11$~g/cm$^{3}$ and $n_{T}^{0,mix}=2.13$~g/cm$^{3}$, and $K_{T}^{0}=11.4$~GPa and $K_{T}^{0,\,mix}=10.4$~GPa. Hence it comes out that dolomite is an ideal mixture on a certain $P$-domain, whereas natrocarbonatite slightly deviates from ideality. Most likely the latter finding can be explained by the fact that the natrocarbonatite includes both uni- and divalent cations. In practice, as the deviation from ideality seems to be fairly small, the EoS of the end-members of the Mg--Ca--Li--Na--K system can be used to estimate with confidence the density of any mixture. 
\begin{table*}[th]
\begin{tabular}{|c|c|c|c|c|}
\cline{2-5} 
\multicolumn{1}{c|}{} & \ce{MgCO3}  & \ce{CaCO3}  & \ce{CaMg(CO3)2} & Natro\tabularnewline
\hline 
$T$ (K) & $1873-2073$ & $1100-2073$ & $1653-2073$ & $823-1600$\tabularnewline
\hline 
$P$ ~(GPa)  & $0-15$ & $0-15$ & $0-15$ & $0-6$\tabularnewline
\hline 
$T_{ref}$ (K) & 1873 & 1623 & 1653 & 1073\tabularnewline
\hline 
$n_{ref}$ ($\text{g/cm}^{3}$)  & 2.23 & 2.25 & 2.26 & 2.14\tabularnewline
\hline 
$\alpha_{0}$ (K$^{-1}$)  & $-1.50 \times 10^{-4}$  & $-1.54 \times 10^{-4}$  & $-1.23 \times 10^{-4}$  & $-2.36 \times 10^{-4}$ \tabularnewline
\hline 
$\alpha_{1}$ (K$^{-2}$)  & $8.22 \times 10^{-8}$  & $0.49 \times 10^{-8}$  & $-7.65 \times 10^{-8}$  & $-0.20 \times 10^{-8}$ \tabularnewline
\hline 
$K_{ref}^{0}$(GPa)  & 10.98  & 12.74 & 12.13 & 11.72\tabularnewline
\hline 
$b_{1}$ (K$^{-1}$)  & $5.0 \times 10^{-4}$  & $6.1 \times 10^{-4}$  & $3.7 \times 10^{-4}$  & $10.2 \times 10^{-4}$ \tabularnewline
\hline 
$b_{2}$ (K$^{-2}$)  & $2.0 \times 10^{-7}$  & $1.5 \times 10^{-7}$  & $5.0 \times 10^{-7}$  & $7.7\times 10^{-7}$ \tabularnewline
\hline 
$K^{\prime0}$ & 9.5 & 7.7 & 8.5 & 8.0\tabularnewline
\hline 
\end{tabular}
\caption{Parameters of the third-order Birch-Murnaghan equation of state and $T-P$ domain of validity. Natro refers to the natrocarbonatite melt of composition Na$_{1.1}$K$_{0.18}$Ca$_{0.36}$CO$_3$ (see text).\\
The reference temperature $T_{ref}$ is defined for numerical purposes and set to the melting temperature for each composition, $n_{ref}^{0}$ and $K_{ref}^{0}$ are the density and bulk modulus at 1-bar and $T_{ref}$. Other empirical parameters ($\alpha_{0}$, $\alpha_{1}$, $b_{1}$ and $b_{2}$) are defined by $n_{T}^{0}=n_{ref}^{0}e^{\int_{T_{ref}}^{T}-(\alpha_{0}+\alpha_{1}T)\,\mathrm{d}T}$ and $K_{T}^{0}={K_{ref}^{0}}/\big(1+b_{1}(T-T_{ref})+b_{2}(T- T_{ref})^{2}\big)$. See \citet{moi2018} for details on the fitting method.}
\label{teo} 
\end{table*}

\subsection{Structure}
\label{sstruct}
\begin{figure*}
\subfloat[\label{fstructa}]{{\includegraphics{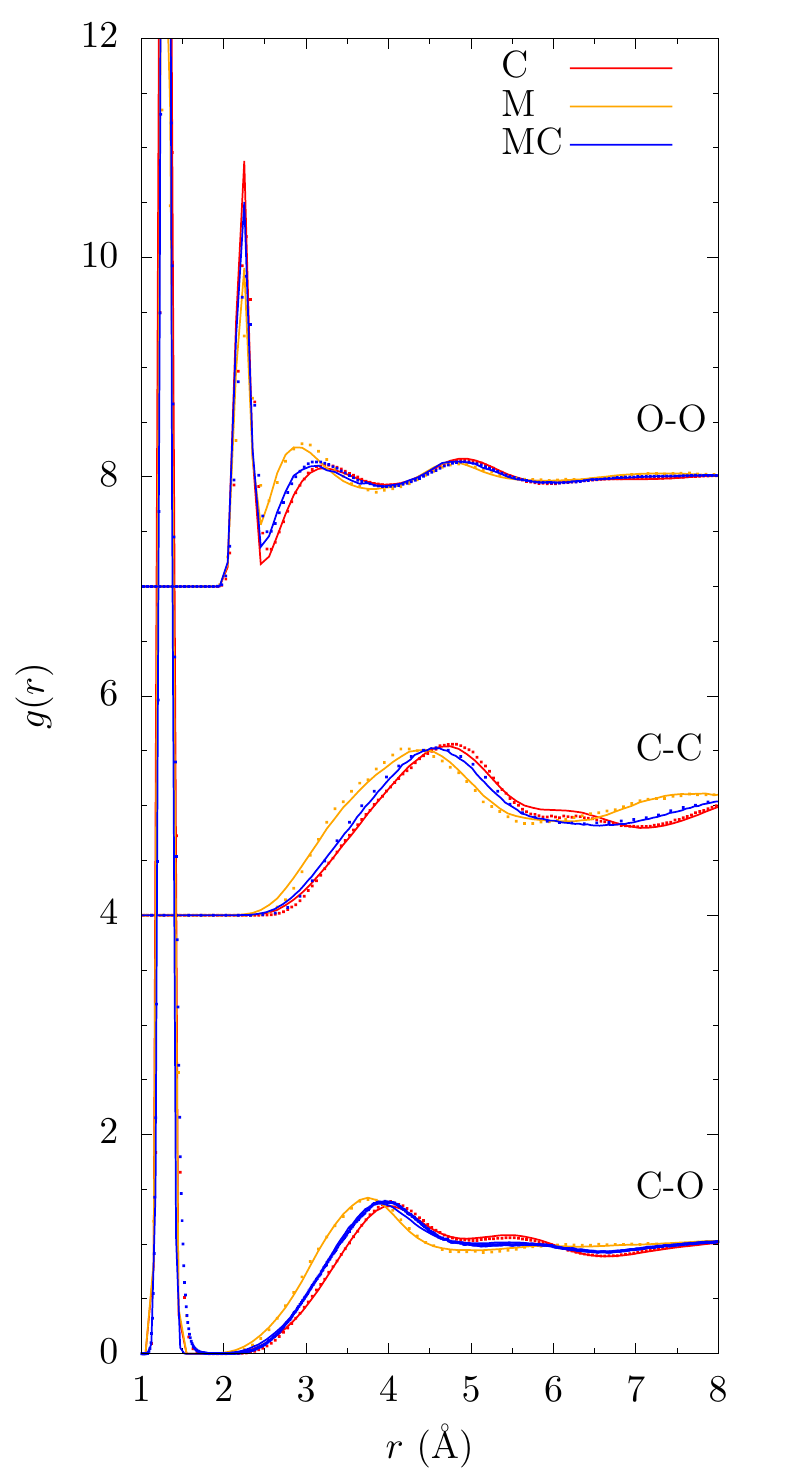}}} 
\subfloat[\label{fstructb}]{{\includegraphics{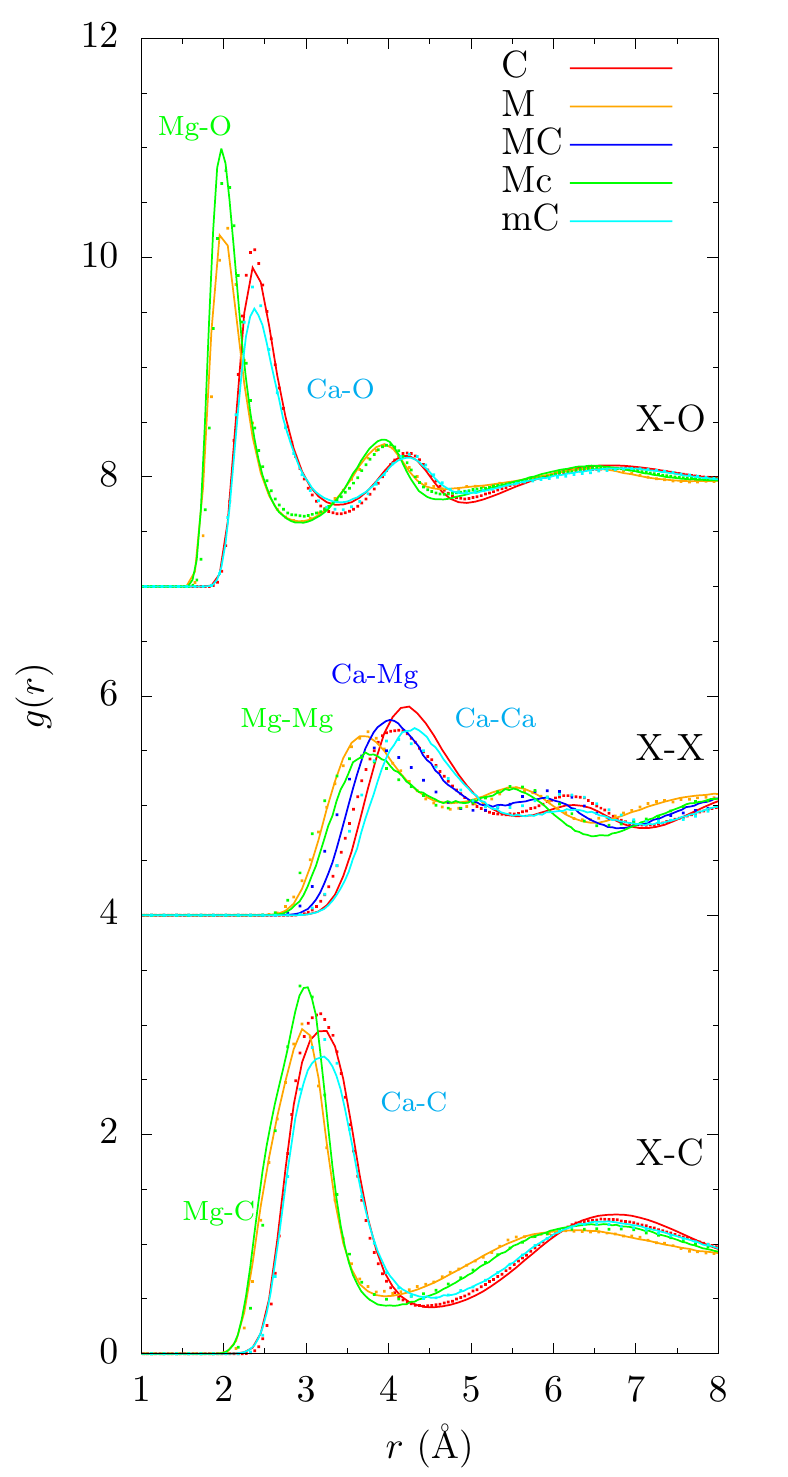}}}
\caption{MD pair distribution functions O$-$O, C$-$C, C$-$O, X$-$O, X$-$X and X$-$C where C and O are carbon and oxygen atoms of the \ce{CO3} ions and X$=$Ca or Mg (see color code) in \ce{MgCO3} (M) at 1873~K and 2.49~g/cm$^3$, \ce{CaCO3} (C) at 1623~K and 2.30~g/cm$^3$ and \ce{CaMg(CO3)2} (MC) at 1773~K and 2.25~g/cm$^3$.  The labels "Mc" and "mC" refer to the \ce{MgCO3} and to the \ce{CaCO3} components of \ce{CaMg(CO3)2}, respectively. The PDFs calculated from the AIMD simulations (this study and Ref.\cite{Vuilleumier2014}) at the same temperature and density are represented by the dotted lines (plain lines for MD). To facilitate visualization, the different PDFs were shifted vertically.}
\label{fstruct} 
\end{figure*}
\sloppy
In the $T-P$ range of this study ($T\leq 2013$~K and $P\leq 15$~GPa), the stability of the internal structure of the carbonate ion has been pointed out by neutron diffraction measurements\cite{Kohara1998}, \emph{in situ} X-ray diffraction measurements\cite{Hudspeth2018} and by AIMD simulations (Ref. \onlinecite{Vuilleumier2014,moi2018} and this study). Moreover the carbonate anion, unlike the \ce{SiO4} units in silicates, cannot share a covalent bond with other atoms. Therefore carbonate melts display a structure contrasting with that of most geological melts (silicate) that are always polymerized to a certain degree. This is why carbonate melts share with molten salts (e.g. NaCl) a high ionic diffusivity and a low viscosity.\\
Until now, among the alkaline-earth carbonates, only the structure of calcite melt has been investigated by classical molecular dynamics simulations\cite{Genge1995,Vuilleumier2014,Hurt2018} and by AIMD simulations.\cite{Vuilleumier2014,Du2018} Fig.~\ref{fstruct} shows the PDFs for \ce{MgCO3} and \ce{CaCO3} at corresponding states (i.e. near melting point: $T=1873$~K, $P\sim 2$~GPa and $n=2.49$~g/cm$^{3}$ for \ce{MgCO3}; $T=1623$~K, $P\sim 0.5$~GPa and $n=2.30$~g/cm$^{3}$ for \ce{CaCO3}). Note that, as expected, they are in very good agreement with the PDFs issued from AIMD (compare the plain lines with the dotted lines). The first peak of $g_{\text{CO}}(r)$ at $r_{\text{CO}}=$ 1.29~\AA\, corresponds to the three oxygen atoms bonded to a same carbon atom in a carbonate unit. Between 1.7~\AA\, and 2.1~\AA\, $g_{\text{CO}}(r)$ = 0 because no C--O bond dissociation occurs during the simulation. On the other hand, the first peak of $g_{\text{OO}}$ represents the O--O intramolecular distance ($r_{\ce{OO}}=2.22$~\AA) in a carbonate ion. The first minimum of $g_{\ce{OO}}$ is non-zero, meaning that the distance between two oxygen atoms of two different \ce{CO3}$^{2-}$ units may be as short as the O--O intramolecular distance. Concerning the anion-anion correlations, each \ce{CO3}$^{2-}$ unit is surrounded by $12-15$ other carbonate ions at $r_{\ce{CC}}\simeq 4-6$~\AA\, on average. The first C--C peak on Fig.~\ref{fstructa} is broad with a shouldering on its low-$r$ flank ($3.5-4$~\AA), that is especially noticeable for \ce{MgCO3}, and a small bump at $\sim~5-7$~\AA\, in the region of the first minimum, this time more pronounced in \ce{CaCO3}. Each  cation (Mg$^{2+}$ or Ca$^{2+}$) ion is surrounded by 6 carbonate groups (against $4.5-4.8$ for alkali), at a mean cation-carbon distance of $3-4$~\AA\, in \ce{MgCO3} and $3.15-4.5$~\AA\, in \ce{CaCO3} (note that these distances are the same in \ce{CaMg(CO3)2} at 1773~K and 2.25~g/cm$^3$). As for the ratio of the coordination numbers $N_{c}^{\text{X-O}}$/$N_{c}^{\text{X-C}}$ (where X = Mg or Ca), indicative of the orientation of the carbonate ions around the cations, it is 1.1 for Mg and 1.3 for Ca (as compared to $1.3-1.6$ for alkali cations). In comparison to pair distribution functions for alkali carbonate melts, the cation-anion PDF $g_{\ce{XC}}(r)$ for alkaline-earth carbonates are more simple as they show no shoulder on the first peak (see Fig.\ref{fstructb} and compare with Fig.~2 in \citet{moi2018}). Moreover, the second peak of $g_{\ce{XC}}(r)$ is broader, especially for Mg. It can also be noted that the shape of the PDFs is less sensitive to the size of the cation. Looking at the cation-cation PDFs, the only difference between Mg and Ca is the amplitude, greater for the Ca--Ca pair with a coordination number around $\sim 10$ instead of $\sim 7$ for Mg--Mg.  It is noteworthy that the PDFs of \ce{CaMg(CO3)2} are intermediate between those of \ce{CaCO3} and \ce{MgCO3}. In fact, most structural features observed in molten calcite and magnesite are similar and can be interpreted as  a simple homothetic transformation upon volumetric change from \ce{CaCO3} to \ce{MgCO3}. Thus the pair distribution functions X--C, X--O and X--X in \ce{CaMg(CO3)2} are almost identical to the corresponding ones observed in \ce{CaCO3} and \ce{MgCO3} (Fig.~\ref{fstructb}) and this remains true even at high pressures (Figs.~S1, S2 and S3 in the supplementary material). Under pressure (and up to 12~GPa), the PDFs shift progressively towards lower distances, reflecting the melt compaction (see Figs.~S1, S2 and S3). Moreover the average number of \ce{CO3}$^{2-}$ anions around cations increases slightly under pressure, from 6 to 7. In contrast, the coordination number of \ce{CO3}$^{2-}$ around \ce{CO3}$^{2-}$ ($\sim 12-15$) does not evolve with pressure, neither does the $N_{c}^{\text{X-O}}$/$N_{c}^{\text{X-C}}$ ratio.

\section{Transport Properties}
\label{strans}
For \ce{MgCO3}, \ce{CaCO3} and \ce{CaMg(CO3)2}, a series of simulations ($\sim$ 15) was performed at different thermodynamic conditions, with a
duration long enough to reach the diffusive regime ($10-20$~ns). From each run we calculated accurately (see below) the self-diffusion
coefficients $D_{s}$ of each chemical species $s=$~Ca, Mg and CO$_{3}^{2-}$, the electrical conductivity $\sigma$ and the viscosity $\eta$, given
by \cite{AllenTild,Hess2002}:  
\begin{eqnarray}
D_{s} & = & \lim_{t\to\infty}\cfrac{1}{6t}\,\cfrac{1}{N_{s}}\sum_{i=1}^{N_{s}}\Big\langle\vert\mathbf{r}_{i}(t)-\mathbf{r}_{i}(0)\vert^{2}\Big\rangle\enspace,\label{eD}\\
\sigma & = & \lim_{t\to\infty}\cfrac{1}{6t}\,\cfrac{e^{2}}{k_{{\rm B}}TV}\Big\langle\Big\vert\sum_{i=1}^{N}z_{i}\big(\mathbf{r}_{i}(t)-\mathbf{r}_{i}(0)\big)\Big\vert^{2}\Big\rangle\enspace,\label{econd}\\
\eta & = & \lim_{t\to\infty}\cfrac{V}{k_{{\rm B}}T}\int_{0}^{t}{\rm d}\tau\Big\langle\Pi_{\alpha\beta}(\tau)\cdot\Pi_{\alpha\beta}(0)\Big\rangle \enspace,
\end{eqnarray}
where $\Pi_{\alpha\beta}(t)$ refers to the off-diagonal pressure tensor components ($\alpha,\beta=x,y,z$, see references \cite{AllenTild,Hess2002}
for more details), $N$ is the total number of ions in the simulation box of volume $V$, $N_{s}$ the number of ions of species $s$, $k_{{\rm B}}$ the Boltzmann constant and $e$ the elementary charge, $m_{i}$ is the mass of ion $i$, $\mathbf{r}_{i}(t)$ its position, $v_{i\alpha}(t)$
the component $\alpha$ of its velocity, $F_{ij\beta}(t)$ is the component $\beta$ of the force exerted by ion $j$ on ion $i$, separated
by a distance $r_{ij}(t)$ at time $t$. The conduction charge $z_{i}$ is taken as the formal charge, which is usual for simple ionic liquids.
\cite{Adams1977} Note from the equations above that the self-diffusion coefficient $D_{s}$, resulting from an average over the $N_{s}$ ions
of a specific species $s$ (sum $\sum_{i=1}^{N_{s}}$ in Eq.~(\ref{eD})), are calculated with a great accuracy (calculation uncertainty of 1\%
in this study). On the other hand, the electrical conductivity and the viscosity, as collective observables, are trickier to estimate due to the slow convergence of the corresponding time correlation functions. In the following the values we present for these quantities have an error bar within $5-10$\%.\\
The temperature and pressure dependence of the transport coefficients can be modeled by an Arrhenius activation law 
 \begin{eqnarray}
D_{s}(P,T) & = & D_{s}^{0}\,e^{-(E_{a}^{D_{s}}+PV_{a}^{D_{s}})/RT}\enspace,\label{eqDArrh}\\
\sigma(P,T) & = & \sigma^{0}\,e^{-(E_{a}^{\sigma}+PV_{a}^{\sigma})/RT}\enspace,\label{eqSigmaArrh}\\
\eta(P,T) & = & \eta^{0}\,e^{(E_{a}^{\eta}+PV_{a}^{\eta})/RT}\enspace,\label{eqViscoArrh}
\end{eqnarray}
where $E_{a}^{X}$ is the activation energy associated to the physical quantity $X$ and $V_{a}^{X}$ is an activation volume accounting for its pressure dependence.\cite{Vuilleumier2014,Corradini2016,moi2018}. The values of $X^{0}$, $E_{a}^{X}$ and $V_{a}^{X}$ were determined by fitting the molecular dynamics data for the three melts \ce{MgCO3}, \ce{CaCO3} and \ce{CaMg(CO3)2} (see Figs.~\ref{fdiffca}, \ref{fdiffmg}, \ref{fdiffdol}, \ref{fcond}, \ref{fvisco}). They are given in Table \ref{tfitdiff} for $D_{s}$ and in Table~\ref{tfittransp} for $\sigma$ and $\eta$, for the pressure and temperature range mentioned in Table~\ref{teo}. 

\subsection{Diffusion coefficient}\label{sdiff}
\begin{figure}
\subfloat[\label{fdiffcaca}]{\includegraphics{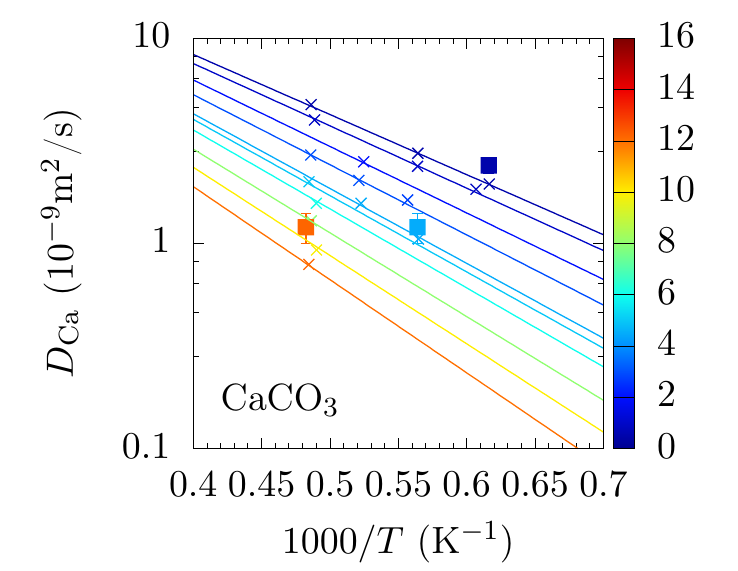}}\\
\subfloat[\label{fdiffcac}]{\includegraphics{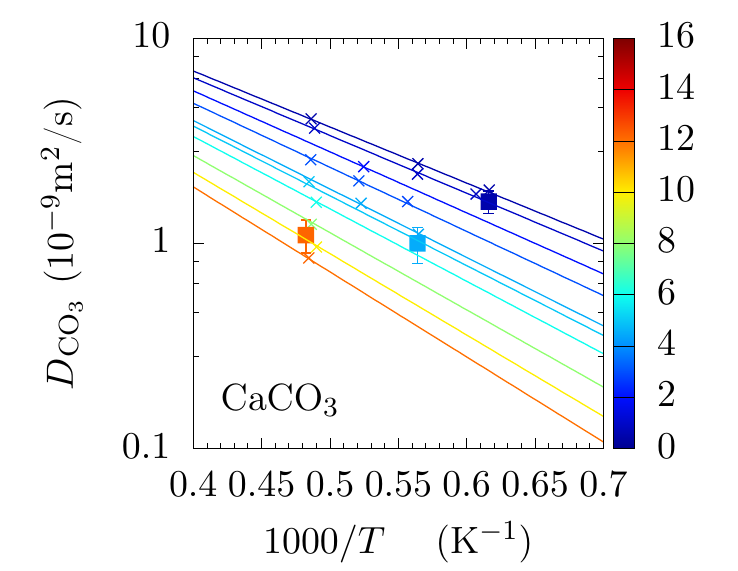}}
\caption{Diffusion coefficients in \ce{CaCO3} calculated by MD (crosses) and isobaric activation laws obtained by fitting all the simulation points. The AIMD calculations of \citet{Vuilleumier2014} are represented by plain squares. The pressures of the isobars are \{0.5, 1, 2, 3, 4.5, 5, 6, 8, 10, 12\} in GPa and are referred to with a color code (vertical scale).}
\label{fdiffca}
\end{figure}

\begin{figure}
\subfloat[\label{fdiffmgm}]{\includegraphics{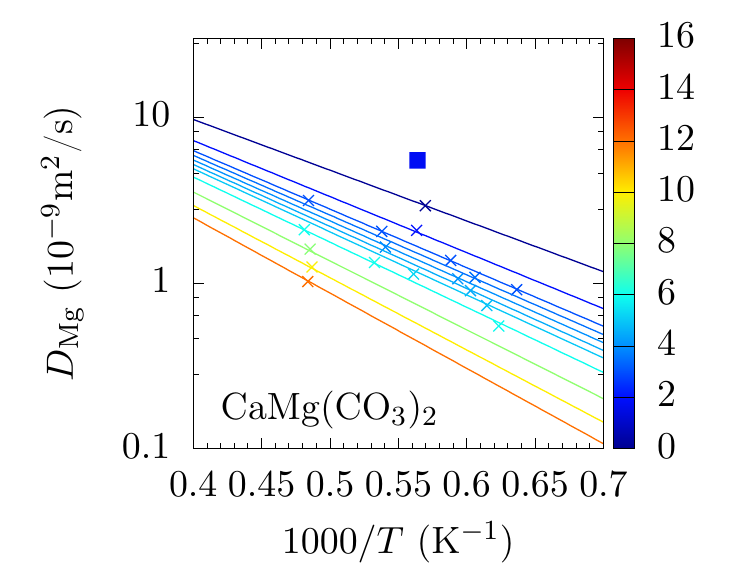}}\\
\subfloat[\label{fidiffmgc}]{\includegraphics{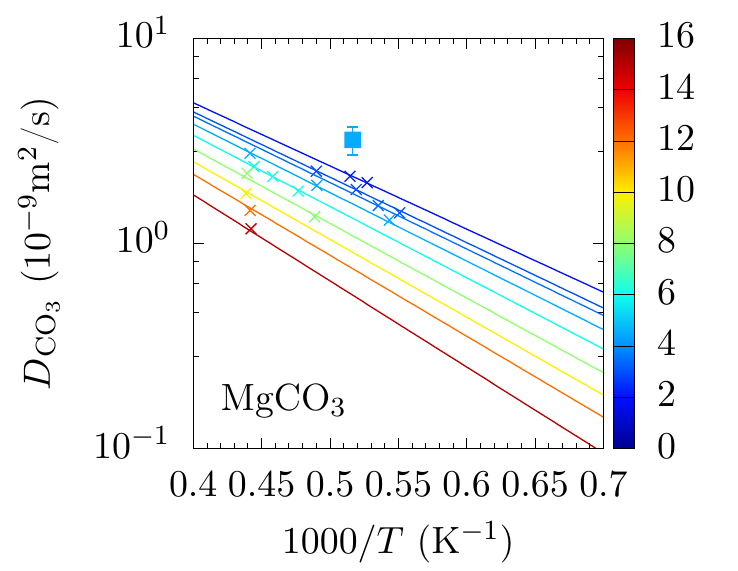}}
\caption{Diffusion coefficients in \ce{MgCO3} calculated by MD (crosses) and isobaric activation laws obtained by fitting all the simulation points. The AIMD calculations (this study) are represented by plain squares. The pressures of the isobars are \{2, 3, 3.5, 4.5, 6, 8, 10, 12, 15\}  in GPa and are referred to with a color code (vertical scale).}
\label{fdiffmg}
\end{figure}

\begin{figure}
\subfloat[\label{fdiffdolm}]{\includegraphics{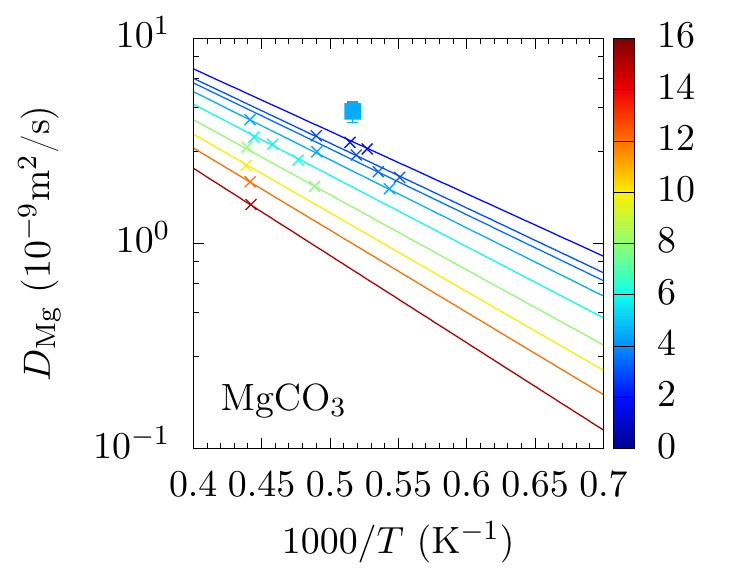}}\\
\subfloat[\label{fdiffdolca}]{\includegraphics{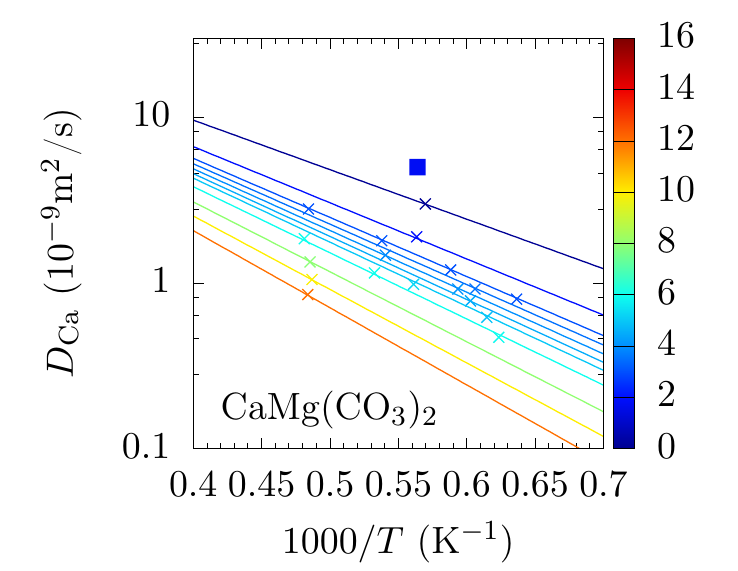}}\\
\subfloat[\label{fdiffdolc}]{\includegraphics{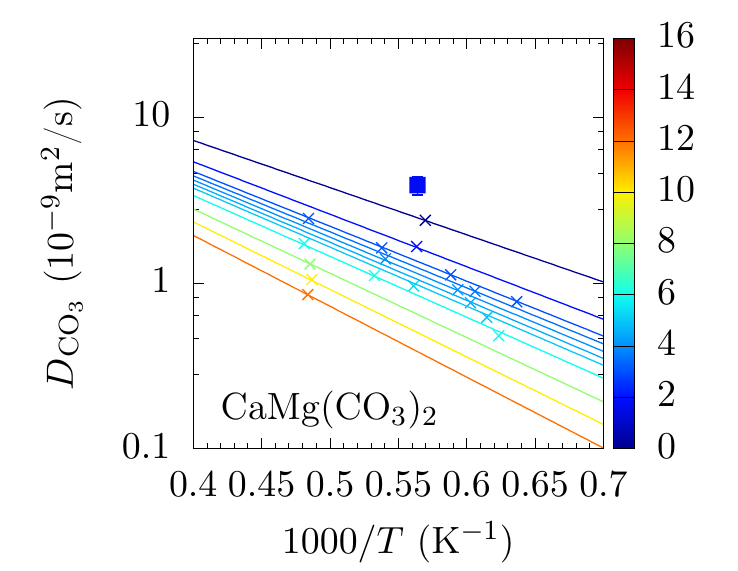}}
\caption{Diffusion coefficients in \ce{CaMg(CO3)2} calculated by MD (crosses) and isobaric activation laws obtained by fitting all the simulation points. The AIMD calculations (this study) are represented by plain squares. The pressures of the isobars are \{0.1, 2, 3, 3.5, 4, 4.5, 5, 6, 8, 10, 12\} in GPa and are referred to with a color code (vertical scale).}
\label{fdiffdol}
\end{figure}

\begingroup
\squeezetable
\begin{table*}
\begin{tabular}{|l|c|c|c|c|}
\hline 
 & \ce{MgCO3}  & \ce{CaCO3}  & \multicolumn{2}{c|}{\ce{CaMg(CO3)2} (left: X=Mg, right: X=Ca)}\tabularnewline
\hline 
${D}_{0,\,\text{X}}$ ($10^{-9}$m$^{2}$/s)  & 117  & 122 & 164 & 150\tabularnewline
\hline 
$E_{a}^{D_{\text{X}}}$ (kJ/mol)   & 54  & 54 & 58 & 57\tabularnewline
\hline 
$V_{a}^{D_{\text{X}}}$ (cm$^{3}$/mol)  & $2.5-0.07 P +0.0016 P^{2}$ & $4.6-0.28 P +0.01006 P^{2}$   & {$3.5-0.13 P +0.0029 P^{2}$ } & $4.6-0.28 P + 0.0100 P^{2}$ \tabularnewline
\hline 
${D}_{0,\,\text{CO}_{3}}$ ($10^{-9}$m$^{2}$/s)
  & 83 & 86 & \multicolumn{2}{c|}{100}\tabularnewline
\hline 
$E_{a}^{D_{\text{CO}_{3}}}$(kJ/mol)  & 54 & 51 & \multicolumn{2}{c|}{54}\tabularnewline
\hline 
$V_{a}^{D_{\text{CO}_{3}}}$(cm$^{3}$/mol)   & $2.6-0.11P+0.0036 P^{2}$  & $3.3-0.08 P +0.0007 P^{2}$  & \multicolumn{2}{c|}{$2.1-0.08 P +0.0085P^{2}$ }\tabularnewline
\hline 
\end{tabular}
\caption{Parameters of the Arrhenius activation law (\ref{eqDArrh}) obtained by the interpolation of all MD simulation points. $P$ is in GPa.}
\label{tfitdiff} 
\end{table*}
\endgroup
 
The diffusivity of calco-magnesian carbonate melts has never been measured and it is only recently that estimations have been provided by the AIMD simulations of \citet{Vuilleumier2014}. According to their calculations the diffusion coefficients of Ca and CO$_3^{2-}$ in \ce{CaCO3} along its melting curve (up to 12~GPa) are of comparable magnitude with the ones in purely alkali melts.\cite{moi2018} The agreement between the data of the present MD study and the AIMD simulations of \citet{Vuilleumier2014} is very good at 0.5 and 4.5~GPa, a bit less at 12~GPa where the values of MD are below those of AIMD (Fig.~\ref{fdiffca}). We have also evaluated the diffusion coefficients from our AIMD simulations of \ce{MgCO3} at 4.5~GPa and \ce{CaMg(CO3)2} at 2~GPa. In the two cases they are greater than the values issued from MD (Figs.~\ref{fdiffmg} and \ref{fdiffdol}). This is consistent with the slight discrepancy between the two models for the equation of state. Indeed at a given $T, P$ point, the densities yielded by MD simulations are systematically greater than the ones yielded by AIMD (Fig.~\ref{feosall}). This means that the free volume of diffusion is smaller in the MD model. By performing a MD simulation of \ce{MgCO3} at the same density as AIMD (1873~K and 2.49~g/cm$^{3}$), and recalculating the coefficients from this run, we get larger values (namely $D\textsubscript{Mg}=3.11$ and $D\textsubscript{CO$_{3}$}=2.13~10^{-9}$~m$^{2}$/s, instead of 1.85 and 1.30), but still below the values from AIMD: $4.4\pm 0.5$ and $3.5 \pm 0.5~10^{-9}$~m$^{2}$/s, respectively. This could indicate that our empirical force field fails to some extent to describe cohesive forces in every detail. However there is no evidence that AIMD sketches them much more accurately.\\
The activation energy ($51-58$~kJ/mol) depends little on the ion species (Ca, Mg or \ce{CO3}). But it is perceptibly higher than that for alkali melts at 1~bar in which the coulombic forces are weaker.\cite{moi2018} The magnitude of the diffusion coefficients are also slightly smaller in Ca-Mg carbonate melts than in their alkali counterparts. Interestingly if we consider the \ce{K2CO3}-\ce{CaCO3} mixture (at 1~bar and $1100-1200$~K) $D$\textsubscript{Ca} is lower than $D$\textsubscript{K} (by a  factor $\sim$2) and very close to $D$\textsubscript{CO$_{3}$}. Moreover, compared to pure \ce{K2CO3}, the presence of the divalent cation decreases $D$\textsubscript{CO$_{3}$} and $D$\textsubscript{K} by a factor 2 and increases their activation energies by $\sim+50\%$.  

\subsection{Electrical Conductivity}
\begin{table*}
\begin{tabular}{|l|c|c|c|}
\hline 
 & \ce{MgCO3}  & \ce{CaCO3}   & \ce{CaMg(CO3)2}  \tabularnewline
\hline 
$\sigma_0$ 	(S/m) & 3842.5 & 2593.7 & 4838.7\tabularnewline
\hline  $E_{a}^{\sigma}$(kJ/mol)	& 38 & 	34 & 42 \tabularnewline
\hline  $V_{a}^{\sigma}$ (cm$^3$/mol) & 1.0+0.076P-0.0036P$^2$ & 2.3+0.048P-0.0041P$^2$ &  2.7+0.090P-0.0043P$^2$ \tabularnewline
\hline  $\eta_0$ (Pa$\cdot$s)	&  3.9$\times$10$^{-4}$  & 2.8$\times$10$^{-4}$ &   1.3$\times$10$^{-4}$	\tabularnewline
\hline  $E_{a}^{\eta}$ (kJ/mol)	& 37 & 39 & 50 \tabularnewline
\hline $V_{a}^{\eta}$	(cm$^3$/mol) & 2.1-0.02P  & 3.6-0.09P & 3.4-0.05P  \tabularnewline
\hline 
\end{tabular}
\caption{Parameters of the Arrhenius activation laws (\ref{eqSigmaArrh}) and (\ref{eqViscoArrh}) for the electrical conductivity and the viscosity, obtained by the interpolation of all MD simulation points. $P$ is in GPa.}
\label{tfittransp}
\end{table*}

\begin{figure}
\includegraphics{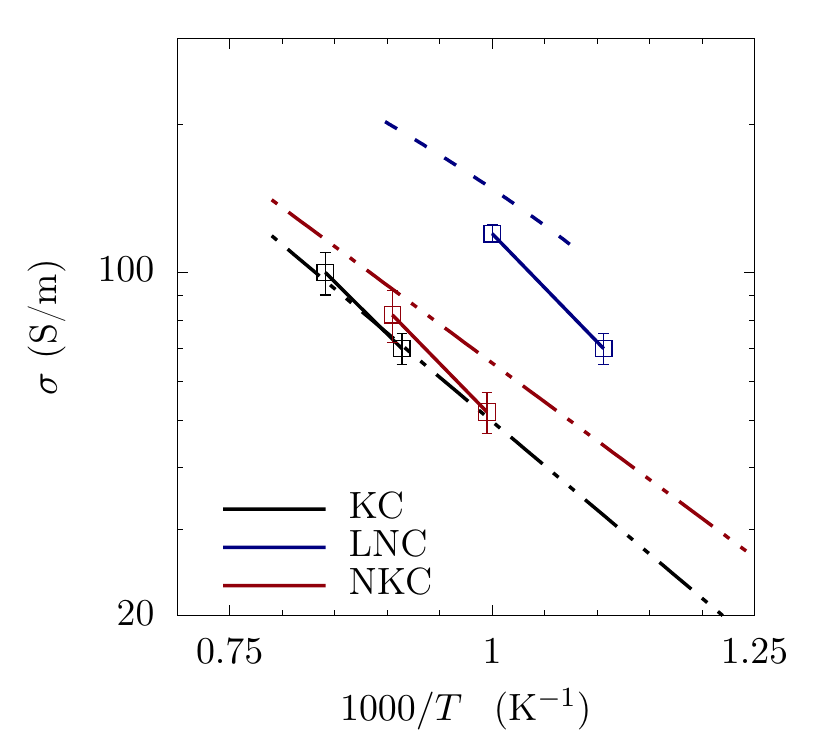}\caption{Electrical conductivity at 1 bar from MD (square and plain lines as
a guide to the eye) and from  the experiments of \citet{Kojima2009} (dashed line) and of \citet{Gaillard2008} (long dashed and double dotted lines) for
\ce{K2Ca(CO3)2} (KC), \ce{Li2Na2Ca(CO3)3} (LNC) and \ce{Na2K2Ca(CO3)3}
(NKC).}
\label{fcompcond} 
\end{figure}

\begin{figure}
\includegraphics{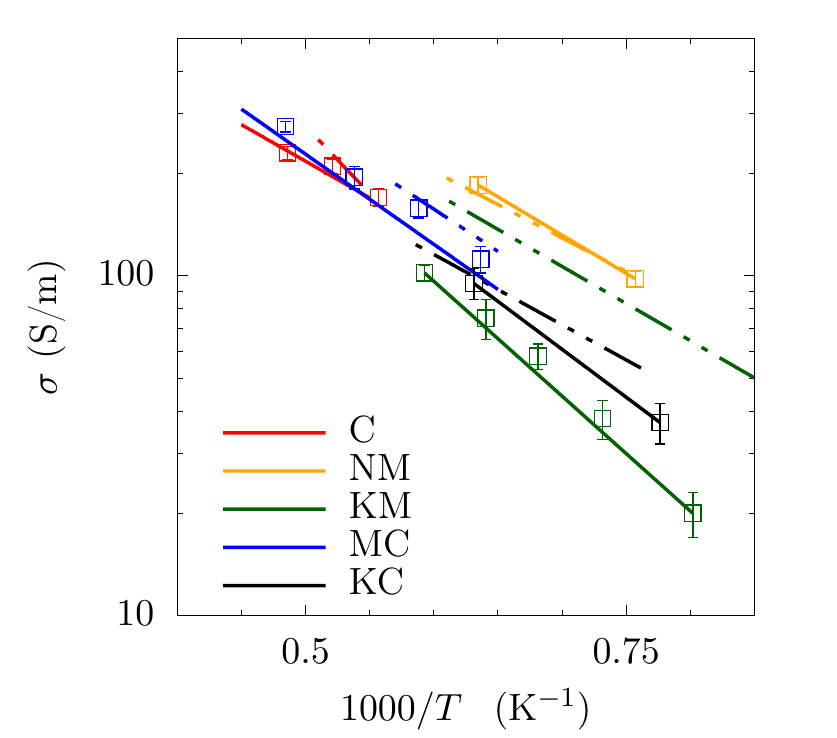}\caption{Electrical conductivity at 3~GPa calculated in MD (square) and measured
by Sifr\'e \emph{et al.}\cite{Sifre2015} (dashed line) for \ce{CaCO3} (C) and several
equimolar mixtures: \ce{Na2Mg(CO3)2} (NM), \ce{K2Mg(CO3)2}
(KM), \ce{CaMg(CO3)2} (MC) and \ce{K2Ca(CO3)2} (KC). Plain
lines are the Arrhenius activation slopes for C and CM (see Table~\ref{tfittransp})
and guides to the eye for NM, KM, NK and KC.}
\label{fcompcond2} 
\end{figure}
\begin{figure}
\subfloat[\label{fcondm}]{\includegraphics{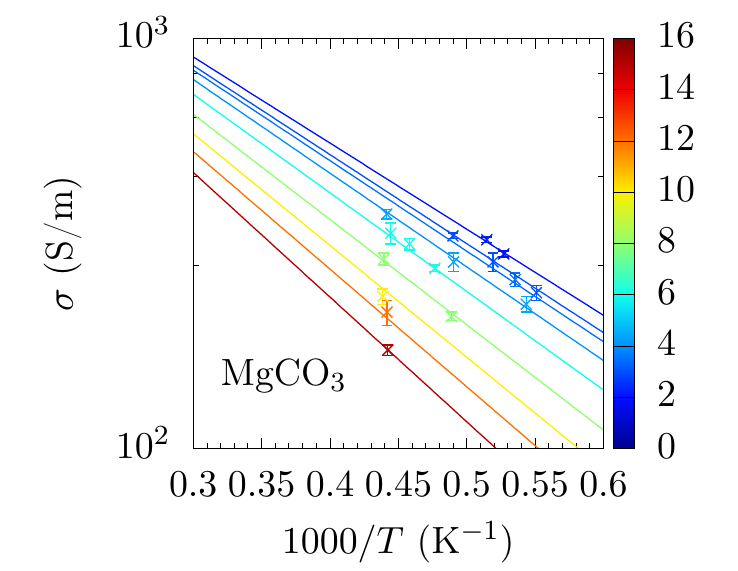}}\\
\subfloat[\label{fcondc}]{\includegraphics{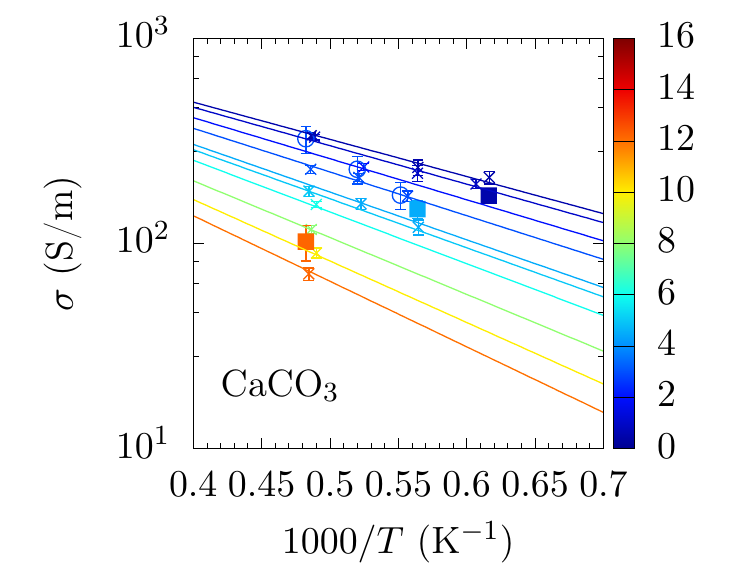}}\\
\subfloat[\label{fcondmc}]{\includegraphics{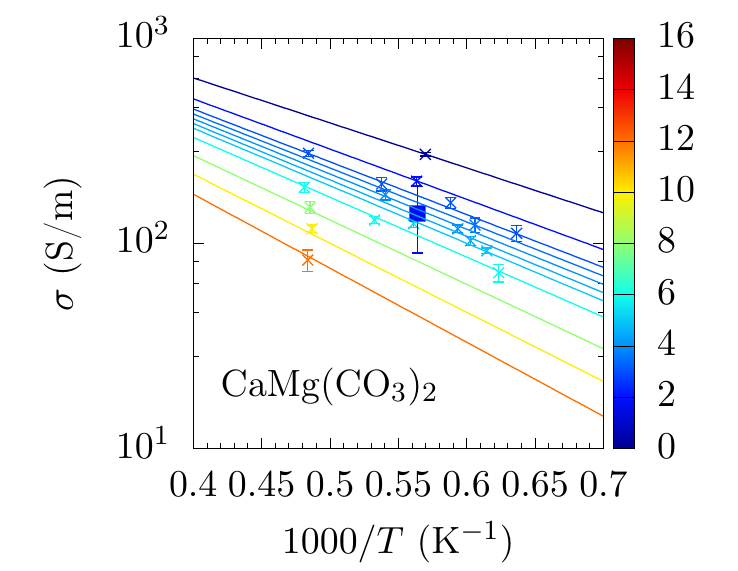}}
\caption{Electrical conductivity from MD (crosses) and isobaric activation laws (plain lines). The pressures (in GPa) of the isobars are \{2, 3, 3.5, 4.5, 6, 8, 10, 12, 15 GPa\} for $\ce{MgCO3}$ (top left), \{0.5, 1, 2, 3, 4.5, 5, 6, 8, 10, 12 GPa\} for $\ce{CaCO3}$ (top right) and \{0.1, 2, 3, 3.5, 4, 4.5, 5, 6, 8, 10, 12\} for $\ce{CaMg(CO3)2}$ (bottom) and are referred to with a color code (vertical scale). Estimates from AIMD calculations (this study for \ce{CaMg(CO3)2} and \citet{Vuilleumier2014} for \ce{CaCO3}) are represented by plain squares.}
\label{fcond}
\end{figure}
As molten salts, carbonate melts are characterized by a high electrical conductivity in the range of $10^1-10^3$~S/m, depending on composition, temperature and pressure, which is up to two orders of magnitude more conductive than silicate melts at the same thermodynamic conditions.\cite{Gaillard2008,Sifre2014,Sifre2015} The knowledge of the electrical conductivity of alkali carbonates is crucial for their industrial applications as electrolytes in fuel cell devices. Many experimental and numerical studies have been devoted to this issue. Hence, the electrical conductivity of the end-members and of binary and ternary mixtures of the system \ce{Li2CO3}--\ce{Na2CO3}--\ce{K2CO3} have been abundantly documented.\cite{Janz1988,Kojima2007,Kojima2008,moi2018}  Because the addition of small amounts of alkaline-earth carbonates improves the performance of fuel cell devices, in particular in terms of durability,\cite{Kojima2003} the electrical conductivity of the Li--K (62--38~mol\%) molten carbonate was measured by impedance spectroscopy and found to decrease linearly with small amounts of \ce{CaCO3} \cite{Lair2012}. The electrical conductivity of carbonate melts is not only important for industrial applications, it is also of fundamental interest to understand the conductivity anomalies in the asthenosphere of the Earth's mantle. To address this question, \citet{Gaillard2008} measured the electrical conductivity of binary and ternary mixtures in the Na--K--Ca system at atmospheric pressure. With regard to Ca-bearing mixtures, MD is in quantitive agreement with their values (deviation of 30\% at most), but with a slightly higher activation energy (Fig.~\ref{fcompcond}). Following the work of \citet{Gaillard2008}, Sifr\'e  \emph{et al.}\cite{Sifre2015} studied (up to 3 GPa) Ca and Mg-bearing carbonate compositions: \ce{CaCO3}, a natural dolomite ($\sim$ \ce{CaMg(CO3)2}), \ce{K2Mg(CO3)2}, \ce{K2Ca(CO3)2} and \ce{Na2Mg(CO3)2}. These studies\cite{Gaillard2008,Sifre2015} show that the electrical conductivity depends slightly on the chemical composition (see Figs.~\ref{fcompcond} and \ref{fcompcond2}). The smaller the cation and the lower its charge, the higher is the electrical conductivity.\cite{Kojima2008,Sifre2015}. As a consequence, the addition of \ce{CaCO3} or \ce{MgCO3} in a pure alkali carbonate reduces somewhat the conductivity.\\
Fig.~\ref{fcompcond2} reports the calculated and measured conductivities for \ce{CaCO3} and for the above mentioned Ca or Mg-bearing mixtures at 3 GPa. The agreement is very good for dolomite and for the \ce{Na2Mg(CO3)2} mixture. Note that for dolomite at 3~GPa and 1800~K, \citet{Yoshino2012} reported an electrical conductivity of 105~S/m, which is almost twice lower than the values of this study and from Sifr\'e  \emph{et al.}.\cite{Sifre2014} Concerning \ce{CaCO3}, the calculated and the measured electrical conductivities overlap within uncertainty (e.g. at 3 GPa and 1923 K MD yields $210 \pm 10$~S/m and Sifr\'e \emph{et al.}\cite{Sifre2015} $230 \pm 25$~S/m). The agreement between MD and experiments is also good for \ce{K2Ca(CO3)2}, although it slightly degrades towards low temperatures. Most striking is our disagreement on the \ce{K2Mg(CO3)2} mixtures (by a factor $\sim 2-3$, Fig.~\ref{fcompcond2}), although our force field satisfactorily reproduced the behavior of other Mg-containing mixtures. An explanation could be that this composition is the most dissymetric one of our study (a small divalent cation coexists with a large monovalent cation in equal proportions). On the other hand, an experimental bias cannot be excluded because this mixture is known for being glass-forming and for easily decomposing \cite{Dobson1996}.\\
As already emphasized in previous studies, the increase of conductivity with temperature is well fitted by an Arrhenius law with activation energies ranging from 34 to 42~kJ/mol at $P = 1$~atm (Table~\ref{tfittransp}). Note that \citet{Gaillard2008} found $E_a^\sigma$ in the range 30 to 35 kJ/mol in alkali-bearing melts and \citet{moi2018} $\sim$ 20 kJ/mol for purely alkali melts by MD. The conductivity decreases weakly with pressure, which can be accounted for by an increase of the activation energy with $P$. Hence, we calculated that this energy is between 42 and 51~kJ/mol at 3~GPa depending on the melt composition, whereas Sifr\'e \emph{et al.}\cite{Sifre2015} reported values from 37 to 48 kJ/mol. For dolomite, these authors found an activation energy of 48~kJ/mol, while our  calculated value is 41 kJ/mol. That is a fairly good agreement considering that the experimental error on each measurement is about 10\%. As for \citet{Yoshino2012} they reported a value of 38~kJ/mol at 3~GPa. With regard to the calcite melt (\ce{CaCO3}), the activation energy calculated by MD (41~kJ/mol at 3~GPa) is much lower than the value reported by Sifr\'e \emph{et al.}\cite{Sifre2015} (76~kJ/mol at the same pressure), but consistent with the ones they reported for other carbonate compositions (in the range $37-48$~kJ/mol).\\
The electrical conductivity estimated from AIMD simulations (this study for \ce{CaMg(CO3)2} and \citet{Vuilleumier2014} for \ce{CaCO3}) are reported on Figs.~\ref{fcond}. For \ce{CaCO3} the agreement between MD and AIMD is good at 0.5 GPa but diminishes at higher pressures. As evoked in the case of diffusion coefficients, most of these discrepancies may be related to the small difference between the equation of state provided by the two models. At a given $T-P$, the density is smaller in the AIMD simulations, hence the ionic mobility and the electrical conductivity are higher. For dolomite, given the large uncertainty on the AIMD value ($\sim 50$\%), it can only be stated that the results of the two simulations models (MD and AIMD) are compatible. 
\begin{table*}
\begin{center}
\begin{tabular}{|c|c|c|c|c|c|c|c|c|c|} \hline
       & $T$(K)  &  $P$ (GPa)& $\sigma$ (S/m) & $H$  & $H_{\rm X_1}$ & $H_{\rm X_2}$ & $H_{\rm CO_3}$ & $H_{\rm X_1-X_2}$& $H_{\rm X-CO_3}$ \\ \hline
\ce{MgCO3} &  1823   &  3.0        &   245          & 0.92 &  $-0.14$ & -- & $-0.32$   & -- & 0.38     \\  \hline
\ce{CaCO3} &  1773   &  1.0        &   220          & 0.81 & $-0.22$ & -- & $-0.35$   & -- & 0.38 \\ \hline
MC  &  1573   &  3.0        &   112    & 0.85 & $-0.08$ & 0 & $-0.36$   & $-0.03$ & 0.34 \\  \hline
NM  &  1573   &  3.0        &   186    & 0.87 & $-0.04$ & $-0.04$ & $-0.29$   & $-0.16$ & 0.40 \\  \hline
KM  &  1573   &  3.0        &   75    & 0.72 & $-0.10$ & $-0.03$ & $-0.23$   & $-0.22$ & 0.29 \\  \hline
KC  &  1573   &  3.0        &   95    & 0.75 & $-0.10$ & $-0.05$ & $-0.25$   & $-0.20$ & 0.35 \\  \hline
\end{tabular}
\caption{Haven ratio and its ionic contributions, where X denotes all cations. Note that the error bar on these quantities is of the order of the error bar on the electrical conductivity: $\sim 5-10$\%. For mixtures, the notation X$_1$ (= Mg, Na or K) and X$_2$ (= Ca or Mg)  is the same as the one appearing in the composition name (first column, with M $=$ Mg, C $=$ Ca, N $=$ Na and K $=$ K).}
\label{tGKvsNE}
\end{center}
\end{table*}

Another advantage of the MD approach is that phenomenological relations for transport properties can be tested. It can be shown that the electrical conductivity $\sigma$ and the diffusion coefficients are linked by a generalized Nernst-Einstein relation, which takes into account the cross correlations between ionic motions, $\sigma=H\sigma^{NE}$ where 
\begin{equation}
\sigma^{NE}=\cfrac{e^2}{k_{\rm B}TV}\sum_{s} N_s z_s^{2}D_s \enspace,
\label{eNE}
\end{equation}
with $N_s$ , $z_s$ and $D_s$ are the number, the formal charge and the self diffusion coefficient of ions of atomic species $s$. The Nernst-Einstein relation assumes that ions move independently from each others. The Haven ratio:
\begin{equation}
H=\cfrac{\sigma}{\sigma^{NE}}=1+ \sum_{s} H_s+ \sum_{s}
\sum_{s'\ne s} H_{ss'} \enspace ,
\end{equation}
accounts for the average cross correlations (through a scalar product) between the displacements of ions of species $s$
\begin{equation}
\begin{split}
H_s=\lim_{t\to \infty} & \cfrac{1}{6t} \cfrac{z_{s}^{2}}{\sum_{s}N_s
  z_{s}^2 D_s}\\&\times\sum_{i=1}^{N_s} \sum_{j\ne i}^{N_s} \langle
\vec{\Delta}_i^{(s)}(t)\cdot\vec{\Delta}_j^{(s)}(t)\rangle \enspace, \enspace \\
 \mathrm{where} \enspace & \vec{\Delta}_i^{(s)}(t)=\vec{r}_i^{(s)}(t)-\vec{r}_i^{(s)}(0)
\end{split}
\label{enes}
\end{equation}
and the average cross correlations between the displacement of an ion $i$ of species $s$ ($\vec{\Delta}_i^{(s)}$) and that of an ion $j$ of another species $s'$ ($\vec{\Delta}_j^{(s')}$)
\begin{equation}
\begin{split}
H_{ss'}= \lim_{t\to \infty} &\cfrac{1}{6t}
\cfrac{z_s z_{s'}}{\sum_{s}N_s z_{s}^2 D_s}
\\&\times \sum_{i=1}^{N_s} \sum_{j=1}^{N_{s'}} \langle
\vec{\Delta}_i^{(s)}(t)\cdot\vec{\Delta}_j^{(s')}(t)\rangle \enspace.
\end{split}
\label{eness}
\end{equation}
The Nernst-Einstein equation is recovered for $H=1$, although it doesn't imply that  $H_{s}$ and $H_{ss'} \simeq 0$, as these two terms can cancel each other (see \citet{moi2018}). 
For instance in \ce{MgCO3} (for further details see Table~\ref{tGKvsNE}), $H$ is close to 1 ($H=0.92$) although its decomposition gives $H_{\rm Mg}=-0.14$, $H_{\rm CO_3}=-0.32$ and $H_{\rm Mg-CO_3}=0.38$. Furthermore the fact that both $H_{\rm Mg}$ and $H_{\rm CO_3}$ contributions are negative implies that ions of a same species have a high probability of moving towards opposite directions (see Eq.~(\ref{enes})), which decreases the conductivity. Regarding the cation-anion correlation term, $H_{\rm Mg-CO_3}=0.38$, it is positive and almost exactly cancels out the $H_{\rm Mg}+H_{\rm CO_3}$ sum. Because Mg and CO$_3$ ions have charges of opposite signs (see Eq.~(\ref{eness})), $H_{\rm Mg-CO_3} > 0 $ is indicative of an anti-correlation of the displacements of the cations and of the anions, that is to say, they are, on average, moving towards opposite directions, which increases the conductivity. For \ce{CaCO3}, an anti-correlation of all the ionic displacements is also observed: a positive contribution of the anion-cation correlation $H_{\rm Ca-CO_3}$ is not fully overbalanced by the correlations between ions of a same species leading to a Haven ratio less than 1 ($H=1+H_{\rm Ca}+H_{\rm CO_3}+H_{\rm Ca-CO_3}=0.81$). \\
We carried on this approach with some binary mixtures (see Table~\ref{tGKvsNE}), including dolomite. It is worth noticing that the displacements of cations of the same species are mostly not correlated to one another ($H_{\rm X_1}$ and $H_{\rm X_2}$ both $\sim 0$, where X$_1$, X$_2$ = Mg, Ca, Na or K). However the cross correlations between cationic displacements (as expressed by the term $H_{\rm X_1-X_2}$) seem to depend on the charge of the cations. Indeed, in dolomite $H_{\rm X_1-X_2}$ is close to 0 ($H_{\rm Mg-Ca}=-0.03$), while in the investigated alkali-alkaline earth mixtures it ranges between -0.16 and -0.22. As for the other terms ($H_{\rm CO_3}$ and $H_{\rm X-CO_3}$), they give opposite contributions to $H$ (see Table~\ref{tGKvsNE}). So, although the Nernst-Einstein equation yields, for the melts studied here, a reasonable estimation of the electrical conductivity ($+ 10-30$\% from the exactly calculated one, $\sigma$), it provides no information on the relevance of its underlying assumption (the ions move independently from each other). In fact, as we show it here, its usefulness relies on a cancellation effect between the different ion-ion correlations. As the dependence of this cancellation effect with composition, $T$ or $P$ is non-trivial, we recommend a circumspect use of the Nernst-Einstein approximation.

\subsection{Viscosity}
\begin{figure}
\subfloat[\label{fviscom}]{\includegraphics{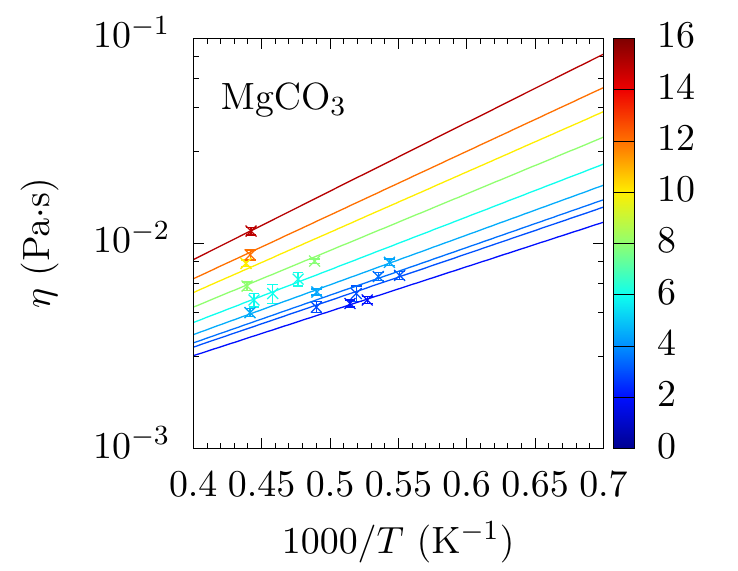}}\\
\subfloat[\label{fviscoc}]{\includegraphics{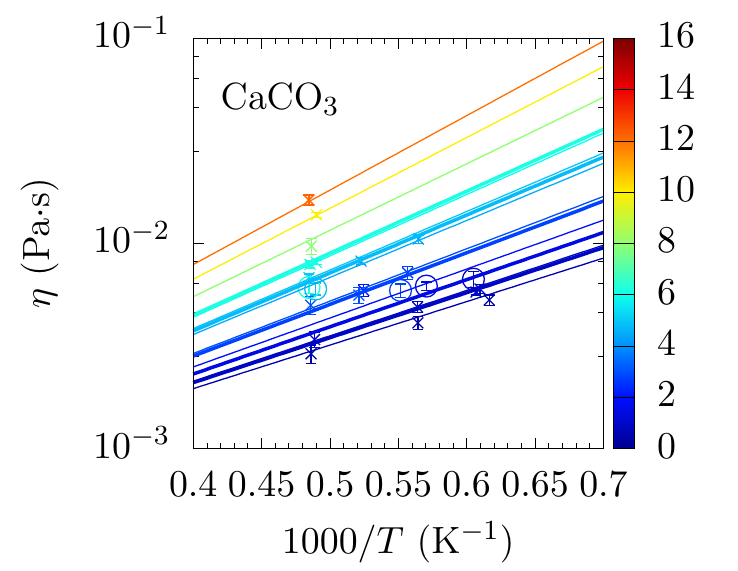}}\\
\subfloat[\label{fviscomc}]{\includegraphics{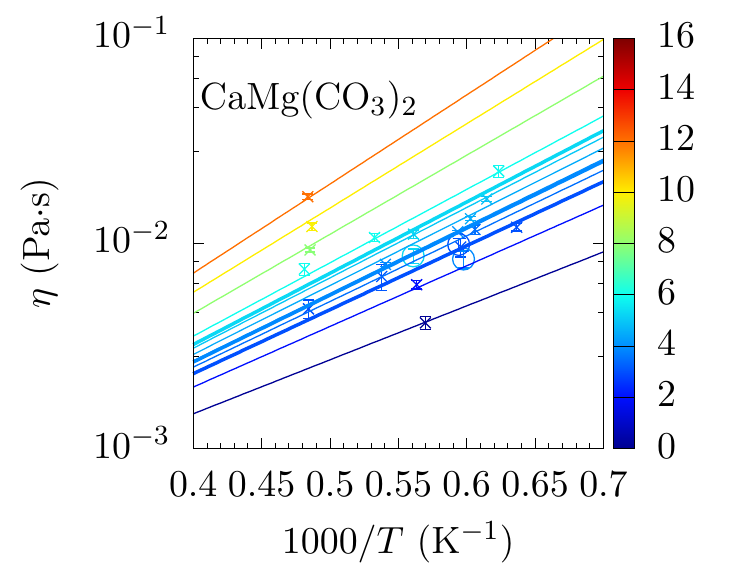}}
\caption{Viscosity from MD (crosses) and isobaric activation laws (plain lines). The pressures (in GPa) of the isobars are \{2, 3, 3.5, 4.5, 6, 8, 10, 12, 15 GPa\} for $\ce{MgCO3}$ (top left), \{0.5, 1, 2, 3, 4.5, 5, 6, 8, 10, 12 GPa\} for $\ce{CaCO3}$ (top right) and \{0.1, 2, 3, 3.5, 4, 4.5, 5, 6, 8, 10, 12\} for $\ce{CaMg(CO3)2}$ (bottom) and are referred to with a color code (vertical scale). The data recently reported by \citet{Kono2014} are plotted as circles, the corresponding isobars (0.9, 1.5, 2.8, 4.8 and 6.2~GPa for \ce{CaCO3} and 3.0, 3.9 and 5.3~GPa for \ce{CaMg(CO3)2}) obtained from the present MD study
are reported for comparison (bold blue lines).}
\label{fvisco}
\end{figure}
\begin{figure}
\includegraphics{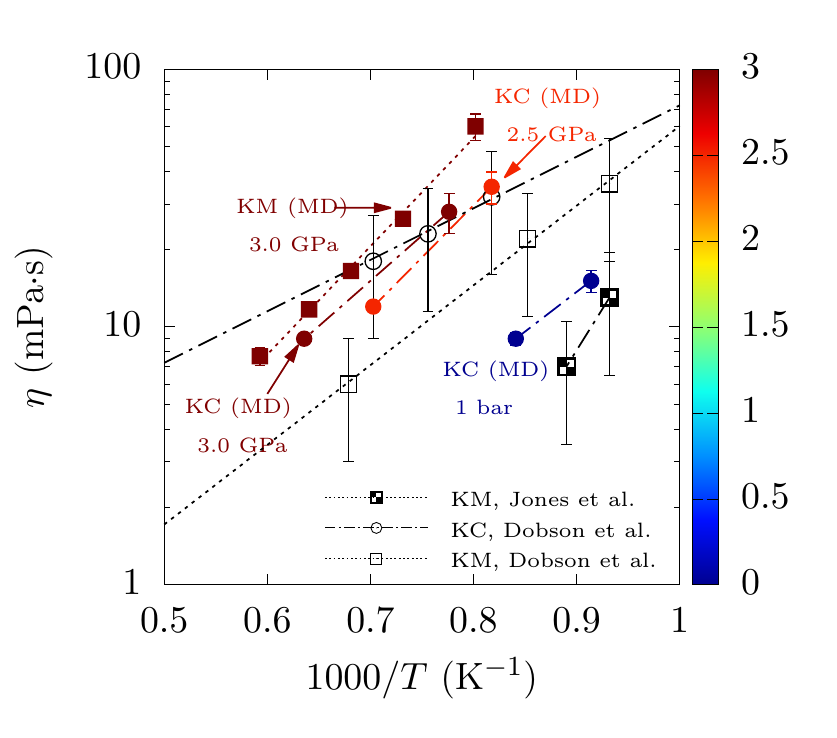}\label{fcompeta2} 
\caption{Viscosity calculated by MD (plain symbols) and measured by \citet{Dobson1996} (empty symbols) and by Jones \emph{et al.}\cite{Jones1995} (half-empty symbols) for \ce{K2Ca(CO3)2} (KC, circles) and  \ce{K2Mg(CO3)2} (KM, squares). The color code refers to the pressure as indicated by the vertical scale (in GPa).}
\label{fcompetaP}  
\end{figure}
\begin{figure}[ht]
\centering
\includegraphics{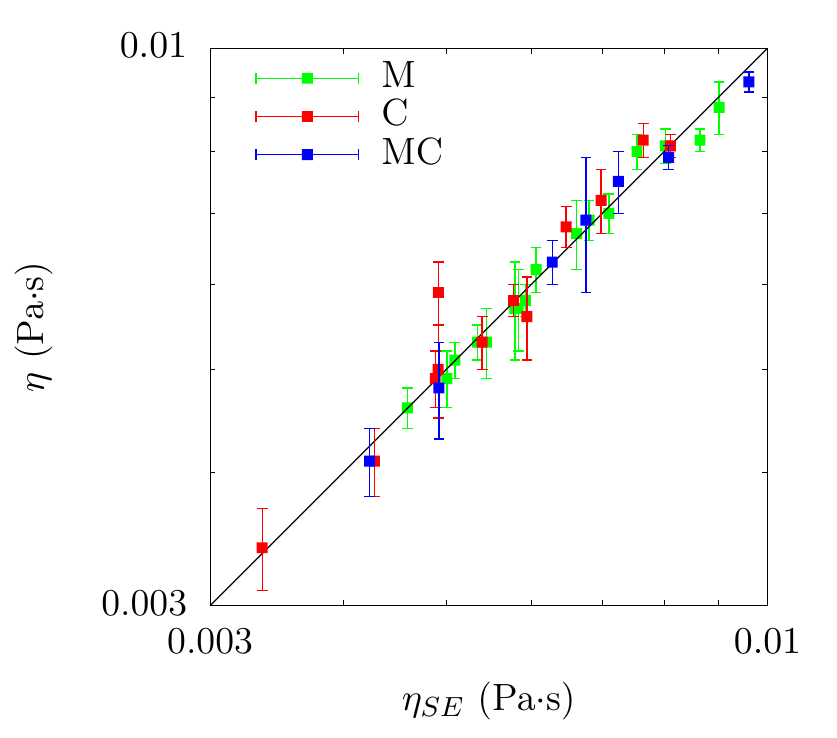}%
\caption{Comparison of the viscosity calculated using the Green-Kubo formula, $\eta$, and the Stokes-Einstein equation, $\eta_{SE}$, at all $T$ and $P$. The $d$ parameter, figuring in Eq.~(\ref{eSE}) has been adjusted for \ce{MgCO3} (M), \ce{CaCO3} (C) and \ce{CaMg(CO3)2} (MC) to 3.2, 3.5 and 3.4 \AA , respectively, so as to align the data on the $\eta=\eta_{SE}$ bisector (black line).}
\label{fSE}
\end{figure}
\begin{figure}
\centering
\includegraphics{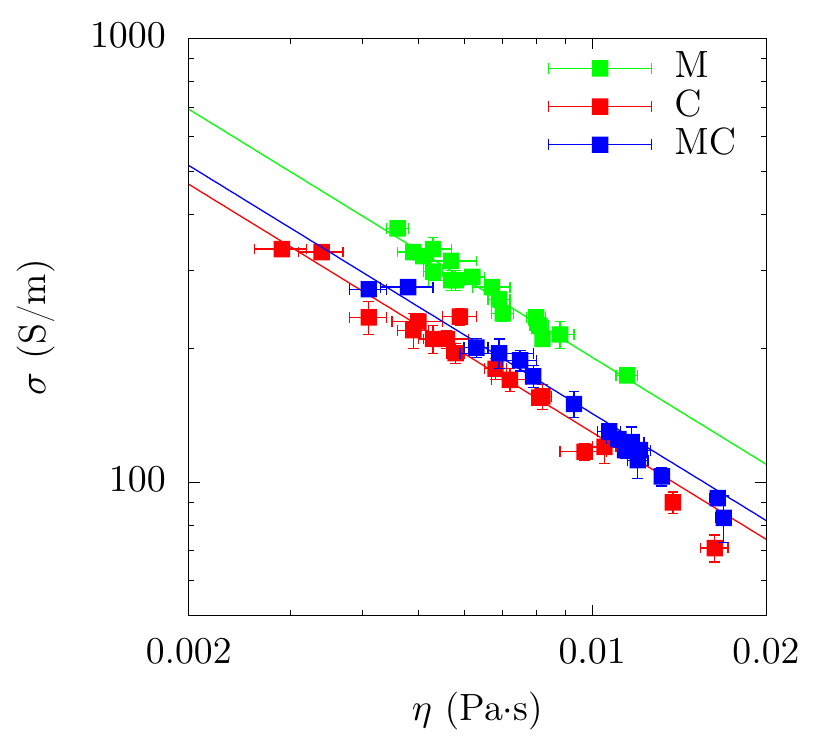}  
\caption{Electrical conductivity as a function of viscosity. The squares are the MD data and the lines are empirical fits of the MD data by equation $\sigma=A/\eta^{0.8}$ (where $\sigma$ is in S/m and $\eta$ in Pa$\cdot$s) with $A=$ 4.78, 3.25 and 3.58 for \ce{CaCO3} (C), \ce{MgCO3} (M) and \ce{CaMg(CO3)2} (MC), respectively.}
\label{sigmaeta}
\end{figure}
Very little is known on the pressure dependence of the viscosity of molten carbonates. Based on molten salt data reviewed by Janz,\cite{Janz1988} \citet{Wolff1994} roughly estimated an order of magnitude ($\sim$ 100 mPa$\cdot$s) for the viscosity of calcium-rich carbonatites at 973 K and ambient pressure. The ex-situ measurements at mantle pressures ($P = 3$~GPa) made  by Sykes \emph{et al.}\cite{Sykes1992} seem to greatly overestimate the viscosity of dolomitic melts (e.g. for 70:30 mol\% \ce{CaCO3}--\ce{MgCO3}, $\eta=600$~mPa.s at 1473~K and 1~GPa) in light of the very high activation energy they obtained \cite{Kono2014}, probably because of an incomplete melting of the sample\cite{Jones1995}. On the other hand, \citet{Dobson1996} performed the first in-situ measurements of the viscosity of \ce{K2Ca(CO3)2} and \ce{K2Mg(CO3)2} melts, up to 5.5~GPa, by using the falling sphere method with X-ray radiography. In this pioneering experiment, the temperature was difficult to control and the relative error on viscosity was typically 50\%, due to incomplete melting of the sample and to convection effects at the highest temperatures. Besides the limited  frame rate used to capture the images of the falling sphere greatly reduces the accuracy of the final velocity measurement, which is crucial for determining the viscosity \cite{Kono2014}. Nevertheless it was found that Ca,~Mg-bearing carbonates under high $T-P$ have a viscosity similar to that of alkali carbonates at 1 bar ($6-36$~mPa$\cdot$s in the range $2.5-5.5$~GPa). This  qualitative observation is consistent with the results of our MD calculations (see Fig.~\ref{fvisco}). However we do not agree with the assertion of the authors\cite{Dobson1996} that the effect of pressure on viscosity is negligible in the pressure range investigated (up to 5.5 GPa) and we think that this observation results from the large error bars of the study ($\sim$ 50\% on $\eta$ and $\pm$ 0.5 GPa on $P$, see also the discrepancy with the first results of Jones \emph{et al.}\cite{Jones1995}). We calculated the viscosity of \ce{K2Ca(CO3)2} at 0, 2.5 and 3 GPa and the viscosity of \ce{K2Mg(CO3)2} at 3 GPa. The viscosity depends strongly on the temperature by following an Arrhenius law with an activation energy a little higher for \ce{K2Mg(CO3)2} than for \ce{K2Ca(CO3)2}. Even at these moderate pressures it is obvious that the activation energy depends on pressure (Fig.~\ref{fcompetaP}), in contrast with the observation of \citet{Dobson1996}. At a given pressure, we obtained similar activation energies for the two compositions \ce{K2Mg(CO3)2} and \ce{K2Ca(CO3)2} ($\sim$~80~kJ/mol at 3~GPa). However these activation energies differ from the ones calculated for the end-members \ce{CaCO3} and \ce{MgCO3} ($43-49$~kJ/mol), for dolomite (60~kJ/mol) and for alkali melts ($31-39$~kJ/mol)\cite{moi2018} at the same pressure. This illustrates that the viscosity not only depends on the temperature, but also on the chemical composition.
\sloppy
Recently, \citet{Kono2014} have reported viscosities of calcite and dolomitic (Mg$_{0.40}$Fe$_{0.09}$Ca$_{0.51}$CO$_3$) melts at temperatures just above the melting point up to 6.2~GPa, by using the falling sphere method with a powerful technique of ultra-fast synchrotron X-ray imaging. The measured viscosities ($6-10$~mPa.s, with an error of 9\%) are of the same order of magnitude as those of \citet{Dobson1996}, although the composition and the temperature differ in the two studies. The values of \citet{Kono2014} (at $0.9-6.2$~GPa and $1653-2063$~K for calcite, $3.0-5.3$~GPa and $1683-1783$~K for dolomite) are reported on Fig.~\ref{fvisco}. For calcite a satisfying agreement (within $10-15$\%, i.e. within the overlap of error bars) is found with our calculations, better than with the values of \citet{Vuilleumier2014} For dolomite the agreement, although comprised between 3 and 30\%, is reasonable once considered some anomalous trend in the experimental data. Indeed the viscosities measured by \citet{Kono2014} at 3.9~GPa and $1000/T\sim 0.6$ is lower than the one measured at 3.0 GPa at the same temperature, whereas the viscosity is expected to increase upon increasing pressure. Incidentally, we do not think that the small content (9 mol\%) of \ce{FeCO3} in the experimental composition has a significant contribution to the viscosity and could account for the MD-experiment discrepancy.\\ 
For natrocarbonatite (Natro, Na$_{1.1}$K$_{0.18}$Ca$_{0.36}$CO$_3$) at 1 bar and 823~K (near eruption conditions at the Ol Doinyo Lengai volcano), we calculated a viscosity of 68~$ \pm$~10~mPa$\cdot$s. We believe this is a more reliable value than the estimation given by \citet{Treiman1983} using molten salt data (5 mPa$\cdot$s), which seems rather low for a Ca-bearing mixture at relatively low temperature. For example compared with the alkali ternary eutectic mixture (Li$_{0.435}$Na$_{0.315}$K$_{0.25}$CO$_{3}$) the viscosity of natrocarbonatite is greater by a factor $\sim 3$ according to our MD calculations.\cite{moi2018} This is consistent with the observations made by both experiments and simulations, of a decrease of the ionic conductivity upon addition of the alkaline-earth cation Ca.\cite{Gaillard2008,Lair2012} Interestingly the calculated viscosity is much lower than the ones measured by \citet{Norton1997} for several carbonatitic melts issued from the eruption of Ol Doinyo Lengai in 1988. At 823 K, the measured viscosities vary a lot ($10^2-10^4$~mPa/s) as a result of varying compositions (with or without a silicate component), crystallinity (presence of crystals of various sizes) and vesicularity (presence of \ce{CO2} bubbles). Consequently, for a composition close to the one we studied (Na$_{1.1}$K$_{0.18}$Ca$_{0.36}$CO$_3$), they reported a viscosity as high as $\sim$ 600 mPa$\cdot$s. At variance with experiments where a complete melting is uneasy to assess, the MD value (68~$ \pm$~10~mPa$\cdot$s) is representative of carbonate melts. Besides, this oil-like viscosity is consistent with field observations where effusion of a very fluid lava is seen (see the videos of \citet{Fischer2009} with the link of reference \onlinecite{nsf2009}).\\
To obtain a phenomenological description of the viscosity of simple liquids, the Stokes-Einstein equation is often used
\begin{equation}
\eta_{SE}=\cfrac{k_{\rm B}T}{2\pi Dd} \enspace ,
\label{eSE} 
\end{equation}
where $d$ is the diameter of the diffusing particle and $D$ its diffusion coefficient. In the present case, the latter is assumed to be given by the arithmetic mean of the diffusion coefficients, $D=\sum_s x_s D_S/\sum_s x_s$ where $x_s$ is the molar fraction of ion of species $s$ in the melt. We chose $d=3.2$, 3.5 and 3.4~\AA \, for \ce{MgCO3}, \ce{CaCO3} and \ce{CaMg(CO3)2} respectively. The behavior of the MD-calculated viscosity with pressure and temperature was very well reproduced by Eq.~(\ref{eSE}) using these parameters (Fig.~\ref{fSE}). To calculate the hydrodynamic diameter in a more grounded framework we used the following expression: $d=x_\text{X}^2d_\text{XX}+x_{\text{CO}_3}^2d_{\text{CO}_3\text{CO}_3}+2x_\text{X}x_{\text{CO}_3}d_{\text{X}\text{CO}_3}$, where $x_\text{X}$ and $x_{\text{CO}_3}$ are the molar fractions of cation X and of anion CO$_3$, respectively, and $d_\text{XX}$, $d_{\text{CO}_3\text{CO}_3}$ and $d_{\text{X}\text{CO}_3}$ the cation-cation, carbonate-carbonate and cation-carbonate distances issued from the closest approach distances indicated by the corresponding PDFs (in fact, the $d$ parameters correspond to the distances at which the integral of the PDFs is equal to 1). This ansatz leads to $d= 3.1$, 3.4 and 3.3~\AA \, for \ce{MgCO3}, \ce{CaCO3} and \ce{CaMg(CO3)2} respectively. These values are almost identical ($-0.1$~\AA) to the ones obtained by fitting the values of $d$ to Eq.~(\ref{eSE}). \\
Finally, the electrical conductivity $\sigma$ (in S/m) and the viscosity $\eta$ (in Pa.s) can be related by the following empirical formula: 
\begin{equation}
\sigma=\cfrac{A}{\eta^{0.8}} \enspace,
\end{equation}
where $A=$  4.80, 3.25 and 3.58 for \ce{MgCO3}, \ce{CaCO3} and \ce{CaMg(CO3)2} respectively (Fig.~\ref{sigmaeta}). A comparable relation between these two transport coefficients has been also highlighted experimentally and computationally for various melt compositions. \cite{Sifre2015,Vuilleumier2014,moi2018}.

\section{Conclusion}
Following our previous work on the \ce{Li2CO3}--\ce{Na2CO3}--\ce{K2CO3} melts, we have studied the thermodynamics, the microscopic structure and the transport properties (diffusion coefficients, electrical conductivity and viscosity) of Ca and Mg-bearing carbonate melts up to 2073~K and 15~GPa. For that we have developed an empirical force field benchmarked on data from experiments (density of the rhombohedral crystal phases at 300~K and up to 4~GPa, and the density and the compressibility of the \ce{K2Ca(CO3)2} melt) and from AIMD simulations (microscopic structure of five liquids). The density and compressibility, evaluated for the metastable melts of \ce{MgCO3} and \ce{CaCO3} at 1100~K and 1~bar, are in very good agreement with the estimates of the experimental literature. Moreover we have shown that alkaline-earth carbonate mixtures behave ideally regarding the density and the compressibility. Based on the example of a Na--Ca--K melt (natrocarbonatite), the assumption of an ideal behavior for alkali-alkaline-earth mixtures seems a little less accurate but still reasonable.  \\
The equations of state of carbonate melts with a composition of prior interest in the study of the Earth's mantle (\ce{MgCO3}, \ce{CaCO3} and \ce{CaMg(CO3)2}), as well as that modeling the natrocarbonatite emitted at the Ol Doinyo Lengai volcano (Na$_{1.10}$K$_{0.18}$Ca$_{0.36}$CO$_3$), were evaluated and modeled by a third-order Birch-Murnaghan formula. Covering a large $T$, $P$ range, these data may help in the debate on the geodynamics of carbonatitic melts relative to silicate melts.\cite{Liu2007}\\
The analysis of the PDFs associated to the dolomitic melt showed an ideal behavior of the microscopic structure, as the PDFs in \ce{CaMg(CO3)2} are similar to the corresponding ones in \ce{CaCO3} and in \ce{MgCO3}. Moreover, no major modification of the structure was observed with increasing pressure.\\
As for the transport properties (diffusion coefficients, electrical conductivity and viscosity), they evolve smoothly (Arrhenius-like) over the studied $T-P$ domain. Thus at the high $T-P$ conditions of the Earth's mantle, Mg and Ca-bearing molten carbonates keep the main features of molten salts, namely the ions are highly mobile. For example the viscosity is comprised in the range $1-100$~mPa$\cdot$s, that is in between the one of water and the one of olive oil at room conditions. However, the composition has a non negligible effect on transport coefficients: calco-magnesian melts are systematically more viscous than their alkali counterparts. These results may provide new insights into magmatic processes implying carbonatitic melts.\cite{Treiman1983,Treiman1989,Wolff1994}.\\
Finally we discussed the reliability of the phenomenological Nernst-Einstein and the Stokes-Einstein equations, that relate the diffusion coefficients to the electrical conductivity and to the viscosity, respectively. These relations are often used to infer one of the transport coefficient from another one that has been measured. According to the present study, both formulas lead to reasonable values. However the underlying assumptions of the relations are not always representative of the transport mechanism itself. We have shown that the fairly good predictions provided by the Nernst-Einstein equation (which assumes that the ions move independently from one another) result from a partial cancellation of interionic dynamic correlations whose dependence with composition, $T$ or $P$ is non-trivial. As a consequence we recommend a circumspect use of the Nernst-Einstein approximation.\\
In summary, the  overall agreement between the results of the MD simulations using this force field and the full set of available experimental data (thermodynamics and transport coefficients) provides evidence of the ability of our FF to describe with accuracy the properties of any melt in the \ce{MgCO3}--\ce{CaCO3}--\ce{Li2CO3}--\ce{Na2CO3}--\ce{K2CO3} system.

\section*{Supplementary Material}
See supplementary material for the density of crystal phases calculated by anisotropic relaxation of the rhombohedral structures up to 4~GPa, the evolution of the PDFs issued by MD with increasing pressure (up to 15~GPa), a summary of all calculated properties with their uncertainties.

\begin{acknowledgments}
The research leading to these results has received funding from the R\'{e}gion Ile-de-France and the European Community’s Seventh Framework Program (FP7/2007-2013) under Grant agreement (ERC, N$^\circ$ 279790). The authors acknowledge GENCI for HPC resources (Grant No. 2015-082309).
\end{acknowledgments}

\bibliography{main}

\begin{thebibliography}{87}%
\makeatletter
\providecommand \@ifxundefined [1]{%
 \@ifx{#1\undefined}
}%
\providecommand \@ifnum [1]{%
 \ifnum #1\expandafter \@firstoftwo
 \else \expandafter \@secondoftwo
 \fi
}%
\providecommand \@ifx [1]{%
 \ifx #1\expandafter \@firstoftwo
 \else \expandafter \@secondoftwo
 \fi
}%
\providecommand \natexlab [1]{#1}%
\providecommand \enquote  [1]{``#1''}%
\providecommand \bibnamefont  [1]{#1}%
\providecommand \bibfnamefont [1]{#1}%
\providecommand \citenamefont [1]{#1}%
\providecommand \href@noop [0]{\@secondoftwo}%
\providecommand \href [0]{\begingroup \@sanitize@url \@href}%
\providecommand \@href[1]{\@@startlink{#1}\@@href}%
\providecommand \@@href[1]{\endgroup#1\@@endlink}%
\providecommand \@sanitize@url [0]{\catcode `\\12\catcode `\$12\catcode
  `\&12\catcode `\#12\catcode `\^12\catcode `\_12\catcode `\%12\relax}%
\providecommand \@@startlink[1]{}%
\providecommand \@@endlink[0]{}%
\providecommand \url  [0]{\begingroup\@sanitize@url \@url }%
\providecommand \@url [1]{\endgroup\@href {#1}{\urlprefix }}%
\providecommand \urlprefix  [0]{URL }%
\providecommand \Eprint [0]{\href }%
\providecommand \doibase [0]{http://dx.doi.org/}%
\providecommand \selectlanguage [0]{\@gobble}%
\providecommand \bibinfo  [0]{\@secondoftwo}%
\providecommand \bibfield  [0]{\@secondoftwo}%
\providecommand \translation [1]{[#1]}%
\providecommand \BibitemOpen [0]{}%
\providecommand \bibitemStop [0]{}%
\providecommand \bibitemNoStop [0]{.\EOS\space}%
\providecommand \EOS [0]{\spacefactor3000\relax}%
\providecommand \BibitemShut  [1]{\csname bibitem#1\endcsname}%
\let\auto@bib@innerbib\@empty
\bibitem [{\citenamefont {Desmaele}\ \emph {et~al.}(2019)\citenamefont
  {Desmaele}, \citenamefont {Sator}, \citenamefont {Vuilleumier},\ and\
  \citenamefont {Guillot}}]{moi2018}%
  \BibitemOpen
  \bibfield  {author} {\bibinfo {author} {\bibfnamefont {E.}~\bibnamefont
  {Desmaele}}, \bibinfo {author} {\bibfnamefont {N.}~\bibnamefont {Sator}},
  \bibinfo {author} {\bibfnamefont {R.}~\bibnamefont {Vuilleumier}}, \ and\
  \bibinfo {author} {\bibfnamefont {B.}~\bibnamefont {Guillot}},\ }\href
  {\doibase 10.1063/1.5082731} {\bibfield  {journal} {\bibinfo  {journal} {J.
  Chem. Phys.}\ }\textbf {\bibinfo {volume} {150}},\ \bibinfo {pages} {094504}
  (\bibinfo {year} {2019})}\BibitemShut {NoStop}%
\bibitem [{\citenamefont {Desmaele}(2017)}]{moi}%
  \BibitemOpen
  \bibfield  {author} {\bibinfo {author} {\bibfnamefont {E.}~\bibnamefont
  {Desmaele}},\ }\emph {\bibinfo {title} {Physicochemical Properties of Molten
  Carbonates from Atomistic Simulations}},\ \href
  {https://hal.archives-ouvertes.fr/tel-01767010/} {Ph.D. thesis},\ \bibinfo
  {school} {Sorbonne Universit\'e} (\bibinfo {year} {2017})\BibitemShut
  {NoStop}%
\bibitem [{\citenamefont {Lair}\ \emph {et~al.}(2012)\citenamefont {Lair},
  \citenamefont {Albin}, \citenamefont {Ringuedé},\ and\ \citenamefont
  {Cassir}}]{Lair2012}%
  \BibitemOpen
  \bibfield  {author} {\bibinfo {author} {\bibfnamefont {V.}~\bibnamefont
  {Lair}}, \bibinfo {author} {\bibfnamefont {V.}~\bibnamefont {Albin}},
  \bibinfo {author} {\bibfnamefont {A.}~\bibnamefont {Ringuedé}}, \ and\
  \bibinfo {author} {\bibfnamefont {M.}~\bibnamefont {Cassir}},\ }\href
  {\doibase 10.1016/j.ijhydene.2011.09.153} {\bibfield  {journal} {\bibinfo
  {journal} {Int. J. Hydrogen Energy}\ }\textbf {\bibinfo {volume} {37}},\
  \bibinfo {pages} {19357} (\bibinfo {year} {2012})}\BibitemShut {NoStop}%
\bibitem [{\citenamefont {Chery}, \citenamefont {Lair},\ and\ \citenamefont
  {Cassir}(2015{\natexlab{a}})}]{Chery2015}%
  \BibitemOpen
  \bibfield  {author} {\bibinfo {author} {\bibfnamefont {D.}~\bibnamefont
  {Chery}}, \bibinfo {author} {\bibfnamefont {V.}~\bibnamefont {Lair}}, \ and\
  \bibinfo {author} {\bibfnamefont {M.}~\bibnamefont {Cassir}},\ }\href
  {\doibase 10.1016/j.electacta.2015.01.216} {\bibfield  {journal} {\bibinfo
  {journal} {Electrochim. Acta}\ }\textbf {\bibinfo {volume} {160}},\ \bibinfo
  {pages} {74} (\bibinfo {year} {2015}{\natexlab{a}})}\BibitemShut {NoStop}%
\bibitem [{\citenamefont {Chery}, \citenamefont {Lair},\ and\ \citenamefont
  {Cassir}(2015{\natexlab{b}})}]{Chery2015rev}%
  \BibitemOpen
  \bibfield  {author} {\bibinfo {author} {\bibfnamefont {D.}~\bibnamefont
  {Chery}}, \bibinfo {author} {\bibfnamefont {V.}~\bibnamefont {Lair}}, \ and\
  \bibinfo {author} {\bibfnamefont {M.}~\bibnamefont {Cassir}},\ }\href
  {\doibase 10.3389/fenrg.2015.00043} {\bibfield  {journal} {\bibinfo
  {journal} {Front. Energy Res.}\ }\textbf {\bibinfo {volume} {3}},\ \bibinfo
  {pages} {43} (\bibinfo {year} {2015}{\natexlab{b}})}\BibitemShut {NoStop}%
\bibitem [{\citenamefont {Cassir}, \citenamefont {McPhail},\ and\ \citenamefont
  {Moreno}(2012)}]{Cassir2012}%
  \BibitemOpen
  \bibfield  {author} {\bibinfo {author} {\bibfnamefont {M.}~\bibnamefont
  {Cassir}}, \bibinfo {author} {\bibfnamefont {S.}~\bibnamefont {McPhail}}, \
  and\ \bibinfo {author} {\bibfnamefont {A.}~\bibnamefont {Moreno}},\ }\href
  {\doibase 10.1016/j.ijhydene.2011.11.006} {\bibfield  {journal} {\bibinfo
  {journal} {Int. J. Hydrogen Energy}\ }\textbf {\bibinfo {volume} {37}},\
  \bibinfo {pages} {19345} (\bibinfo {year} {2012})}\BibitemShut {NoStop}%
\bibitem [{\citenamefont {Cassir}, \citenamefont {Ringuedé},\ and\
  \citenamefont {Lair}(2013)}]{Cassir2013}%
  \BibitemOpen
  \bibfield  {author} {\bibinfo {author} {\bibfnamefont {M.}~\bibnamefont
  {Cassir}}, \bibinfo {author} {\bibfnamefont {A.}~\bibnamefont {Ringuedé}}, \
  and\ \bibinfo {author} {\bibfnamefont {V.}~\bibnamefont {Lair}},\ }in\ \href
  {\doibase 10.1016/B978-0-12-398538-5.00017-2} {\emph {\bibinfo {booktitle}
  {Molten Salts Chemistry}}},\ \bibinfo {editor} {edited by\ \bibinfo {editor}
  {\bibfnamefont {F.}~\bibnamefont {Lantelme}}\ and\ \bibinfo {editor}
  {\bibfnamefont {H.}~\bibnamefont {Groult}}}\ (\bibinfo  {publisher}
  {Elsevier},\ \bibinfo {address} {Oxford},\ \bibinfo {year}
  {2013})\BibitemShut {NoStop}%
\bibitem [{\citenamefont {Dasgupta}\ and\ \citenamefont
  {Hirschmann}(2006)}]{Dasgupta2006}%
  \BibitemOpen
  \bibfield  {author} {\bibinfo {author} {\bibfnamefont {R.}~\bibnamefont
  {Dasgupta}}\ and\ \bibinfo {author} {\bibfnamefont {M.~M.}\ \bibnamefont
  {Hirschmann}},\ }\href {\doibase 10.1038/nature04612} {\bibfield  {journal}
  {\bibinfo  {journal} {Nature}\ }\textbf {\bibinfo {volume} {440}},\ \bibinfo
  {pages} {659} (\bibinfo {year} {2006})}\BibitemShut {NoStop}%
\bibitem [{\citenamefont {Gaillard}\ \emph {et~al.}(2008)\citenamefont
  {Gaillard}, \citenamefont {Malki}, \citenamefont {Iacono-Marziano},
  \citenamefont {Pichavant},\ and\ \citenamefont {Scaillet}}]{Gaillard2008}%
  \BibitemOpen
  \bibfield  {author} {\bibinfo {author} {\bibfnamefont {F.}~\bibnamefont
  {Gaillard}}, \bibinfo {author} {\bibfnamefont {M.}~\bibnamefont {Malki}},
  \bibinfo {author} {\bibfnamefont {G.}~\bibnamefont {Iacono-Marziano}},
  \bibinfo {author} {\bibfnamefont {M.}~\bibnamefont {Pichavant}}, \ and\
  \bibinfo {author} {\bibfnamefont {B.}~\bibnamefont {Scaillet}},\ }\href
  {\doibase 10.1126/science.1164446} {\bibfield  {journal} {\bibinfo  {journal}
  {Science}\ }\textbf {\bibinfo {volume} {322}},\ \bibinfo {pages} {1363}
  (\bibinfo {year} {2008})}\BibitemShut {NoStop}%
\bibitem [{\citenamefont {Dasgupta}\ and\ \citenamefont
  {Hirschmann}(2010)}]{Dasgupta2010}%
  \BibitemOpen
  \bibfield  {author} {\bibinfo {author} {\bibfnamefont {R.}~\bibnamefont
  {Dasgupta}}\ and\ \bibinfo {author} {\bibfnamefont {M.~M.}\ \bibnamefont
  {Hirschmann}},\ }\href {\doibase 10.1016/j.epsl.2010.06.039} {\bibfield
  {journal} {\bibinfo  {journal} {Earth Planet. Sci. Lett.}\ }\textbf {\bibinfo
  {volume} {298}},\ \bibinfo {pages} {1} (\bibinfo {year} {2010})}\BibitemShut
  {NoStop}%
\bibitem [{\citenamefont {Dasgupta}(2013)}]{Dasgupta2013}%
  \BibitemOpen
  \bibfield  {author} {\bibinfo {author} {\bibfnamefont {R.}~\bibnamefont
  {Dasgupta}},\ }\href {\doibase 10.2138/rmg.2013.75.7} {\bibfield  {journal}
  {\bibinfo  {journal} {Rev. Mineral. Geochem.}\ }\textbf {\bibinfo {volume}
  {75}},\ \bibinfo {pages} {183} (\bibinfo {year} {2013})}\BibitemShut
  {NoStop}%
\bibitem [{\citenamefont {Hammouda}\ and\ \citenamefont
  {Keshav}(2015)}]{Hammouda2015}%
  \BibitemOpen
  \bibfield  {author} {\bibinfo {author} {\bibfnamefont {T.}~\bibnamefont
  {Hammouda}}\ and\ \bibinfo {author} {\bibfnamefont {S.}~\bibnamefont
  {Keshav}},\ }\href {\doibase 10.1016/j.chemgeo.2015.05.018} {\bibfield
  {journal} {\bibinfo  {journal} {Chem. Geol.}\ }\textbf {\bibinfo {volume}
  {418}},\ \bibinfo {pages} {171} (\bibinfo {year} {2015})}\BibitemShut
  {NoStop}%
\bibitem [{\citenamefont {Mitchell}(2005)}]{Mitchell2005}%
  \BibitemOpen
  \bibfield  {author} {\bibinfo {author} {\bibfnamefont {R.~H.}\ \bibnamefont
  {Mitchell}},\ }\href {\doibase 10.2113/gscanmin.43.6.2049} {\bibfield
  {journal} {\bibinfo  {journal} {Can. Mineral.}\ }\textbf {\bibinfo {volume}
  {43}},\ \bibinfo {pages} {2049} (\bibinfo {year} {2005})}\BibitemShut
  {NoStop}%
\bibitem [{\citenamefont {Woolley}\ and\ \citenamefont
  {Kjarsgaard}(2008)}]{Woolley2008}%
  \BibitemOpen
  \bibfield  {author} {\bibinfo {author} {\bibfnamefont {A.}~\bibnamefont
  {Woolley}}\ and\ \bibinfo {author} {\bibfnamefont {B.}~\bibnamefont
  {Kjarsgaard}},\ }\href {\doibase 10.4095/225115} {\enquote {\bibinfo {title}
  {{Carbonatite Occurrences of the World: Map and Database. Geological Survey
  of Canada. Open File 5796}},}\ } (\bibinfo {year} {2008})\BibitemShut
  {NoStop}%
\bibitem [{\citenamefont {Keller}\ and\ \citenamefont
  {Zaitsev}(2012)}]{Keller2012}%
  \BibitemOpen
  \bibfield  {author} {\bibinfo {author} {\bibfnamefont {J.}~\bibnamefont
  {Keller}}\ and\ \bibinfo {author} {\bibfnamefont {A.~N.}\ \bibnamefont
  {Zaitsev}},\ }\href {\doibase 10.1016/j.lithos.2012.05.022} {\bibfield
  {journal} {\bibinfo  {journal} {Lithos}\ }\textbf {\bibinfo {volume} {148}},\
  \bibinfo {pages} {45} (\bibinfo {year} {2012})}\BibitemShut {NoStop}%
\bibitem [{\citenamefont {Woolley}\ and\ \citenamefont
  {Church}(2005)}]{Woolley2005}%
  \BibitemOpen
  \bibfield  {author} {\bibinfo {author} {\bibfnamefont {A.~R.}\ \bibnamefont
  {Woolley}}\ and\ \bibinfo {author} {\bibfnamefont {A.~A.}\ \bibnamefont
  {Church}},\ }\href {\doibase 10.1016/j.lithos.2005.03.018} {\bibfield
  {journal} {\bibinfo  {journal} {Lithos}\ }\textbf {\bibinfo {volume} {85}},\
  \bibinfo {pages} {1} (\bibinfo {year} {2005})}\BibitemShut {NoStop}%
\bibitem [{\citenamefont {Spivak}\ \emph {et~al.}(2012)\citenamefont {Spivak},
  \citenamefont {Litvin}, \citenamefont {Ovsyannikov}, \citenamefont
  {Dubrovinskaia},\ and\ \citenamefont {Dubrovinsky}}]{Spivak2012}%
  \BibitemOpen
  \bibfield  {author} {\bibinfo {author} {\bibfnamefont {A.~V.}\ \bibnamefont
  {Spivak}}, \bibinfo {author} {\bibfnamefont {Y.~A.}\ \bibnamefont {Litvin}},
  \bibinfo {author} {\bibfnamefont {S.~V.}\ \bibnamefont {Ovsyannikov}},
  \bibinfo {author} {\bibfnamefont {N.~A.}\ \bibnamefont {Dubrovinskaia}}, \
  and\ \bibinfo {author} {\bibfnamefont {L.~S.}\ \bibnamefont {Dubrovinsky}},\
  }\href {\doibase 10.1016/j.jssc.2012.02.041} {\bibfield  {journal} {\bibinfo
  {journal} {J. Solid State Chem.}\ }\textbf {\bibinfo {volume} {191}},\
  \bibinfo {pages} {102} (\bibinfo {year} {2012})}\BibitemShut {NoStop}%
\bibitem [{\citenamefont {Solopova}\ \emph {et~al.}(2013)\citenamefont
  {Solopova}, \citenamefont {Litvin}, \citenamefont {Spivak}, \citenamefont
  {Dubrovinskaia}, \citenamefont {Dubrovinsky},\ and\ \citenamefont
  {Urusov}}]{Solopova2013}%
  \BibitemOpen
  \bibfield  {author} {\bibinfo {author} {\bibfnamefont {N.~A.}\ \bibnamefont
  {Solopova}}, \bibinfo {author} {\bibfnamefont {Y.~A.}\ \bibnamefont
  {Litvin}}, \bibinfo {author} {\bibfnamefont {A.~V.}\ \bibnamefont {Spivak}},
  \bibinfo {author} {\bibfnamefont {N.~A.}\ \bibnamefont {Dubrovinskaia}},
  \bibinfo {author} {\bibfnamefont {L.~S.}\ \bibnamefont {Dubrovinsky}}, \ and\
  \bibinfo {author} {\bibfnamefont {V.~S.}\ \bibnamefont {Urusov}},\ }\href
  {\doibase 10.1134/S1028334X13110068} {\bibfield  {journal} {\bibinfo
  {journal} {Dokl. Earth Sci.}\ }\textbf {\bibinfo {volume} {453}},\ \bibinfo
  {pages} {1106} (\bibinfo {year} {2013})}\BibitemShut {NoStop}%
\bibitem [{\citenamefont {Solopova}\ \emph {et~al.}(2015)\citenamefont
  {Solopova}, \citenamefont {Dubrovinsky}, \citenamefont {Spivak},
  \citenamefont {Litvin},\ and\ \citenamefont {Dubrovinskaia}}]{Solopova2015}%
  \BibitemOpen
  \bibfield  {author} {\bibinfo {author} {\bibfnamefont {N.~A.}\ \bibnamefont
  {Solopova}}, \bibinfo {author} {\bibfnamefont {L.}~\bibnamefont
  {Dubrovinsky}}, \bibinfo {author} {\bibfnamefont {A.~V.}\ \bibnamefont
  {Spivak}}, \bibinfo {author} {\bibfnamefont {Y.~A.}\ \bibnamefont {Litvin}},
  \ and\ \bibinfo {author} {\bibfnamefont {N.}~\bibnamefont {Dubrovinskaia}},\
  }\href {\doibase 10.1007/s00269-014-0701-1} {\bibfield  {journal} {\bibinfo
  {journal} {Phys. Chem. Miner.}\ }\textbf {\bibinfo {volume} {42}},\ \bibinfo
  {pages} {73} (\bibinfo {year} {2015})}\BibitemShut {NoStop}%
\bibitem [{\citenamefont {Irving}\ and\ \citenamefont
  {Wyllie}(1975)}]{Irving1975}%
  \BibitemOpen
  \bibfield  {author} {\bibinfo {author} {\bibfnamefont {A.~J.}\ \bibnamefont
  {Irving}}\ and\ \bibinfo {author} {\bibfnamefont {P.~J.}\ \bibnamefont
  {Wyllie}},\ }\href {\doibase 10.1016/0016-7037(75)90183-0} {\bibfield
  {journal} {\bibinfo  {journal} {Geochim. Cosmochim. Acta}\ }\textbf {\bibinfo
  {volume} {39}},\ \bibinfo {pages} {35} (\bibinfo {year} {1975})}\BibitemShut
  {NoStop}%
\bibitem [{\citenamefont {Suito}\ \emph {et~al.}(2001)\citenamefont {Suito},
  \citenamefont {Namba}, \citenamefont {Horikawa}, \citenamefont {Taniguchi},
  \citenamefont {Sakurai}, \citenamefont {Kobayashi}, \citenamefont {Onodera},
  \citenamefont {Shimomura},\ and\ \citenamefont {Kikegawa}}]{Suito2001}%
  \BibitemOpen
  \bibfield  {author} {\bibinfo {author} {\bibfnamefont {K.}~\bibnamefont
  {Suito}}, \bibinfo {author} {\bibfnamefont {J.}~\bibnamefont {Namba}},
  \bibinfo {author} {\bibfnamefont {T.}~\bibnamefont {Horikawa}}, \bibinfo
  {author} {\bibfnamefont {Y.}~\bibnamefont {Taniguchi}}, \bibinfo {author}
  {\bibfnamefont {N.}~\bibnamefont {Sakurai}}, \bibinfo {author} {\bibfnamefont
  {M.}~\bibnamefont {Kobayashi}}, \bibinfo {author} {\bibfnamefont
  {A.}~\bibnamefont {Onodera}}, \bibinfo {author} {\bibfnamefont
  {O.}~\bibnamefont {Shimomura}}, \ and\ \bibinfo {author} {\bibfnamefont
  {T.}~\bibnamefont {Kikegawa}},\ }\href {\doibase 10.2138/am-2001-8-906}
  {\bibfield  {journal} {\bibinfo  {journal} {Am. Mineral.}\ }\textbf {\bibinfo
  {volume} {86}},\ \bibinfo {pages} {997} (\bibinfo {year} {2001})}\BibitemShut
  {NoStop}%
\bibitem [{\citenamefont {Kojima}(2009)}]{Kojima2009}%
  \BibitemOpen
  \bibfield  {author} {\bibinfo {author} {\bibfnamefont {T.}~\bibnamefont
  {Kojima}},\ }\emph {\bibinfo {title} {Physical and Chemical Properties of
  Molten Carbonates}},\ \href
  {http://www.lib.kobe-u.ac.jp/repository/thesis/d1/D1004596.pdf} {Ph.D.
  thesis},\ \bibinfo  {school} {Kobe University} (\bibinfo {year}
  {2009})\BibitemShut {NoStop}%
\bibitem [{\citenamefont {Dobson}\ \emph {et~al.}(1996)\citenamefont {Dobson},
  \citenamefont {Jones}, \citenamefont {Rabe}, \citenamefont {Sekine},
  \citenamefont {Kurita}, \citenamefont {Taniguchi}, \citenamefont {Kondo},
  \citenamefont {Kato}, \citenamefont {Shimomura},\ and\ \citenamefont
  {Urakawa}}]{Dobson1996}%
  \BibitemOpen
  \bibfield  {author} {\bibinfo {author} {\bibfnamefont {D.~P.}\ \bibnamefont
  {Dobson}}, \bibinfo {author} {\bibfnamefont {A.~P.}\ \bibnamefont {Jones}},
  \bibinfo {author} {\bibfnamefont {R.}~\bibnamefont {Rabe}}, \bibinfo {author}
  {\bibfnamefont {T.}~\bibnamefont {Sekine}}, \bibinfo {author} {\bibfnamefont
  {K.}~\bibnamefont {Kurita}}, \bibinfo {author} {\bibfnamefont
  {T.}~\bibnamefont {Taniguchi}}, \bibinfo {author} {\bibfnamefont
  {T.}~\bibnamefont {Kondo}}, \bibinfo {author} {\bibfnamefont
  {T.}~\bibnamefont {Kato}}, \bibinfo {author} {\bibfnamefont {O.}~\bibnamefont
  {Shimomura}}, \ and\ \bibinfo {author} {\bibfnamefont {S.}~\bibnamefont
  {Urakawa}},\ }\href {\doibase 10.1016/0012-821X(96)00139-2} {\bibfield
  {journal} {\bibinfo  {journal} {Earth Planet. Sci. Lett.}\ }\textbf {\bibinfo
  {volume} {143}},\ \bibinfo {pages} {207} (\bibinfo {year}
  {1996})}\BibitemShut {NoStop}%
\bibitem [{\citenamefont {Kono}\ \emph {et~al.}(2014)\citenamefont {Kono},
  \citenamefont {Kenney-Benson}, \citenamefont {Hummer}, \citenamefont
  {Ohfuji}, \citenamefont {Park}, \citenamefont {Shen}, \citenamefont {Wang},
  \citenamefont {Kavner},\ and\ \citenamefont {Manning~C.}}]{Kono2014}%
  \BibitemOpen
  \bibfield  {author} {\bibinfo {author} {\bibfnamefont {Y.}~\bibnamefont
  {Kono}}, \bibinfo {author} {\bibfnamefont {C.}~\bibnamefont {Kenney-Benson}},
  \bibinfo {author} {\bibfnamefont {D.}~\bibnamefont {Hummer}}, \bibinfo
  {author} {\bibfnamefont {H.}~\bibnamefont {Ohfuji}}, \bibinfo {author}
  {\bibfnamefont {C.}~\bibnamefont {Park}}, \bibinfo {author} {\bibfnamefont
  {G.}~\bibnamefont {Shen}}, \bibinfo {author} {\bibfnamefont {Y.}~\bibnamefont
  {Wang}}, \bibinfo {author} {\bibfnamefont {A.}~\bibnamefont {Kavner}}, \ and\
  \bibinfo {author} {\bibfnamefont {E.}~\bibnamefont {Manning~C.}},\ }\href
  {\doibase 10.1038/ncomms6091} {\bibfield  {journal} {\bibinfo  {journal}
  {Nat. Commun.}\ }\textbf {\bibinfo {volume} {5}},\ \bibinfo {pages} {5091}
  (\bibinfo {year} {2014})}\BibitemShut {NoStop}%
\bibitem [{\citenamefont {Sifr\'e}, \citenamefont {Hashim},\ and\ \citenamefont
  {Gaillard}(2015)}]{Sifre2015}%
  \BibitemOpen
  \bibfield  {author} {\bibinfo {author} {\bibfnamefont {D.}~\bibnamefont
  {Sifr\'e}}, \bibinfo {author} {\bibfnamefont {L.}~\bibnamefont {Hashim}}, \
  and\ \bibinfo {author} {\bibfnamefont {F.}~\bibnamefont {Gaillard}},\ }\href
  {\doibase 10.1016/j.chemgeo.2014.09.022} {\bibfield  {journal} {\bibinfo
  {journal} {Chem. Geol.}\ }\textbf {\bibinfo {volume} {418}},\ \bibinfo
  {pages} {189} (\bibinfo {year} {2015})}\BibitemShut {NoStop}%
\bibitem [{\citenamefont {Hurt}\ and\ \citenamefont {Lange}(2019)}]{Hurt2019}%
  \BibitemOpen
  \bibfield  {author} {\bibinfo {author} {\bibfnamefont {S.~M.}\ \bibnamefont
  {Hurt}}\ and\ \bibinfo {author} {\bibfnamefont {R.~A.}\ \bibnamefont
  {Lange}},\ }\href {\doibase 10.1016/j.gca.2018.12.031} {\bibfield  {journal}
  {\bibinfo  {journal} {Geochim. Cosmochim. Acta}\ ,\ \bibinfo {pages} {123}}
  (\bibinfo {year} {2019})}\BibitemShut {NoStop}%
\bibitem [{\citenamefont {Zhang}\ and\ \citenamefont {Liu}(2015)}]{Zhang2015}%
  \BibitemOpen
  \bibfield  {author} {\bibinfo {author} {\bibfnamefont {Z.}~\bibnamefont
  {Zhang}}\ and\ \bibinfo {author} {\bibfnamefont {Z.}~\bibnamefont {Liu}},\
  }\href {\doibase 10.1007/s11631-015-0036-8} {\bibfield  {journal} {\bibinfo
  {journal} {Chin. J. Geochem.}\ }\textbf {\bibinfo {volume} {34}},\ \bibinfo
  {pages} {13} (\bibinfo {year} {2015})}\BibitemShut {NoStop}%
\bibitem [{\citenamefont {Li}\ \emph {et~al.}(2017)\citenamefont {Li},
  \citenamefont {Li}, \citenamefont {Lange}, \citenamefont {Liu},\ and\
  \citenamefont {Militzer}}]{Li2017}%
  \BibitemOpen
  \bibfield  {author} {\bibinfo {author} {\bibfnamefont {Z.}~\bibnamefont
  {Li}}, \bibinfo {author} {\bibfnamefont {J.}~\bibnamefont {Li}}, \bibinfo
  {author} {\bibfnamefont {R.}~\bibnamefont {Lange}}, \bibinfo {author}
  {\bibfnamefont {J.}~\bibnamefont {Liu}}, \ and\ \bibinfo {author}
  {\bibfnamefont {B.}~\bibnamefont {Militzer}},\ }\href {\doibase
  10.1016/j.epsl.2016.10.027} {\bibfield  {journal} {\bibinfo  {journal} {Earth
  Planet. Sci. Lett.}\ }\textbf {\bibinfo {volume} {457}},\ \bibinfo {pages}
  {395} (\bibinfo {year} {2017})}\BibitemShut {NoStop}%
\bibitem [{\citenamefont {Vuilleumier}\ \emph {et~al.}(2014)\citenamefont
  {Vuilleumier}, \citenamefont {Seitsonen}, \citenamefont {Sator},\ and\
  \citenamefont {Guillot}}]{Vuilleumier2014}%
  \BibitemOpen
  \bibfield  {author} {\bibinfo {author} {\bibfnamefont {R.}~\bibnamefont
  {Vuilleumier}}, \bibinfo {author} {\bibfnamefont {A.}~\bibnamefont
  {Seitsonen}}, \bibinfo {author} {\bibfnamefont {N.}~\bibnamefont {Sator}}, \
  and\ \bibinfo {author} {\bibfnamefont {B.}~\bibnamefont {Guillot}},\ }\href
  {\doibase 10.1016/j.gca.2014.06.037} {\bibfield  {journal} {\bibinfo
  {journal} {Geochim. Cosmochim. Acta}\ }\textbf {\bibinfo {volume} {141}},\
  \bibinfo {pages} {547} (\bibinfo {year} {2014})}\BibitemShut {NoStop}%
\bibitem [{\citenamefont {Du}\ \emph {et~al.}(2018)\citenamefont {Du},
  \citenamefont {Wu}, \citenamefont {Tse},\ and\ \citenamefont {Pan}}]{Du2018}%
  \BibitemOpen
  \bibfield  {author} {\bibinfo {author} {\bibfnamefont {X.}~\bibnamefont
  {Du}}, \bibinfo {author} {\bibfnamefont {M.}~\bibnamefont {Wu}}, \bibinfo
  {author} {\bibfnamefont {J.~S.}\ \bibnamefont {Tse}}, \ and\ \bibinfo
  {author} {\bibfnamefont {Y.}~\bibnamefont {Pan}},\ }\href {\doibase
  10.1021/acsearthspacechem.7b00100} {\bibfield  {journal} {\bibinfo  {journal}
  {ACS Earth and Space Chem.}\ }\textbf {\bibinfo {volume} {2}},\ \bibinfo
  {pages} {1} (\bibinfo {year} {2018})}\BibitemShut {NoStop}%
\bibitem [{\citenamefont {Corradini}, \citenamefont {F.-X.},\ and\
  \citenamefont {Vuilleumier}(2016)}]{Corradini2016}%
  \BibitemOpen
  \bibfield  {author} {\bibinfo {author} {\bibfnamefont {D.}~\bibnamefont
  {Corradini}}, \bibinfo {author} {\bibfnamefont {C.}~\bibnamefont {F.-X.}}, \
  and\ \bibinfo {author} {\bibfnamefont {R.}~\bibnamefont {Vuilleumier}},\
  }\href {\doibase 10.1063/1.4943392} {\bibfield  {journal} {\bibinfo
  {journal} {J. Chem. Phys.}\ }\textbf {\bibinfo {volume} {144}},\ \bibinfo
  {pages} {104507} (\bibinfo {year} {2016})}\BibitemShut {NoStop}%
\bibitem [{\citenamefont {Liu}\ and\ \citenamefont {Lange}(2003)}]{Liu2003}%
  \BibitemOpen
  \bibfield  {author} {\bibinfo {author} {\bibfnamefont {Q.}~\bibnamefont
  {Liu}}\ and\ \bibinfo {author} {\bibfnamefont {R.}~\bibnamefont {Lange}},\
  }\href {\doibase 10.1007/s00410-003-0505-7} {\bibfield  {journal} {\bibinfo
  {journal} {Contrib. Mineral. Petrol.}\ }\textbf {\bibinfo {volume} {146}},\
  \bibinfo {pages} {370} (\bibinfo {year} {2003})}\BibitemShut {NoStop}%
\bibitem [{\citenamefont {Hudspeth}, \citenamefont {Sanloup},\ and\
  \citenamefont {Kono}(2018)}]{Hudspeth2018}%
  \BibitemOpen
  \bibfield  {author} {\bibinfo {author} {\bibfnamefont {J.}~\bibnamefont
  {Hudspeth}}, \bibinfo {author} {\bibfnamefont {C.}~\bibnamefont {Sanloup}}, \
  and\ \bibinfo {author} {\bibfnamefont {Y.}~\bibnamefont {Kono}},\ }\href
  {\doibase 10.7185/geochemlet.1813} {\bibfield  {journal} {\bibinfo  {journal}
  {Geochem. Perspect. Lett.}\ }\textbf {\bibinfo {volume} {7}},\ \bibinfo
  {pages} {17} (\bibinfo {year} {2018})}\BibitemShut {NoStop}%
\bibitem [{\citenamefont {Yuen}, \citenamefont {Lister},\ and\ \citenamefont
  {Nyburg}(1978)}]{Yuen1978}%
  \BibitemOpen
  \bibfield  {author} {\bibinfo {author} {\bibfnamefont {P.~S.}\ \bibnamefont
  {Yuen}}, \bibinfo {author} {\bibfnamefont {M.~W.}\ \bibnamefont {Lister}}, \
  and\ \bibinfo {author} {\bibfnamefont {S.~C.}\ \bibnamefont {Nyburg}},\
  }\href {\doibase 10.1063/1.435920} {\bibfield  {journal} {\bibinfo  {journal}
  {J. Chem. Phys.}\ }\textbf {\bibinfo {volume} {68}},\ \bibinfo {pages} {1936}
  (\bibinfo {year} {1978})}\BibitemShut {NoStop}%
\bibitem [{\citenamefont {Dove}\ \emph {et~al.}(1992)\citenamefont {Dove},
  \citenamefont {Winkler}, \citenamefont {Leslie}, \citenamefont {Harris},\
  and\ \citenamefont {Salje}}]{Dove1992}%
  \BibitemOpen
  \bibfield  {author} {\bibinfo {author} {\bibfnamefont {M.~T.}\ \bibnamefont
  {Dove}}, \bibinfo {author} {\bibfnamefont {B.}~\bibnamefont {Winkler}},
  \bibinfo {author} {\bibfnamefont {M.}~\bibnamefont {Leslie}}, \bibinfo
  {author} {\bibfnamefont {M.~J.}\ \bibnamefont {Harris}}, \ and\ \bibinfo
  {author} {\bibfnamefont {E.~K.~H.}\ \bibnamefont {Salje}},\ }\href
  {https://pubs.geoscienceworld.org/msa/ammin/article-abstract/77/3-4/244/42613/a-new-interatomic-potential-model-for-calcite?redirectedFrom=fulltext}
  {\bibfield  {journal} {\bibinfo  {journal} {Am. Mineral.}\ }\textbf {\bibinfo
  {volume} {77}},\ \bibinfo {pages} {244} (\bibinfo {year} {1992})}\BibitemShut
  {NoStop}%
\bibitem [{\citenamefont {Pavese}\ \emph {et~al.}(1992)\citenamefont {Pavese},
  \citenamefont {Catti}, \citenamefont {Price},\ and\ \citenamefont
  {Jackson}}]{Pavese1992}%
  \BibitemOpen
  \bibfield  {author} {\bibinfo {author} {\bibfnamefont {A.}~\bibnamefont
  {Pavese}}, \bibinfo {author} {\bibfnamefont {M.}~\bibnamefont {Catti}},
  \bibinfo {author} {\bibfnamefont {G.~D.}\ \bibnamefont {Price}}, \ and\
  \bibinfo {author} {\bibfnamefont {R.~A.}\ \bibnamefont {Jackson}},\ }\href
  {\doibase 10.1007/BF00198605} {\bibfield  {journal} {\bibinfo  {journal}
  {Phys. Chem. Miner.}\ }\textbf {\bibinfo {volume} {19}},\ \bibinfo {pages}
  {80} (\bibinfo {year} {1992})}\BibitemShut {NoStop}%
\bibitem [{\citenamefont {Pavese}\ \emph {et~al.}(1996)\citenamefont {Pavese},
  \citenamefont {Catti}, \citenamefont {Parker},\ and\ \citenamefont
  {Wall}}]{Pavese1996}%
  \BibitemOpen
  \bibfield  {author} {\bibinfo {author} {\bibfnamefont {A.}~\bibnamefont
  {Pavese}}, \bibinfo {author} {\bibfnamefont {M.}~\bibnamefont {Catti}},
  \bibinfo {author} {\bibfnamefont {S.~C.}\ \bibnamefont {Parker}}, \ and\
  \bibinfo {author} {\bibfnamefont {A.}~\bibnamefont {Wall}},\ }\href {\doibase
  10.1007/BF00202303} {\bibfield  {journal} {\bibinfo  {journal} {Phys. Chem.
  Miner.}\ }\textbf {\bibinfo {volume} {23}},\ \bibinfo {pages} {89} (\bibinfo
  {year} {1996})}\BibitemShut {NoStop}%
\bibitem [{\citenamefont {Fisler}, \citenamefont {Gale},\ and\ \citenamefont
  {Cygan}(2000)}]{Fisler2000}%
  \BibitemOpen
  \bibfield  {author} {\bibinfo {author} {\bibfnamefont {D.~K.}\ \bibnamefont
  {Fisler}}, \bibinfo {author} {\bibfnamefont {J.~D.}\ \bibnamefont {Gale}}, \
  and\ \bibinfo {author} {\bibfnamefont {R.~T.}\ \bibnamefont {Cygan}},\ }\href
  {\doibase 10.2138/am-2000-0121} {\bibfield  {journal} {\bibinfo  {journal}
  {Am. Mineral.}\ }\textbf {\bibinfo {volume} {85}},\ \bibinfo {pages} {217}
  (\bibinfo {year} {2000})}\BibitemShut {NoStop}%
\bibitem [{\citenamefont {Archer}\ \emph {et~al.}(2003)\citenamefont {Archer},
  \citenamefont {Birse}, \citenamefont {Dove}, \citenamefont {Redfern},
  \citenamefont {Gale},\ and\ \citenamefont {Cygan}}]{Archer2003}%
  \BibitemOpen
  \bibfield  {author} {\bibinfo {author} {\bibfnamefont {T.~D.}\ \bibnamefont
  {Archer}}, \bibinfo {author} {\bibfnamefont {S.~E.~A.}\ \bibnamefont
  {Birse}}, \bibinfo {author} {\bibfnamefont {M.~T.}\ \bibnamefont {Dove}},
  \bibinfo {author} {\bibfnamefont {S.~A.~T.}\ \bibnamefont {Redfern}},
  \bibinfo {author} {\bibfnamefont {J.~D.}\ \bibnamefont {Gale}}, \ and\
  \bibinfo {author} {\bibfnamefont {R.~T.}\ \bibnamefont {Cygan}},\ }\href
  {\doibase 10.1007/s00269-002-0269-z} {\bibfield  {journal} {\bibinfo
  {journal} {Phys. Chem. Miner.}\ }\textbf {\bibinfo {volume} {30}},\ \bibinfo
  {pages} {416} (\bibinfo {year} {2003})}\BibitemShut {NoStop}%
\bibitem [{\citenamefont {Raiteri}\ \emph {et~al.}(2010)\citenamefont
  {Raiteri}, \citenamefont {Gale}, \citenamefont {Quigley},\ and\ \citenamefont
  {Rodger}}]{Raiteri2010}%
  \BibitemOpen
  \bibfield  {author} {\bibinfo {author} {\bibfnamefont {P.}~\bibnamefont
  {Raiteri}}, \bibinfo {author} {\bibfnamefont {J.~D.}\ \bibnamefont {Gale}},
  \bibinfo {author} {\bibfnamefont {D.}~\bibnamefont {Quigley}}, \ and\
  \bibinfo {author} {\bibfnamefont {P.~M.}\ \bibnamefont {Rodger}},\ }\href
  {\doibase 10.1021/jp910977a} {\bibfield  {journal} {\bibinfo  {journal} {J.
  Phys. Chem. C}\ }\textbf {\bibinfo {volume} {114}},\ \bibinfo {pages} {5997}
  (\bibinfo {year} {2010})}\BibitemShut {NoStop}%
\bibitem [{\citenamefont {Born}\ and\ \citenamefont {Huang}(1954)}]{Born1954}%
  \BibitemOpen
  \bibfield  {author} {\bibinfo {author} {\bibfnamefont {M.}~\bibnamefont
  {Born}}\ and\ \bibinfo {author} {\bibfnamefont {K.}~\bibnamefont {Huang}},\
  }\href@noop {} {\emph {\bibinfo {title} {{Dynamical Theory of Crystal
  Lattices}}}}\ (\bibinfo  {publisher} {Clarendon press},\ \bibinfo {year}
  {1954})\BibitemShut {NoStop}%
\bibitem [{\citenamefont {Genge}, \citenamefont {Price},\ and\ \citenamefont
  {Jones}(1995)}]{Genge1995}%
  \BibitemOpen
  \bibfield  {author} {\bibinfo {author} {\bibfnamefont {M.~J.}\ \bibnamefont
  {Genge}}, \bibinfo {author} {\bibfnamefont {G.~D.}\ \bibnamefont {Price}}, \
  and\ \bibinfo {author} {\bibfnamefont {A.~P.}\ \bibnamefont {Jones}},\ }\href
  {http://www3.imperial.ac.uk/pls/portallive/docs/1/6831920.PDF} {\bibfield
  {journal} {\bibinfo  {journal} {Earth Planet. Sci. Lett.}\ }\textbf {\bibinfo
  {volume} {131}},\ \bibinfo {pages} {225} (\bibinfo {year}
  {1995})}\BibitemShut {NoStop}%
\bibitem [{\citenamefont {Hurt}\ and\ \citenamefont {Wolf}(2018)}]{Hurt2018}%
  \BibitemOpen
  \bibfield  {author} {\bibinfo {author} {\bibfnamefont {S.~M.}\ \bibnamefont
  {Hurt}}\ and\ \bibinfo {author} {\bibfnamefont {A.~S.}\ \bibnamefont
  {Wolf}},\ }\href {\doibase 10.1007/s00269-018-0995-5} {\bibfield  {journal}
  {\bibinfo  {journal} {Phys. Chem. Miner.}\ }\textbf {\bibinfo {volume}
  {46}},\ \bibinfo {pages} {165} (\bibinfo {year} {2018})}\BibitemShut
  {NoStop}%
\bibitem [{\citenamefont {VandeVondele}\ \emph
  {et~al.}(2005{\natexlab{a}})\citenamefont {VandeVondele}, \citenamefont
  {Krack}, \citenamefont {Mohamed}, \citenamefont {Parrinello}, \citenamefont
  {Chassaing},\ and\ \citenamefont {Hutter}}]{VandeVondele2005a}%
  \BibitemOpen
  \bibfield  {author} {\bibinfo {author} {\bibfnamefont {J.}~\bibnamefont
  {VandeVondele}}, \bibinfo {author} {\bibfnamefont {M.}~\bibnamefont {Krack}},
  \bibinfo {author} {\bibfnamefont {F.}~\bibnamefont {Mohamed}}, \bibinfo
  {author} {\bibfnamefont {M.}~\bibnamefont {Parrinello}}, \bibinfo {author}
  {\bibfnamefont {T.}~\bibnamefont {Chassaing}}, \ and\ \bibinfo {author}
  {\bibfnamefont {J.}~\bibnamefont {Hutter}},\ }\href {\doibase
  10.1016/j.cpc.2004.12.014} {\bibfield  {journal} {\bibinfo  {journal}
  {Comput. Phys. Commun.}\ }\textbf {\bibinfo {volume} {167}},\ \bibinfo
  {pages} {103} (\bibinfo {year} {2005}{\natexlab{a}})}\BibitemShut {NoStop}%
\bibitem [{\citenamefont {Lippert}, \citenamefont {Hutter},\ and\ \citenamefont
  {Parrinello}(1997)}]{Lippert1997}%
  \BibitemOpen
  \bibfield  {author} {\bibinfo {author} {\bibfnamefont {G.}~\bibnamefont
  {Lippert}}, \bibinfo {author} {\bibfnamefont {J.}~\bibnamefont {Hutter}}, \
  and\ \bibinfo {author} {\bibfnamefont {M.}~\bibnamefont {Parrinello}},\
  }\href {\doibase 10.1080/002689797170220} {\bibfield  {journal} {\bibinfo
  {journal} {Mol. Phys.}\ }\textbf {\bibinfo {volume} {92}},\ \bibinfo {pages}
  {477} (\bibinfo {year} {1997})}\BibitemShut {NoStop}%
\bibitem [{\citenamefont {VandeVondele}\ \emph
  {et~al.}(2005{\natexlab{b}})\citenamefont {VandeVondele}, \citenamefont
  {Mohamed}, \citenamefont {Krack}, \citenamefont {Hutter}, \citenamefont
  {Sprik},\ and\ \citenamefont {Parrinello}}]{VandeVondele2005b}%
  \BibitemOpen
  \bibfield  {author} {\bibinfo {author} {\bibfnamefont {J.}~\bibnamefont
  {VandeVondele}}, \bibinfo {author} {\bibfnamefont {F.}~\bibnamefont
  {Mohamed}}, \bibinfo {author} {\bibfnamefont {M.}~\bibnamefont {Krack}},
  \bibinfo {author} {\bibfnamefont {J.}~\bibnamefont {Hutter}}, \bibinfo
  {author} {\bibfnamefont {M.}~\bibnamefont {Sprik}}, \ and\ \bibinfo {author}
  {\bibfnamefont {M.}~\bibnamefont {Parrinello}},\ }\href {\doibase
  10.1063/1.1828433} {\bibfield  {journal} {\bibinfo  {journal} {J. Chem.
  Phys.}\ }\textbf {\bibinfo {volume} {122}},\ \bibinfo {pages} {101} (\bibinfo
  {year} {2005}{\natexlab{b}})}\BibitemShut {NoStop}%
\bibitem [{\citenamefont {VandeVondele}\ and\ \citenamefont
  {Hutter}(2007)}]{VdVHutter2007}%
  \BibitemOpen
  \bibfield  {author} {\bibinfo {author} {\bibfnamefont {J.}~\bibnamefont
  {VandeVondele}}\ and\ \bibinfo {author} {\bibfnamefont {J.}~\bibnamefont
  {Hutter}},\ }\href {\doibase 10.1063/1.1828433} {\bibfield  {journal}
  {\bibinfo  {journal} {J. Chem. Phys.}\ }\textbf {\bibinfo {volume} {127}},\
  \bibinfo {pages} {114105} (\bibinfo {year} {2007})}\BibitemShut {NoStop}%
\bibitem [{\citenamefont {Goedecker}, \citenamefont {Teter},\ and\
  \citenamefont {Hutter}(1996)}]{Goedecker1996}%
  \BibitemOpen
  \bibfield  {author} {\bibinfo {author} {\bibfnamefont {S.}~\bibnamefont
  {Goedecker}}, \bibinfo {author} {\bibfnamefont {M.}~\bibnamefont {Teter}}, \
  and\ \bibinfo {author} {\bibfnamefont {J.}~\bibnamefont {Hutter}},\ }\href
  {\doibase 10.1103/PhysRevB.54.1703} {\bibfield  {journal} {\bibinfo
  {journal} {Phys. Rev. B}\ }\textbf {\bibinfo {volume} {54}},\ \bibinfo
  {pages} {1703} (\bibinfo {year} {1996})}\BibitemShut {NoStop}%
\bibitem [{\citenamefont {Hartwigsen}, \citenamefont {Goedecker},\ and\
  \citenamefont {Hutter}(1998)}]{Hartwigsen1998}%
  \BibitemOpen
  \bibfield  {author} {\bibinfo {author} {\bibfnamefont {C.}~\bibnamefont
  {Hartwigsen}}, \bibinfo {author} {\bibfnamefont {S.}~\bibnamefont
  {Goedecker}}, \ and\ \bibinfo {author} {\bibfnamefont {J.}~\bibnamefont
  {Hutter}},\ }\href {\doibase 10.1103/PhysRevB.58.3641} {\bibfield  {journal}
  {\bibinfo  {journal} {Phys. Rev. B}\ }\textbf {\bibinfo {volume} {58}},\
  \bibinfo {pages} {3641} (\bibinfo {year} {1998})}\BibitemShut {NoStop}%
\bibitem [{\citenamefont {Krack}(2005)}]{Krack2005}%
  \BibitemOpen
  \bibfield  {author} {\bibinfo {author} {\bibfnamefont {M.}~\bibnamefont
  {Krack}},\ }\href {\doibase 10.1007/s00214-005-0655-y} {\bibfield  {journal}
  {\bibinfo  {journal} {Theor. Chem. Acc.}\ }\textbf {\bibinfo {volume}
  {114}},\ \bibinfo {pages} {145} (\bibinfo {year} {2005})}\BibitemShut
  {NoStop}%
\bibitem [{\citenamefont {Becke~A.}(1988)}]{Becke1988}%
  \BibitemOpen
  \bibfield  {author} {\bibinfo {author} {\bibfnamefont {D.}~\bibnamefont
  {Becke~A.}},\ }\href {\doibase 10.1103/PhysRevA.38.3098} {\bibfield
  {journal} {\bibinfo  {journal} {Phys. Rev. A}\ }\textbf {\bibinfo {volume}
  {38}},\ \bibinfo {pages} {3098} (\bibinfo {year} {1988})}\BibitemShut
  {NoStop}%
\bibitem [{\citenamefont {Lee}, \citenamefont {Yang},\ and\ \citenamefont
  {Parr~R.}(1988)}]{Lee1988}%
  \BibitemOpen
  \bibfield  {author} {\bibinfo {author} {\bibfnamefont {C.}~\bibnamefont
  {Lee}}, \bibinfo {author} {\bibfnamefont {W.}~\bibnamefont {Yang}}, \ and\
  \bibinfo {author} {\bibfnamefont {G.}~\bibnamefont {Parr~R.}},\ }\href
  {\doibase 10.1103/PhysRevB.37.785} {\bibfield  {journal} {\bibinfo  {journal}
  {Phys. Rev. B}\ }\textbf {\bibinfo {volume} {37}},\ \bibinfo {pages} {785}
  (\bibinfo {year} {1988})}\BibitemShut {NoStop}%
\bibitem [{\citenamefont {Grimme}\ \emph {et~al.}(2010)\citenamefont {Grimme},
  \citenamefont {Antony}, \citenamefont {Ehrlich},\ and\ \citenamefont
  {Krieg}}]{Grimme2010}%
  \BibitemOpen
  \bibfield  {author} {\bibinfo {author} {\bibfnamefont {S.}~\bibnamefont
  {Grimme}}, \bibinfo {author} {\bibfnamefont {J.}~\bibnamefont {Antony}},
  \bibinfo {author} {\bibfnamefont {S.}~\bibnamefont {Ehrlich}}, \ and\
  \bibinfo {author} {\bibfnamefont {H.}~\bibnamefont {Krieg}},\ }\href
  {\doibase 10.1063/1.3382344} {\bibfield  {journal} {\bibinfo  {journal} {J.
  Chem. Phys.}\ }\textbf {\bibinfo {volume} {132}},\ \bibinfo {pages} {154104}
  (\bibinfo {year} {2010})}\BibitemShut {NoStop}%
\bibitem [{\citenamefont {Nos\'e}(1984{\natexlab{a}})}]{Nose1984a}%
  \BibitemOpen
  \bibfield  {author} {\bibinfo {author} {\bibfnamefont {S.}~\bibnamefont
  {Nos\'e}},\ }\href {\doibase 10.1080/00268978400101201} {\bibfield  {journal}
  {\bibinfo  {journal} {Mol. Phys.}\ }\textbf {\bibinfo {volume} {52}},\
  \bibinfo {pages} {255} (\bibinfo {year} {1984}{\natexlab{a}})}\BibitemShut
  {NoStop}%
\bibitem [{\citenamefont {Nos\'e}(1984{\natexlab{b}})}]{Nose1984b}%
  \BibitemOpen
  \bibfield  {author} {\bibinfo {author} {\bibfnamefont {S.}~\bibnamefont
  {Nos\'e}},\ }\href {\doibase 10.1063/1.447334} {\bibfield  {journal}
  {\bibinfo  {journal} {J. Chem. Phys.}\ }\textbf {\bibinfo {volume} {81}},\
  \bibinfo {pages} {511} (\bibinfo {year} {1984}{\natexlab{b}})}\BibitemShut
  {NoStop}%
\bibitem [{\citenamefont {Smith}\ and\ \citenamefont
  {Forester}(1996)}]{Smith1996}%
  \BibitemOpen
  \bibfield  {author} {\bibinfo {author} {\bibfnamefont {W.}~\bibnamefont
  {Smith}}\ and\ \bibinfo {author} {\bibfnamefont {T.}~\bibnamefont
  {Forester}},\ }\href {\doibase 10.1016/S0263-7855(96)00043-4} {\bibfield
  {journal} {\bibinfo  {journal} {J. Mol. Graphics}\ }\textbf {\bibinfo
  {volume} {14}},\ \bibinfo {pages} {136} (\bibinfo {year} {1996})}\BibitemShut
  {NoStop}%
\bibitem [{\citenamefont {Redfern}, \citenamefont {Wood},\ and\ \citenamefont
  {Henderson}(1993)}]{Redfern1993}%
  \BibitemOpen
  \bibfield  {author} {\bibinfo {author} {\bibfnamefont {S.~A.~T.}\
  \bibnamefont {Redfern}}, \bibinfo {author} {\bibfnamefont {B.~J.}\
  \bibnamefont {Wood}}, \ and\ \bibinfo {author} {\bibfnamefont {C.~M.~B.}\
  \bibnamefont {Henderson}},\ }\href {\doibase 10.1029/93GL02507} {\bibfield
  {journal} {\bibinfo  {journal} {Geophys. Res. Lett.}\ }\textbf {\bibinfo
  {volume} {20}},\ \bibinfo {pages} {2099} (\bibinfo {year}
  {1993})}\BibitemShut {NoStop}%
\bibitem [{\citenamefont {Fiquet}, \citenamefont {Guyot},\ and\ \citenamefont
  {Itie}(1994)}]{Fiquet1994}%
  \BibitemOpen
  \bibfield  {author} {\bibinfo {author} {\bibfnamefont {G.}~\bibnamefont
  {Fiquet}}, \bibinfo {author} {\bibfnamefont {F.}~\bibnamefont {Guyot}}, \
  and\ \bibinfo {author} {\bibfnamefont {J.-P.}\ \bibnamefont {Itie}},\ }\href
  {http://ammin.geoscienceworld.org/content/79/1-2/15.short} {\bibfield
  {journal} {\bibinfo  {journal} {Am. Mineral.}\ }\textbf {\bibinfo {volume}
  {79}},\ \bibinfo {pages} {15} (\bibinfo {year} {1994})}\BibitemShut {NoStop}%
\bibitem [{\citenamefont {Zhang}\ \emph {et~al.}(1997)\citenamefont {Zhang},
  \citenamefont {Martinez}, \citenamefont {Guyot}, \citenamefont {Gillet},\
  and\ \citenamefont {Saxena}}]{Zhang1997}%
  \BibitemOpen
  \bibfield  {author} {\bibinfo {author} {\bibfnamefont {J.}~\bibnamefont
  {Zhang}}, \bibinfo {author} {\bibfnamefont {I.}~\bibnamefont {Martinez}},
  \bibinfo {author} {\bibfnamefont {F.}~\bibnamefont {Guyot}}, \bibinfo
  {author} {\bibfnamefont {P.}~\bibnamefont {Gillet}}, \ and\ \bibinfo {author}
  {\bibfnamefont {S.~K.}\ \bibnamefont {Saxena}},\ }\href {\doibase
  10.1007/s002690050025} {\bibfield  {journal} {\bibinfo  {journal} {Phys.
  Chem. Miner.}\ }\textbf {\bibinfo {volume} {24}},\ \bibinfo {pages} {122}
  (\bibinfo {year} {1997})}\BibitemShut {NoStop}%
\bibitem [{\citenamefont {Ross}(1997)}]{Ross1997}%
  \BibitemOpen
  \bibfield  {author} {\bibinfo {author} {\bibfnamefont {N.~L.}\ \bibnamefont
  {Ross}},\ }\href {\doibase 10.2138/am-1997-7-805} {\bibfield  {journal}
  {\bibinfo  {journal} {Am. Mineral.}\ }\textbf {\bibinfo {volume} {82}},\
  \bibinfo {pages} {682} (\bibinfo {year} {1997})}\BibitemShut {NoStop}%
\bibitem [{\citenamefont {O'Leary}, \citenamefont {Lange},\ and\ \citenamefont
  {Ai}(2015)}]{OLeary2015}%
  \BibitemOpen
  \bibfield  {author} {\bibinfo {author} {\bibfnamefont {M.~C.}\ \bibnamefont
  {O'Leary}}, \bibinfo {author} {\bibfnamefont {R.~A.}\ \bibnamefont {Lange}},
  \ and\ \bibinfo {author} {\bibfnamefont {Y.}~\bibnamefont {Ai}},\ }\href
  {\doibase 10.1007/s00410-015-1157-0} {\bibfield  {journal} {\bibinfo
  {journal} {Contrib. Mineral. Petrol.}\ }\textbf {\bibinfo {volume} {170}},\
  \bibinfo {pages} {3} (\bibinfo {year} {2015})}\BibitemShut {NoStop}%
\bibitem [{\citenamefont {Antao}\ and\ \citenamefont
  {Hassan}(2010)}]{Antao2010}%
  \BibitemOpen
  \bibfield  {author} {\bibinfo {author} {\bibfnamefont {S.~M.}\ \bibnamefont
  {Antao}}\ and\ \bibinfo {author} {\bibfnamefont {I.}~\bibnamefont {Hassan}},\
  }\href {\doibase 10.3749/canmin.48.5.1225} {\bibfield  {journal} {\bibinfo
  {journal} {Can. Mineral.}\ }\textbf {\bibinfo {volume} {48}},\ \bibinfo
  {pages} {1225} (\bibinfo {year} {2010})}\BibitemShut {NoStop}%
\bibitem [{\citenamefont {Buob}\ \emph {et~al.}(2006)\citenamefont {Buob},
  \citenamefont {Luth}, \citenamefont {Schmidt},\ and\ \citenamefont
  {Ulmer}}]{Buob2006}%
  \BibitemOpen
  \bibfield  {author} {\bibinfo {author} {\bibfnamefont {A.}~\bibnamefont
  {Buob}}, \bibinfo {author} {\bibfnamefont {R.~W.}\ \bibnamefont {Luth}},
  \bibinfo {author} {\bibfnamefont {M.~W.}\ \bibnamefont {Schmidt}}, \ and\
  \bibinfo {author} {\bibfnamefont {P.}~\bibnamefont {Ulmer}},\ }\href
  {\doibase 10.2138/am.2006.1910} {\bibfield  {journal} {\bibinfo  {journal}
  {Am. Mineral.}\ }\textbf {\bibinfo {volume} {91}},\ \bibinfo {pages} {435}
  (\bibinfo {year} {2006})}\BibitemShut {NoStop}%
\bibitem [{\citenamefont {Shatskiy}, \citenamefont {Litasov},\ and\
  \citenamefont {Palyanov}(2015)}]{Shat2015}%
  \BibitemOpen
  \bibfield  {author} {\bibinfo {author} {\bibfnamefont {A.}~\bibnamefont
  {Shatskiy}}, \bibinfo {author} {\bibfnamefont {K.}~\bibnamefont {Litasov}}, \
  and\ \bibinfo {author} {\bibfnamefont {Y.}~\bibnamefont {Palyanov}},\ }\href
  {\doibase 10.1016/j.rgg.2015.01.007} {\bibfield  {journal} {\bibinfo
  {journal} {Russ. Geol. Geophys.}\ }\textbf {\bibinfo {volume} {56}},\
  \bibinfo {pages} {113} (\bibinfo {year} {2015})}\BibitemShut {NoStop}%
\bibitem [{\citenamefont {Janz}(1988)}]{Janz1988}%
  \BibitemOpen
  \bibfield  {author} {\bibinfo {author} {\bibfnamefont {G.~J.}\ \bibnamefont
  {Janz}},\ }\href {https://srd.nist.gov/JPCRD/jpcrdS2Vol17.pdf} {\bibfield
  {journal} {\bibinfo  {journal} {J. Phys. Chem. Ref. Data}\ }\textbf {\bibinfo
  {volume} {17 (Suppl. 2)}} (\bibinfo {year} {1988})}\BibitemShut {NoStop}%
\bibitem [{\citenamefont {Markgraf}\ and\ \citenamefont
  {Reeder}(1985)}]{Markgraf1985}%
  \BibitemOpen
  \bibfield  {author} {\bibinfo {author} {\bibfnamefont {S.~A.}\ \bibnamefont
  {Markgraf}}\ and\ \bibinfo {author} {\bibfnamefont {R.~J.}\ \bibnamefont
  {Reeder}},\ }\href {http://rruff.info/doclib/am/vol70/AM70_590.pdf}
  {\bibfield  {journal} {\bibinfo  {journal} {Am. Mineral.}\ }\textbf {\bibinfo
  {volume} {70}},\ \bibinfo {pages} {590} (\bibinfo {year} {1985})}\BibitemShut
  {NoStop}%
\bibitem [{\citenamefont {Hazen}\ \emph {et~al.}(2013)\citenamefont {Hazen},
  \citenamefont {Downs}, \citenamefont {Jones},\ and\ \citenamefont
  {Kah}}]{Hazen20132}%
  \BibitemOpen
  \bibfield  {author} {\bibinfo {author} {\bibfnamefont {R.~M.}\ \bibnamefont
  {Hazen}}, \bibinfo {author} {\bibfnamefont {R.~T.}\ \bibnamefont {Downs}},
  \bibinfo {author} {\bibfnamefont {A.~P.}\ \bibnamefont {Jones}}, \ and\
  \bibinfo {author} {\bibfnamefont {L.}~\bibnamefont {Kah}},\ }\href {\doibase
  10.2138/rmg.2013.75.2} {\bibfield  {journal} {\bibinfo  {journal} {Rev.
  Mineral. Geochem.}\ }\textbf {\bibinfo {volume} {75}},\ \bibinfo {pages} {7}
  (\bibinfo {year} {2013})}\BibitemShut {NoStop}%
\bibitem [{\citenamefont {O'Leary}, \citenamefont {Lange},\ and\ \citenamefont
  {Ai}(2009)}]{OLeary2009}%
  \BibitemOpen
  \bibfield  {author} {\bibinfo {author} {\bibfnamefont {M.~C.}\ \bibnamefont
  {O'Leary}}, \bibinfo {author} {\bibfnamefont {R.~A.}\ \bibnamefont {Lange}},
  \ and\ \bibinfo {author} {\bibfnamefont {Y.}~\bibnamefont {Ai}},\ }in\ \href
  {http://adsabs.harvard.edu/abs/2009AGUFM.V11D1990O} {\emph {\bibinfo
  {booktitle} {AGU Fall Meeting Abstracts}}},\ Vol.~\bibinfo {volume} {1}\
  (\bibinfo {year} {2009})\BibitemShut {NoStop}%
\bibitem [{\citenamefont {Birch}(1947)}]{Birch1947}%
  \BibitemOpen
  \bibfield  {author} {\bibinfo {author} {\bibfnamefont {F.}~\bibnamefont
  {Birch}},\ }\href {\doibase 10.1103/PhysRev.71.809} {\bibfield  {journal}
  {\bibinfo  {journal} {Phys. Rev.}\ }\textbf {\bibinfo {volume} {71}},\
  \bibinfo {pages} {809} (\bibinfo {year} {1947})}\BibitemShut {NoStop}%
\bibitem [{\citenamefont {Kohara}\ \emph {et~al.}(1998)\citenamefont {Kohara},
  \citenamefont {Badyal~Y.}, \citenamefont {Koura}, \citenamefont {Idemoto},
  \citenamefont {Takahashi}, \citenamefont {Curtiss},\ and\ \citenamefont
  {Saboungi}}]{Kohara1998}%
  \BibitemOpen
  \bibfield  {author} {\bibinfo {author} {\bibfnamefont {S.}~\bibnamefont
  {Kohara}}, \bibinfo {author} {\bibfnamefont {S.}~\bibnamefont {Badyal~Y.}},
  \bibinfo {author} {\bibfnamefont {N.}~\bibnamefont {Koura}}, \bibinfo
  {author} {\bibfnamefont {Y.}~\bibnamefont {Idemoto}}, \bibinfo {author}
  {\bibfnamefont {S.}~\bibnamefont {Takahashi}}, \bibinfo {author}
  {\bibfnamefont {L.~A.}\ \bibnamefont {Curtiss}}, \ and\ \bibinfo {author}
  {\bibfnamefont {M.-L.}\ \bibnamefont {Saboungi}},\ }\href {\doibase
  10.1088/0953-8984/10/15/007} {\bibfield  {journal} {\bibinfo  {journal} {J.
  Phys.: Condens. Matter}\ }\textbf {\bibinfo {volume} {10}},\ \bibinfo {pages}
  {3301} (\bibinfo {year} {1998})}\BibitemShut {NoStop}%
\bibitem [{\citenamefont {Allen}\ and\ \citenamefont
  {Tildesley}(1989)}]{AllenTild}%
  \BibitemOpen
  \bibfield  {author} {\bibinfo {author} {\bibfnamefont {M.~P.}\ \bibnamefont
  {Allen}}\ and\ \bibinfo {author} {\bibfnamefont {D.~J.}\ \bibnamefont
  {Tildesley}},\ }\href@noop {} {\emph {\bibinfo {title} {Computer Simulation
  of Liquids}}}\ (\bibinfo  {publisher} {Clarendon Press},\ \bibinfo {address}
  {New York, NY, USA},\ \bibinfo {year} {1989})\BibitemShut {NoStop}%
\bibitem [{\citenamefont {Hess}(2002)}]{Hess2002}%
  \BibitemOpen
  \bibfield  {author} {\bibinfo {author} {\bibfnamefont {B.}~\bibnamefont
  {Hess}},\ }\href {\doibase 10.1063/1.1421362} {\bibfield  {journal} {\bibinfo
   {journal} {J. Chem. Phys.}\ }\textbf {\bibinfo {volume} {116}},\ \bibinfo
  {pages} {209} (\bibinfo {year} {2002})}\BibitemShut {NoStop}%
\bibitem [{\citenamefont {Adams}, \citenamefont {McDonald},\ and\ \citenamefont
  {Singer}(1977)}]{Adams1977}%
  \BibitemOpen
  \bibfield  {author} {\bibinfo {author} {\bibfnamefont {E.~M.}\ \bibnamefont
  {Adams}}, \bibinfo {author} {\bibfnamefont {I.~R.}\ \bibnamefont {McDonald}},
  \ and\ \bibinfo {author} {\bibfnamefont {K.}~\bibnamefont {Singer}},\ }\href
  {\doibase 10.1098/rspa.1977.0154} {\bibfield  {journal} {\bibinfo  {journal}
  {Proc. R. Soc. A}\ }\textbf {\bibinfo {volume} {357}},\ \bibinfo {pages} {37}
  (\bibinfo {year} {1977})}\BibitemShut {NoStop}%
\bibitem [{\citenamefont {Sifr{\'e}}\ \emph {et~al.}(2014)\citenamefont
  {Sifr{\'e}}, \citenamefont {Gard{\'e}s}, \citenamefont {Massuyeau},
  \citenamefont {Hashim}, \citenamefont {Hier-Majumder},\ and\ \citenamefont
  {Gaillard}}]{Sifre2014}%
  \BibitemOpen
  \bibfield  {author} {\bibinfo {author} {\bibfnamefont {D.}~\bibnamefont
  {Sifr{\'e}}}, \bibinfo {author} {\bibfnamefont {E.}~\bibnamefont
  {Gard{\'e}s}}, \bibinfo {author} {\bibfnamefont {M.}~\bibnamefont
  {Massuyeau}}, \bibinfo {author} {\bibfnamefont {L.}~\bibnamefont {Hashim}},
  \bibinfo {author} {\bibfnamefont {S.}~\bibnamefont {Hier-Majumder}}, \ and\
  \bibinfo {author} {\bibfnamefont {F.}~\bibnamefont {Gaillard}},\ }\href
  {\doibase 10.1038/nature13245} {\bibfield  {journal} {\bibinfo  {journal}
  {Nature}\ }\textbf {\bibinfo {volume} {509}},\ \bibinfo {pages} {81}
  (\bibinfo {year} {2014})}\BibitemShut {NoStop}%
\bibitem [{\citenamefont {Kojima}\ \emph {et~al.}(2007)\citenamefont {Kojima},
  \citenamefont {Miyazaki}, \citenamefont {Nomura},\ and\ \citenamefont
  {Tanimoto}}]{Kojima2007}%
  \BibitemOpen
  \bibfield  {author} {\bibinfo {author} {\bibfnamefont {T.}~\bibnamefont
  {Kojima}}, \bibinfo {author} {\bibfnamefont {Y.}~\bibnamefont {Miyazaki}},
  \bibinfo {author} {\bibfnamefont {K.}~\bibnamefont {Nomura}}, \ and\ \bibinfo
  {author} {\bibfnamefont {K.}~\bibnamefont {Tanimoto}},\ }\href {\doibase
  10.1149/1.2789389} {\bibfield  {journal} {\bibinfo  {journal} {J.
  Electrochem. Soc.}\ }\textbf {\bibinfo {volume} {154}},\ \bibinfo {pages}
  {F222} (\bibinfo {year} {2007})}\BibitemShut {NoStop}%
\bibitem [{\citenamefont {Kojima}\ \emph {et~al.}(2008)\citenamefont {Kojima},
  \citenamefont {Miyazaki}, \citenamefont {Nomura},\ and\ \citenamefont
  {Tanimoto}}]{Kojima2008}%
  \BibitemOpen
  \bibfield  {author} {\bibinfo {author} {\bibfnamefont {T.}~\bibnamefont
  {Kojima}}, \bibinfo {author} {\bibfnamefont {Y.}~\bibnamefont {Miyazaki}},
  \bibinfo {author} {\bibfnamefont {K.}~\bibnamefont {Nomura}}, \ and\ \bibinfo
  {author} {\bibfnamefont {K.}~\bibnamefont {Tanimoto}},\ }\href {\doibase
  10.1149/1.2917212} {\bibfield  {journal} {\bibinfo  {journal} {J.
  Electrochem. Soc.}\ }\textbf {\bibinfo {volume} {155}},\ \bibinfo {pages}
  {F150} (\bibinfo {year} {2008})}\BibitemShut {NoStop}%
\bibitem [{\citenamefont {Kojima}\ \emph {et~al.}(2003)\citenamefont {Kojima},
  \citenamefont {Miyazaki}, \citenamefont {Nomura},\ and\ \citenamefont
  {Tanimoto}}]{Kojima2003}%
  \BibitemOpen
  \bibfield  {author} {\bibinfo {author} {\bibfnamefont {T.}~\bibnamefont
  {Kojima}}, \bibinfo {author} {\bibfnamefont {Y.}~\bibnamefont {Miyazaki}},
  \bibinfo {author} {\bibfnamefont {K.}~\bibnamefont {Nomura}}, \ and\ \bibinfo
  {author} {\bibfnamefont {K.}~\bibnamefont {Tanimoto}},\ }\href {\doibase
  10.1149/1.1611494} {\bibfield  {journal} {\bibinfo  {journal} {J.
  Electrochem. Soc.}\ }\textbf {\bibinfo {volume} {150}},\ \bibinfo {pages}
  {E535} (\bibinfo {year} {2003})}\BibitemShut {NoStop}%
\bibitem [{\citenamefont {Yoshino}\ \emph {et~al.}(2012)\citenamefont
  {Yoshino}, \citenamefont {McIsaac}, \citenamefont {Laumonier},\ and\
  \citenamefont {Katsura}}]{Yoshino2012}%
  \BibitemOpen
  \bibfield  {author} {\bibinfo {author} {\bibfnamefont {T.}~\bibnamefont
  {Yoshino}}, \bibinfo {author} {\bibfnamefont {E.}~\bibnamefont {McIsaac}},
  \bibinfo {author} {\bibfnamefont {M.}~\bibnamefont {Laumonier}}, \ and\
  \bibinfo {author} {\bibfnamefont {T.}~\bibnamefont {Katsura}},\ }\href
  {\doibase 10.1016/j.pepi.2012.01.005} {\bibfield  {journal} {\bibinfo
  {journal} {Phys. Earth Planet. Inter.}\ }\textbf {\bibinfo {volume}
  {194–195}},\ \bibinfo {pages} {1} (\bibinfo {year} {2012})}\BibitemShut
  {NoStop}%
\bibitem [{\citenamefont {Jones}, \citenamefont {Dobson},\ and\ \citenamefont
  {Genge}(1995)}]{Jones1995}%
  \BibitemOpen
  \bibfield  {author} {\bibinfo {author} {\bibfnamefont {A.~P.}\ \bibnamefont
  {Jones}}, \bibinfo {author} {\bibfnamefont {D.}~\bibnamefont {Dobson}}, \
  and\ \bibinfo {author} {\bibfnamefont {M.}~\bibnamefont {Genge}},\ }\href
  {\doibase 10.1017/S0016756800011481} {\bibfield  {journal} {\bibinfo
  {journal} {Geol. Mag.}\ }\textbf {\bibinfo {volume} {132}},\ \bibinfo {pages}
  {121} (\bibinfo {year} {1995})}\BibitemShut {NoStop}%
\bibitem [{\citenamefont {Wolff}(1994)}]{Wolff1994}%
  \BibitemOpen
  \bibfield  {author} {\bibinfo {author} {\bibfnamefont {J.~A.}\ \bibnamefont
  {Wolff}},\ }\href {\doibase 10.1017/S0016756800010682} {\bibfield  {journal}
  {\bibinfo  {journal} {Geol. Mag.}\ }\textbf {\bibinfo {volume} {131}},\
  \bibinfo {pages} {145} (\bibinfo {year} {1994})}\BibitemShut {NoStop}%
\bibitem [{\citenamefont {Sykes}, \citenamefont {Baker},\ and\ \citenamefont
  {Wyllie}(1992)}]{Sykes1992}%
  \BibitemOpen
  \bibfield  {author} {\bibinfo {author} {\bibfnamefont {D.}~\bibnamefont
  {Sykes}}, \bibinfo {author} {\bibfnamefont {M.~B.}\ \bibnamefont {Baker}}, \
  and\ \bibinfo {author} {\bibfnamefont {P.~J.}\ \bibnamefont {Wyllie}},\ }in\
  \href {\doibase 10.1029/91EO10131} {\emph {\bibinfo {booktitle} {Abstracts,
  AGU Spring Meeting}}},\ Vol.~\bibinfo {volume} {73}\ (\bibinfo {year}
  {1992})\ p.\ \bibinfo {pages} {372}\BibitemShut {NoStop}%
\bibitem [{\citenamefont {Treiman}\ and\ \citenamefont
  {Schedl}(1983)}]{Treiman1983}%
  \BibitemOpen
  \bibfield  {author} {\bibinfo {author} {\bibfnamefont {A.~H.}\ \bibnamefont
  {Treiman}}\ and\ \bibinfo {author} {\bibfnamefont {A.}~\bibnamefont
  {Schedl}},\ }\href {\doibase 10.1086/628789} {\bibfield  {journal} {\bibinfo
  {journal} {J. Geol.}\ }\textbf {\bibinfo {volume} {91}},\ \bibinfo {pages}
  {437} (\bibinfo {year} {1983})}\BibitemShut {NoStop}%
\bibitem [{\citenamefont {Norton}\ and\ \citenamefont
  {Pinkerton}(1997)}]{Norton1997}%
  \BibitemOpen
  \bibfield  {author} {\bibinfo {author} {\bibfnamefont {G.}~\bibnamefont
  {Norton}}\ and\ \bibinfo {author} {\bibfnamefont {H.}~\bibnamefont
  {Pinkerton}},\ }\href {\doibase 10.1127/ejm/9/2/0351} {\bibfield  {journal}
  {\bibinfo  {journal} {Eur. J. Mineral.}\ }\textbf {\bibinfo {volume} {9}},\
  \bibinfo {pages} {351} (\bibinfo {year} {1997})}\BibitemShut {NoStop}%
\bibitem [{\citenamefont {Fischer}\ \emph {et~al.}(2009)\citenamefont
  {Fischer}, \citenamefont {Burnard}, \citenamefont {Marty}, \citenamefont
  {Hilton}, \citenamefont {Füri}, \citenamefont {Palhol}, \citenamefont
  {Sharp},\ and\ \citenamefont {Mangasini}}]{Fischer2009}%
  \BibitemOpen
  \bibfield  {author} {\bibinfo {author} {\bibfnamefont {T.}~\bibnamefont
  {Fischer}}, \bibinfo {author} {\bibfnamefont {P.}~\bibnamefont {Burnard}},
  \bibinfo {author} {\bibfnamefont {B.}~\bibnamefont {Marty}}, \bibinfo
  {author} {\bibfnamefont {D.}~\bibnamefont {Hilton}}, \bibinfo {author}
  {\bibfnamefont {E.}~\bibnamefont {Füri}}, \bibinfo {author} {\bibfnamefont
  {F.}~\bibnamefont {Palhol}}, \bibinfo {author} {\bibfnamefont
  {Z.}~\bibnamefont {Sharp}}, \ and\ \bibinfo {author} {\bibfnamefont
  {F.}~\bibnamefont {Mangasini}},\ }\href {\doibase 10.1038/nature07977}
  {\bibfield  {journal} {\bibinfo  {journal} {Nature}\ }\textbf {\bibinfo
  {volume} {459}},\ \bibinfo {pages} {77} (\bibinfo {year} {2009})}\BibitemShut
  {NoStop}%
\bibitem [{\citenamefont {NSF}(2009)}]{nsf2009}%
  \BibitemOpen
  \bibfield  {author} {\bibinfo {author} {\bibnamefont {NSF}},\ }\href
  {{https://www.nsf.gov/news/news_summ.jsp?cntn_id=114703}} {\enquote {\bibinfo
  {title} {{Alchemy in Tanzania? Gas Becomes Solid at Surface of Oldoinyo
  Lengai Volcano}},}\ } (\bibinfo {year} {2009}),\ \bibinfo {note} {[accessed
  03.25.2019]}\BibitemShut {NoStop}%
\bibitem [{\citenamefont {Liu}, \citenamefont {Tenner},\ and\ \citenamefont
  {Lange}(2007)}]{Liu2007}%
  \BibitemOpen
  \bibfield  {author} {\bibinfo {author} {\bibfnamefont {Q.}~\bibnamefont
  {Liu}}, \bibinfo {author} {\bibfnamefont {T.~J.}\ \bibnamefont {Tenner}}, \
  and\ \bibinfo {author} {\bibfnamefont {R.~A.}\ \bibnamefont {Lange}},\ }\href
  {\doibase 10.1007/s00410-006-0134-z} {\bibfield  {journal} {\bibinfo
  {journal} {Contrib. Mineral. Petrol.}\ }\textbf {\bibinfo {volume} {153}},\
  \bibinfo {pages} {55} (\bibinfo {year} {2007})}\BibitemShut {NoStop}%
\bibitem [{\citenamefont {Treiman}(1989)}]{Treiman1989}%
  \BibitemOpen
  \bibfield  {author} {\bibinfo {author} {\bibfnamefont {A.~H.}\ \bibnamefont
  {Treiman}},\ }\enquote {\bibinfo {title} {{Carbonatites - Genesis and
  Evolution}},}\ \ (\bibinfo  {publisher} {Unwin-Hyman, London},\ \bibinfo
  {year} {1989})\ Chap.\ \bibinfo {chapter} {{Carbonatite Magma: Properties and
  Processes}}, p.~\bibinfo {pages} {89}\BibitemShut {NoStop}%
\end{thebibliography}%


\begin{thebibliography}{6}%
\makeatletter
\providecommand \@ifxundefined [1]{%
 \@ifx{#1\undefined}
}%
\providecommand \@ifnum [1]{%
 \ifnum #1\expandafter \@firstoftwo
 \else \expandafter \@secondoftwo
 \fi
}%
\providecommand \@ifx [1]{%
 \ifx #1\expandafter \@firstoftwo
 \else \expandafter \@secondoftwo
 \fi
}%
\providecommand \natexlab [1]{#1}%
\providecommand \enquote  [1]{``#1''}%
\providecommand \bibnamefont  [1]{#1}%
\providecommand \bibfnamefont [1]{#1}%
\providecommand \citenamefont [1]{#1}%
\providecommand \href@noop [0]{\@secondoftwo}%
\providecommand \href [0]{\begingroup \@sanitize@url \@href}%
\providecommand \@href[1]{\@@startlink{#1}\@@href}%
\providecommand \@@href[1]{\endgroup#1\@@endlink}%
\providecommand \@sanitize@url [0]{\catcode `\\12\catcode `\$12\catcode
  `\&12\catcode `\#12\catcode `\^12\catcode `\_12\catcode `\%12\relax}%
\providecommand \@@startlink[1]{}%
\providecommand \@@endlink[0]{}%
\providecommand \url  [0]{\begingroup\@sanitize@url \@url }%
\providecommand \@url [1]{\endgroup\@href {#1}{\urlprefix }}%
\providecommand \urlprefix  [0]{URL }%
\providecommand \Eprint [0]{\href }%
\providecommand \doibase [0]{http://dx.doi.org/}%
\providecommand \selectlanguage [0]{\@gobble}%
\providecommand \bibinfo  [0]{\@secondoftwo}%
\providecommand \bibfield  [0]{\@secondoftwo}%
\providecommand \translation [1]{[#1]}%
\providecommand \BibitemOpen [0]{}%
\providecommand \bibitemStop [0]{}%
\providecommand \bibitemNoStop [0]{.\EOS\space}%
\providecommand \EOS [0]{\spacefactor3000\relax}%
\providecommand \BibitemShut  [1]{\csname bibitem#1\endcsname}%
\let\auto@bib@innerbib\@empty
\bibitem [{\citenamefont {Markgraf}\ and\ \citenamefont
  {Reeder}(1985)}]{Markgraf1985}%
  \BibitemOpen
  \bibfield  {author} {\bibinfo {author} {\bibfnamefont {S.~A.}\ \bibnamefont
  {Markgraf}}\ and\ \bibinfo {author} {\bibfnamefont {R.~J.}\ \bibnamefont
  {Reeder}},\ }\href {http://rruff.info/doclib/am/vol70/AM70_590.pdf}
  {\bibfield  {journal} {\bibinfo  {journal} {Am. Mineral.}\ }\textbf {\bibinfo
  {volume} {70}},\ \bibinfo {pages} {590} (\bibinfo {year} {1985})}\BibitemShut
  {NoStop}%
\bibitem [{\citenamefont {Redfern}, \citenamefont {Wood},\ and\ \citenamefont
  {Henderson}(1993)}]{Redfern1993}%
  \BibitemOpen
  \bibfield  {author} {\bibinfo {author} {\bibfnamefont {S.~A.~T.}\
  \bibnamefont {Redfern}}, \bibinfo {author} {\bibfnamefont {B.~J.}\
  \bibnamefont {Wood}}, \ and\ \bibinfo {author} {\bibfnamefont {C.~M.~B.}\
  \bibnamefont {Henderson}},\ }\href {\doibase 10.1029/93GL02507} {\bibfield
  {journal} {\bibinfo  {journal} {Geophys. Res. Lett.}\ }\textbf {\bibinfo
  {volume} {20}},\ \bibinfo {pages} {2099} (\bibinfo {year}
  {1993})}\BibitemShut {NoStop}%
\bibitem [{\citenamefont {Fiquet}, \citenamefont {Guyot},\ and\ \citenamefont
  {Itie}(1994)}]{Fiquet1994}%
  \BibitemOpen
  \bibfield  {author} {\bibinfo {author} {\bibfnamefont {G.}~\bibnamefont
  {Fiquet}}, \bibinfo {author} {\bibfnamefont {F.}~\bibnamefont {Guyot}}, \
  and\ \bibinfo {author} {\bibfnamefont {J.-P.}\ \bibnamefont {Itie}},\ }\href
  {http://ammin.geoscienceworld.org/content/79/1-2/15.short} {\bibfield
  {journal} {\bibinfo  {journal} {Am. Mineral.}\ }\textbf {\bibinfo {volume}
  {79}},\ \bibinfo {pages} {15} (\bibinfo {year} {1994})}\BibitemShut {NoStop}%
\bibitem [{\citenamefont {Zhang}\ \emph {et~al.}(1997)\citenamefont {Zhang},
  \citenamefont {Martinez}, \citenamefont {Guyot}, \citenamefont {Gillet},\
  and\ \citenamefont {Saxena}}]{Zhang1997}%
  \BibitemOpen
  \bibfield  {author} {\bibinfo {author} {\bibfnamefont {J.}~\bibnamefont
  {Zhang}}, \bibinfo {author} {\bibfnamefont {I.}~\bibnamefont {Martinez}},
  \bibinfo {author} {\bibfnamefont {F.}~\bibnamefont {Guyot}}, \bibinfo
  {author} {\bibfnamefont {P.}~\bibnamefont {Gillet}}, \ and\ \bibinfo {author}
  {\bibfnamefont {S.~K.}\ \bibnamefont {Saxena}},\ }\href {\doibase
  10.1007/s002690050025} {\bibfield  {journal} {\bibinfo  {journal} {Phys.
  Chem. Miner.}\ }\textbf {\bibinfo {volume} {24}},\ \bibinfo {pages} {122}
  (\bibinfo {year} {1997})}\BibitemShut {NoStop}%
\bibitem [{\citenamefont {Ross}(1997)}]{Ross1997}%
  \BibitemOpen
  \bibfield  {author} {\bibinfo {author} {\bibfnamefont {N.~L.}\ \bibnamefont
  {Ross}},\ }\href {\doibase 10.2138/am-1997-7-805} {\bibfield  {journal}
  {\bibinfo  {journal} {Am. Mineral.}\ }\textbf {\bibinfo {volume} {82}},\
  \bibinfo {pages} {682} (\bibinfo {year} {1997})}\BibitemShut {NoStop}%
\bibitem [{\citenamefont {Vuilleumier}\ \emph {et~al.}(2014)\citenamefont
  {Vuilleumier}, \citenamefont {Seitsonen}, \citenamefont {Sator},\ and\
  \citenamefont {Guillot}}]{Vuilleumier2014}%
  \BibitemOpen
  \bibfield  {author} {\bibinfo {author} {\bibfnamefont {R.}~\bibnamefont
  {Vuilleumier}}, \bibinfo {author} {\bibfnamefont {A.}~\bibnamefont
  {Seitsonen}}, \bibinfo {author} {\bibfnamefont {N.}~\bibnamefont {Sator}}, \
  and\ \bibinfo {author} {\bibfnamefont {B.}~\bibnamefont {Guillot}},\ }\href
  {\doibase 10.1016/j.gca.2014.06.037} {\bibfield  {journal} {\bibinfo
  {journal} {GCA}\ }\textbf {\bibinfo {volume} {141}},\ \bibinfo {pages} {547}
  (\bibinfo {year} {2014})}\BibitemShut {NoStop}%
\end{thebibliography}%

\end{document}


\title{Supplementary Material for "The \ce{MgCO3}--\ce{CaCO3}--\ce{Li2CO3}--\ce{Na2CO3}--\ce{K2CO3} Carbonate Melts: Thermodynamics and Transport Properties by Atomistic Simulations"}

\author{Elsa Desmaele}\email{elsa.desmaele@gmail.com}
\affiliation{Sorbonne Universit\'e, CNRS, Laboratoire de Physique Th\'eorique de la Mati\`{e}re Condens\'ee, LPTMC, F75005, Paris, France}
\author{Nicolas Sator}
\affiliation{Sorbonne Universit\'e, CNRS, Laboratoire de Physique Th\'eorique de la Mati\`{e}re Condens\'ee, LPTMC, F75005, Paris, France}
\author{Rodolphe Vuilleumier}
\affiliation{PASTEUR, D\'epartement de chimie, \'Ecole normale sup\'erieure, PSL University, Sorbonne Universit\'e, CNRS, 75005 Paris, France}
\author{Bertrand Guillot}\email{guillot@lptmc.jussieu.fr}
\affiliation{Sorbonne Universit\'e, CNRS, Laboratoire de Physique Th\'eorique de la Mati\`{e}re Condens\'ee, LPTMC, F75005, Paris, France}

\date{\today}

\maketitle

\onecolumngrid
 \tableofcontents
 \listoffigures
 \listoftables
 
\clearpage

\begin{table*}[h]
    \begin{tabular}{l lc c  c c c } 
		& $P$(GPa)	& $n$ (g/cm$^{3}$)			& a (\AA) 	& 	b (\AA)	& c (\AA) & $\gamma$ ($^\circ$)\\ \hline	
\ce{MgCO3}\\
MD & 0 & 	3.04	& 4.60	& 4.60 & 15.07 & 120.0	\\
Exp.\cite{Markgraf1985} & 0 & 3.01 & \multicolumn{2}{c}{4.64} & 15.02 & 120 \\ 
Exp.\cite{Redfern1993} & 0 & 3.02 & \multicolumn{2}{c}{4.64} & 14.96 & 120 \\ 
Exp.\cite{Fiquet1994} & 0 & 3.00 & \multicolumn{2}{c}{4.64} & 15.03 & 120 \\ 
Exp.\cite{Zhang1997} & 0 & 3.01 & \multicolumn{2}{c}{4.64} & 15.01 & 120 \\
Exp.\cite{Ross1997} & 0 &  3.01 & \multicolumn{2}{c}{4.63} & 15.02 & 120 \\
\\
MD  & 2 & 	3.07	& 4.58	& 4.58 & 14.91 & 120.0	\\ 	
Exp.\cite{Zhang1997} & 2 & 3.06  & \multicolumn{2}{c}{4.62} & 14.91 & 120 \\
Exp.\cite{Redfern1993} & 2 & 3.03 & \multicolumn{2}{c}{4.63} & 14.95 & 120 \\ 
Exp.\cite{Fiquet1994} & 2 & 3.06 & \multicolumn{2}{c}{4.62} & 14.91 & 120 \\ 
Exp.\cite{Ross1997} & 2 &  3.06 & \multicolumn{2}{c}{4.62} & 14.88 & 120
\\
\\
MD  & 4 & 	3.15	& 4.57	& 4.57 & 14.77 & 119.8	\\
Exp.\cite{Zhang1997} & 4 & 3.12  & \multicolumn{2}{c}{4.59} & 14.75 & 120 \\ 
Exp.\cite{Redfern1993} & 4 & 3.06 & \multicolumn{2}{c}{4.62} & 14.80 & 120 \\ 
Exp.\cite{Fiquet1994} & 4 & 3.09 & \multicolumn{2}{c}{4.60} & 14.82 & 120 \\ 
Exp.\cite{Ross1997} & 4 &  3.11 & \multicolumn{2}{c}{4.60} & 14.75 & 120
\\
\hline	
\\	
\end{tabular}
\caption[Crystal structure parameters for \ce{MgCO3}]{Crystal structure parameters for rhombohedral \ce{MgCO3} at 300~K from the experimental literature (Exp.) and calculated by anisotropic MD simulations ($NST$ ensemble).}
\label{tsupcryst}
\end{table*}

\begin{table*}[h]
    \begin{tabular}{l lc c  c c c } 
		& $P$(GPa)	& $n$ (g/cm$^{3}$)			& a (\AA) 	& 	b (\AA)	& c (\AA) & $\gamma$ ($^\circ$)\\ \hline
\ce{CaCO3}\\
MD  & 0 & 	2.63	&	5.04 & 5.04 & 17.23 & 120.0	\\	
Exp.\cite{Markgraf1985} & 0 & 2.71 & \multicolumn{2}{c}{4.99} & 17.06 & 120 \\ 
\hline 
\ce{CaMg(CO3)2} \\
MD  & 0 & 	2.84	& 4.84	& 4.84 & 16.00 & 120.0	\\
Exp.\cite{Fiquet1994} & 0 & 2.84 & \multicolumn{2}{c}{4.83} & 16.01 & 120 \\ 
\\
MD  & 2 & 	2.91	& 4.81	& 4.81 & 15.74 & 120.0	\\ 
Exp.\cite{Fiquet1994} & 2 & 2.89 & \multicolumn{2}{c}{4.81} & 15.87 & 120 \\ 
\\	
MD  & 4 & 2.97		& 4.80	&  4.80 & 15.51 & 120.2	\\
Exp.\cite{Fiquet1994} & 4 & 2.94 & \multicolumn{2}{c}{4.79} & 15.74 & 120 \\
 \hline	
\end{tabular}
\caption[Crystal structure parameters for \ce{CaCO3} and \ce{CaMg(CO3)2}]{Crystal structure parameters for rhombohedral \ce{CaCO3} and \ce{CaMg(CO3)2} at 300~K from the experimental literature (Exp.) and calculated by anisotropic MD simulations ($NST$ ensemble).}
\label{tsupcryst2}
\end{table*}

\clearpage

\begin{figure*}
\subfloat[\label{fpdfma}]{{\includegraphics{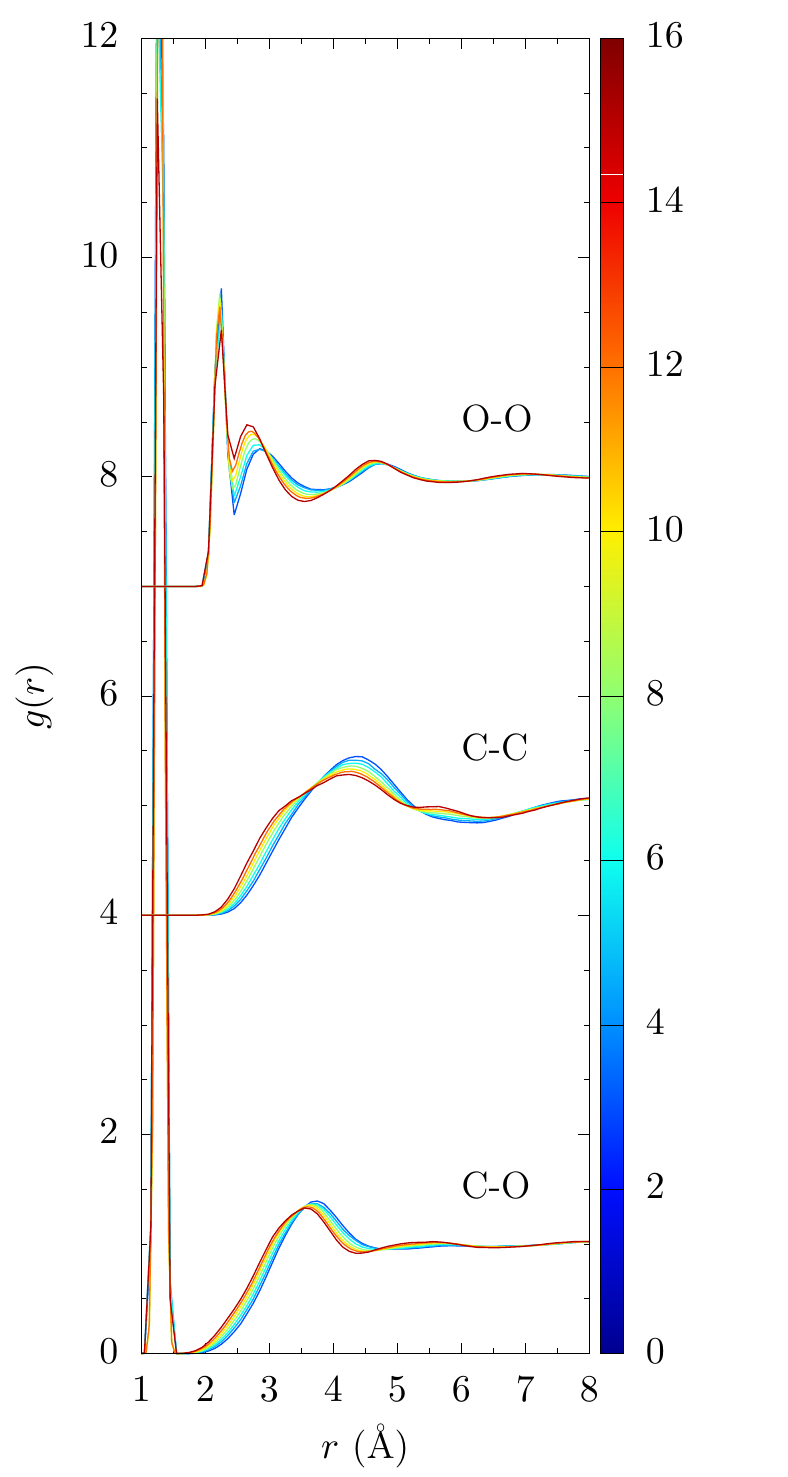}}} 
\subfloat[\label{fpdfmb}]{{\includegraphics{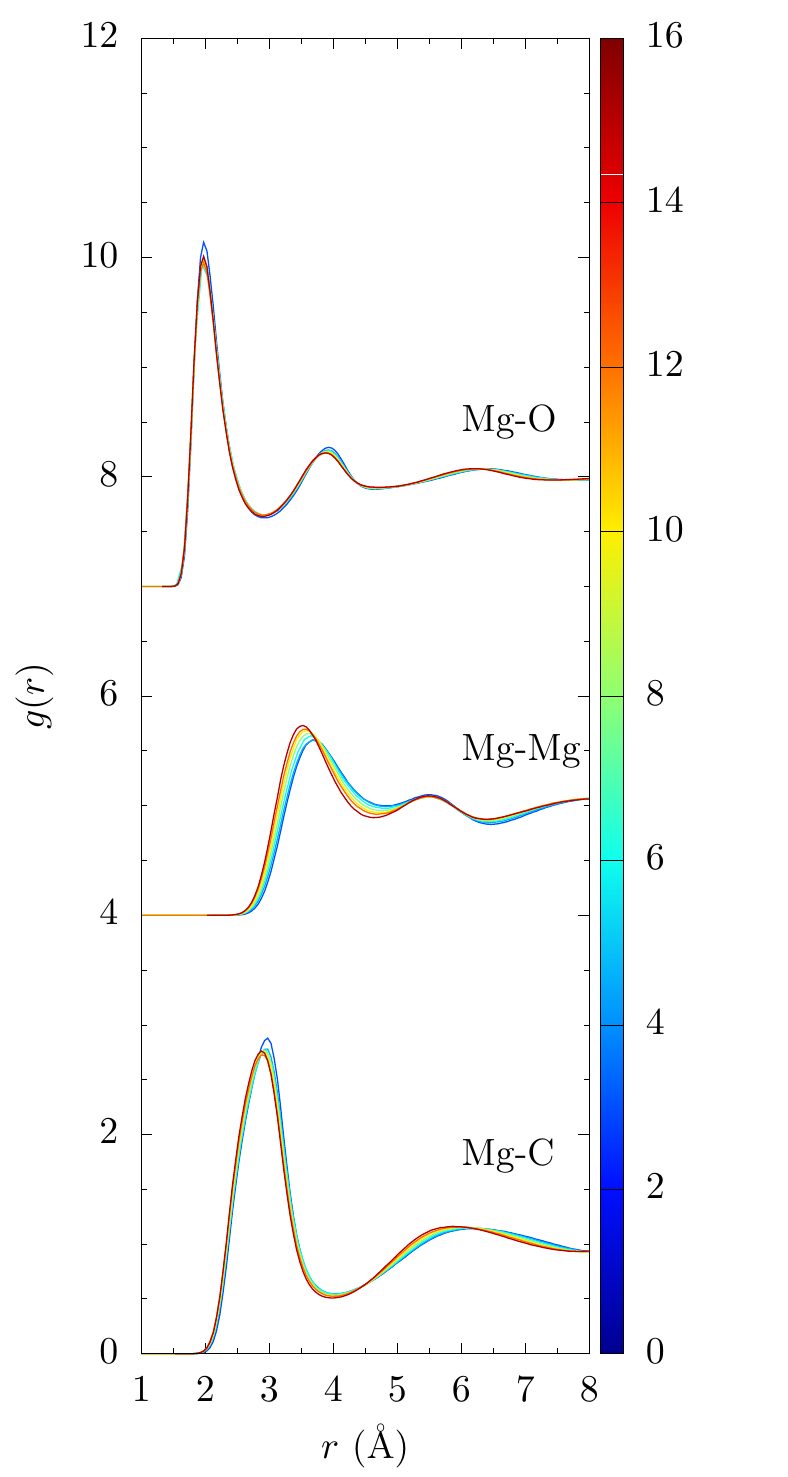}}}
\caption[Pressure evolution of PDFs for \ce{MgCO3}]{Pressure evolution of MD pair distribution functions in $\mathbf{MgCO_3}$ at 2073~K (3~GPa) and  at 2273~K (4.5, 6, 8, 10, 12 and 15~GPa). The color scale refers to the pressure (in GPa). To facilitate visualization, the different PDFs were shifted vertically.}
\label{fpdfm} 
\end{figure*}

\begin{figure*}
\subfloat[\label{fpdfca}]{{\includegraphics{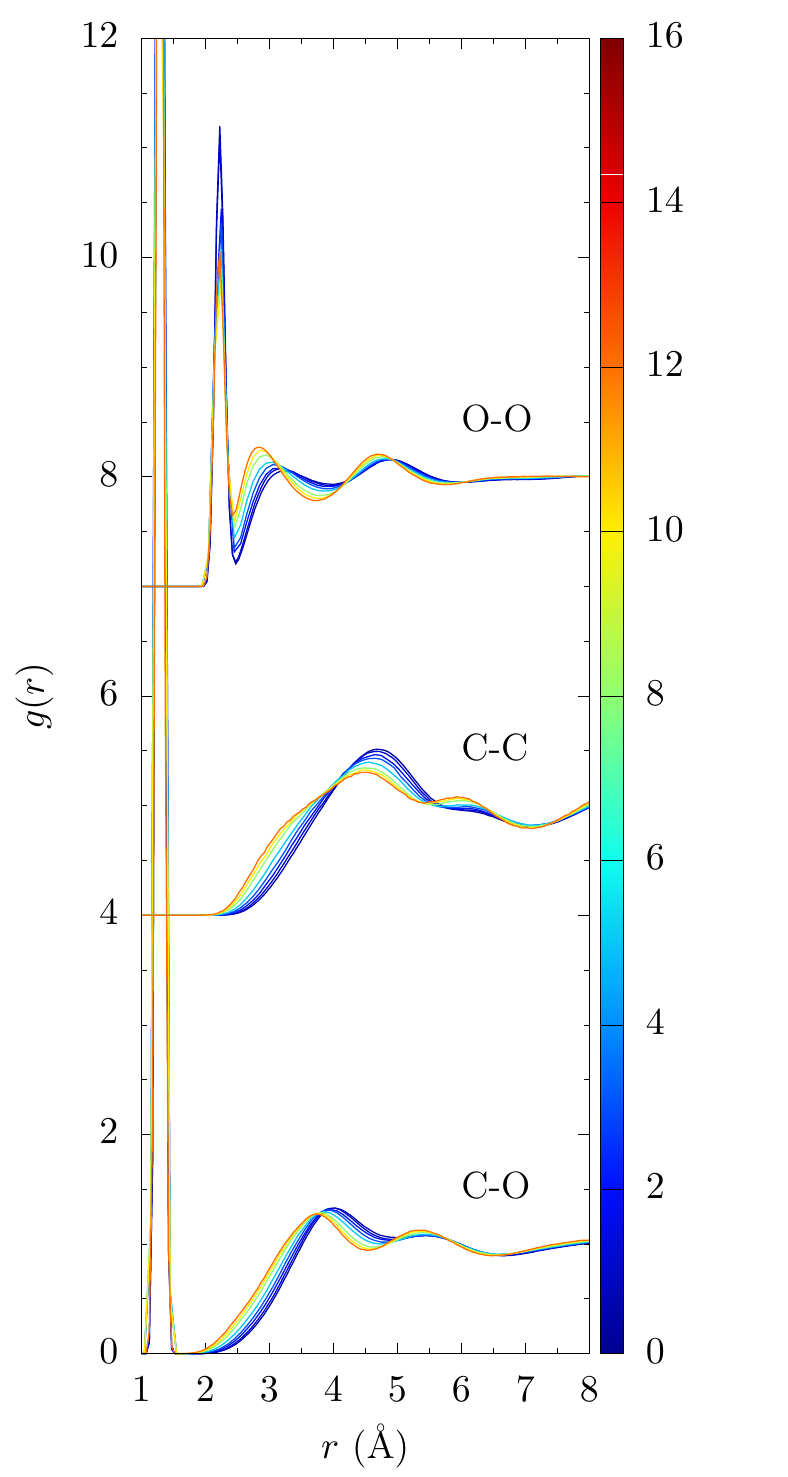}}} 
\subfloat[\label{fpdfcb}]{{\includegraphics{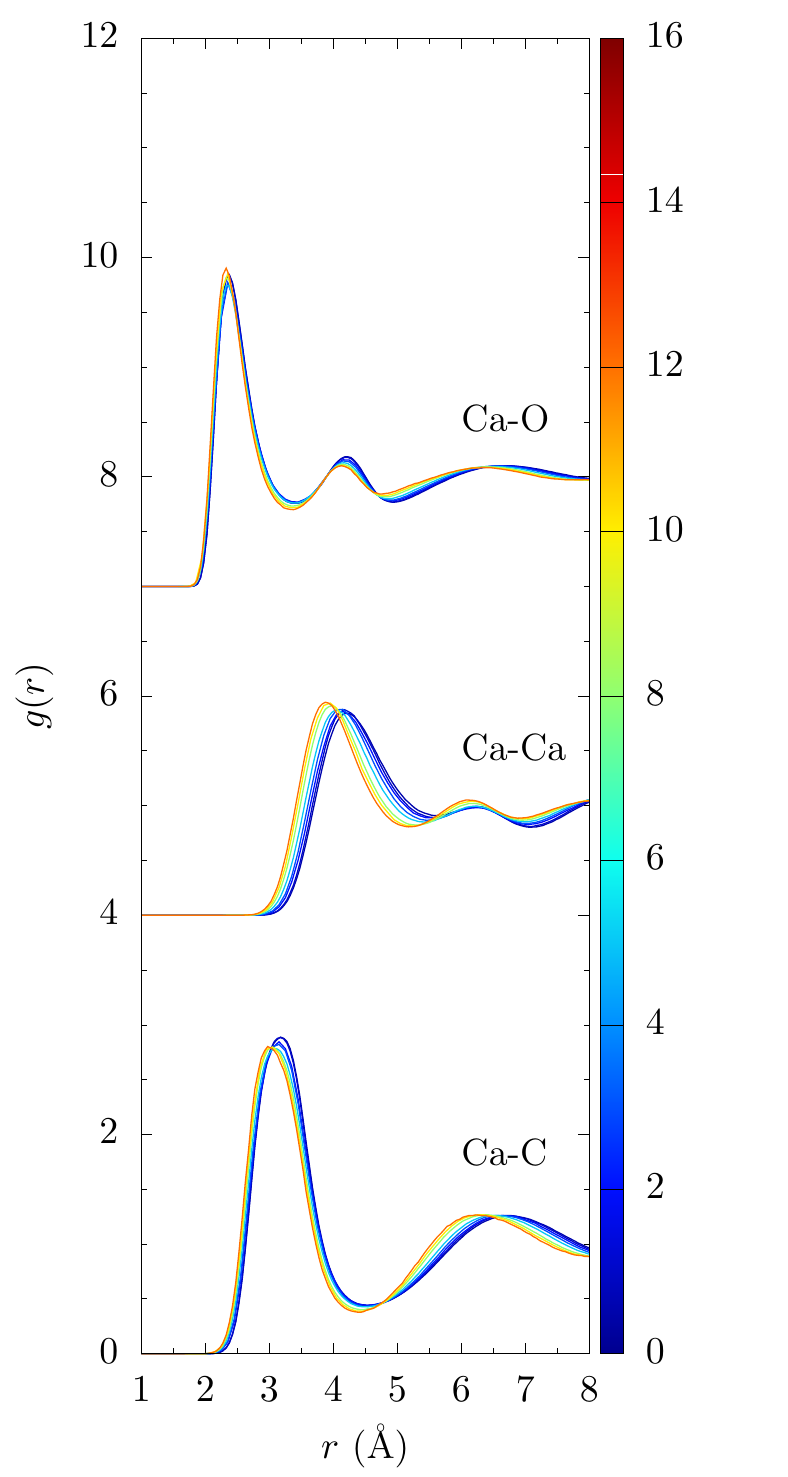}}}
\caption[Pressure evolution of PDFs for \ce{CaCO3}]{Pressure evolution of MD pair distribution functions in $\mathbf{CaCO_3}$ at 1773~K (0.5 and 1~GPa), 1923~K (2 and 3~GPa) and 2073~K (5, 8, 10 and 12~GPa). The color scale refers to the pressure (in GPa). To facilitate visualization, the different PDFs were shifted vertically.}
\label{fpdfc} 
\end{figure*}

\begin{figure*}
\subfloat[\label{fpdfcma}]{{\includegraphics{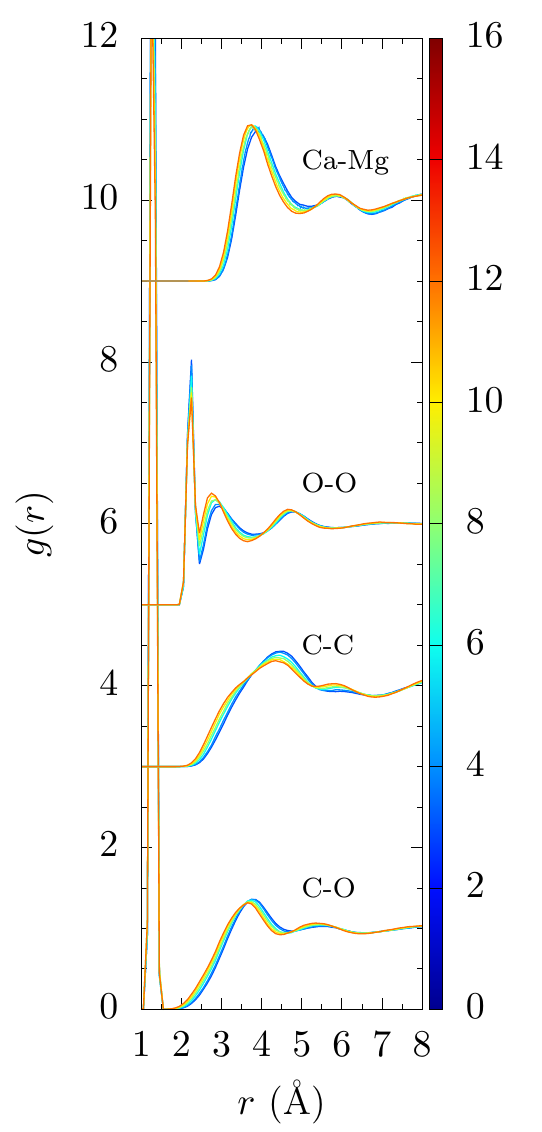}}} 
\subfloat[\label{fpdfcmb}]{{\includegraphics{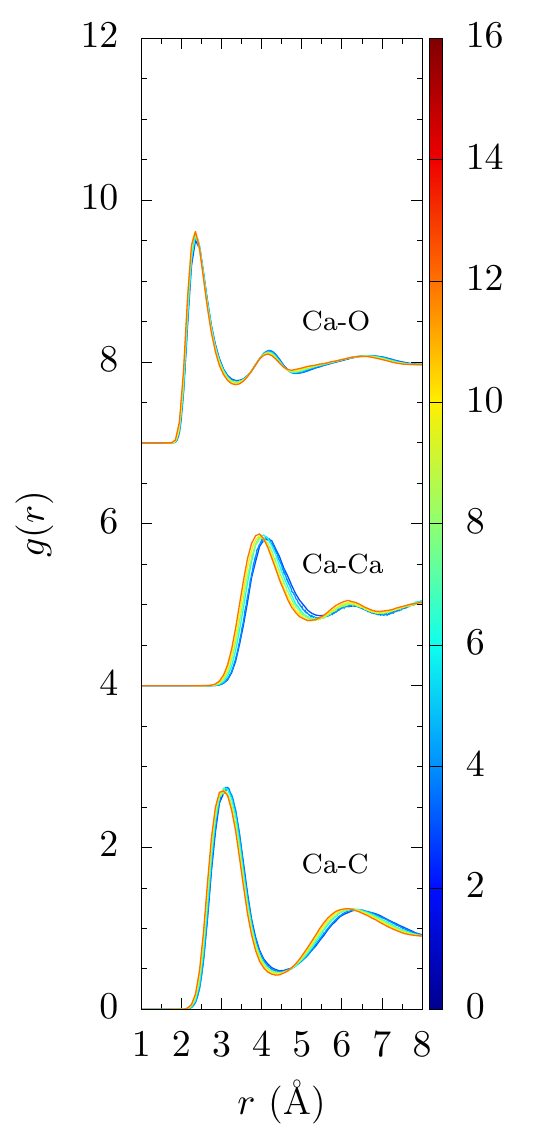}}}
\subfloat[\label{fpdfcmc}]{{\includegraphics{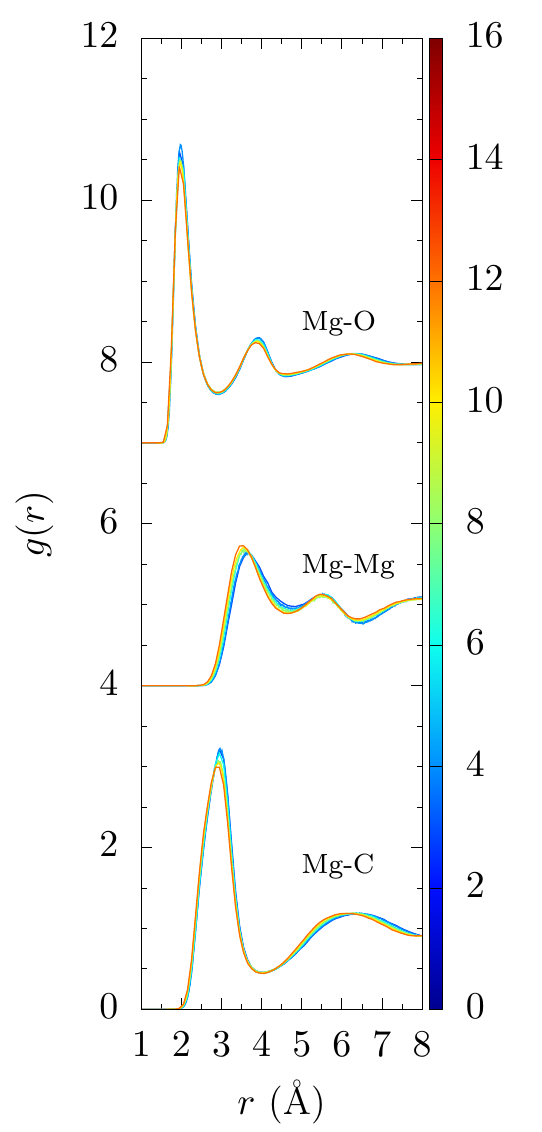}}}
\caption[Pressure evolution of PDFs for \ce{CaMg(CO3)2}]{Pressure evolution of MD pair distribution functions in $\mathbf{CaMg(CO_3)_2}$ at 1873~K (3, 4 and 6~GPa) and 2073~K (8, 10 and 12~GPa). The color scale refers to the pressure (in GPa). To facilitate visualization, the different PDFs were shifted vertically.}
\label{fpdfmc} 
\end{figure*}

\clearpage

\begingroup
\squeezetable
\begin{table}
\begin{center}
\begin{tabular}{l c c c c c c c c} 
		&$T$ 	& $\langle T \rangle$ & 	$n$ 		& $\langle P\rangle$ 	&	$\sigma$	&	$\eta$						   &  ${D}_{\text{Mg}}$ 	&${D}_{\text{CO}_3}$	\\ 
		&(K) 	&			(K)	      &(g/cm$^{3}$)	& (kbar)	    &  (S/m) 			& (mPa$\cdot$s) & ($10^{-9}$m$^2$/s) & ($10^{-9}$m$^2$/s)\\ \hline
\\
\ce{MgCO3}\\	
	& 1940  & 1897 	& 2.49 		& 20	& 298 $\pm$ 5	 & 5.3 $\pm$ 0.2 	& 2.89 & 1.98\\
	& 1940  & 1943  & 2.49		& 22	& 323 $\pm$ 5	 & 5.1 $\pm$ 0.2	& 3.11 & 2.13 \\
\texttt{AIMD} & 1873 	& 1936	& 2.49 	& 45 $\pm$ 15	&  /		& /	& 4.4 $\pm$ 0.5 & 3.2 $\pm$ 0.5  \\
\\	
	& 1823   & 1815  & 2.60 	& 31  	& 240 $\pm$ 10	& 7.0$\pm$ 0.3	& 2.10	& 1.41				\\
	& 1923  & 1926  & 2.56 	& 30	& 285 $\pm$ 15 	& 5.7 $\pm$ 0.5	& 2.70 	& 1.83	 \\ 
	& 2073  & 2041  & 2.53	 	& 29	& 330 $\pm$ 5   & 4.9 $\pm$ 0.3	& 3.35	& 2.25	 \\ 
\\	
	& 1873	& 1868 & 2.61		& 35 & 258 $\pm$ 10	& 6.9 $\pm$ 0.3 & 2.24 & 1.53 \\ 	
\\
	& 1873  & 1840  & 2.68		& 44	& 225 $\pm$ 10   & 8.1 $\pm$ 0.3	& 1.85	& 1.30	 \\ 
	& 2073	& 2039	& 2.64	& 44	& 285 $\pm$ 15	& 	5.8 $\pm$ 0.2 			& 2.8 	& 1.92	\\
	& 2273	& 2265	& 2.60	& 45	& 373 $\pm$ 10	&  	4.6 $\pm$ 0.2 				& 4.02 	&  2.75	\\	
\\
    & 2073 	& 2098	& 2.73	& 61	& 275 $\pm$5 	& 6.7 $\pm$0.5		& 2.55	& 1.80	 \\
    & 2173 	& 2183	& 2.71	& 60 	& 315 $\pm$ 10	& 5.7 $\pm$ 0.6		& 3.05	& 2.12 \\ 
    & 2273  & 2250	& 2.69  & 59    & 335 $\pm$ 20  & 5.3$\pm$ 0.4 		& 3.30	& 2.37 \\
\\
    & 2073  & 2046	& 2.83  & 79   	& 210 $\pm$ 5  	& 8.2 $\pm$ 0.2 	& 1.90	& 1.35 \\    
    & 2273  & 2276	& 2.79  & 80	& 290 $\pm$ 10 	& 6.2 $\pm$ 0.3  	& 2.95	& 2.20 \\ 
\\
    & 2273  & 2280	& 2.88  & 100   & 235 $\pm$ 10 	& 8.0 $\pm$ 0.3 	& 2.40	& 1.75 \\
    \\
    & 2273  & 2264	& 2.96  & 120   & 215 $\pm$ 15 	& 8.8 $\pm$ 0.5 	& 2.00	& 1.45 \\
\\	
	& 2273	& 2261	& 3.06  & 149 	& 174 $\pm$ 5	& 11.5 $\pm$ 0.5	& 1.55	& 1.18	\\  
\\
\hline	
\\
\end{tabular}
\end{center}
\caption[Summary of the calculated properties for \ce{MgCO3}]{Thermodynamic averages and transport coefficients issued from all  MD simulations of $\mathbf{MgCO_3}$. Diffusion coefficients estimated from the AIMD simulations are also mentioned (\texttt{AIMD}).}
\label{tantrans1}
\end{table}
\endgroup

\begingroup
\squeezetable
\begin{table}
\begin{center}
\begin{tabular}{l c c c c c c c c} 
		&$T$ 	& $\langle T \rangle$ & 	$n$ 		& $\langle P\rangle$ 	&	$\sigma$	&	$\eta$						   &  ${D}_{\text{Ca}}$ 	&${D}_{\text{CO}_3}$	\\ 
		&(K) 	&			(K)	      &(g/cm$^{3}$)	& (kbar)	    &  (S/m) 			& (mPa$\cdot$s) & ($10^{-9}$m$^2$/s) & ($10^{-9}$m$^2$/s)\\ \hline
\\
\ce{CaCO3}\\		
	& 1623  & 1622  & 2.33	& 4.9	& 210 $\pm$ 15   & 5.3 $\pm$ 0.3	& 1.95	& 1.82	 \\ 	
\texttt{AIMD} (Ref. \onlinecite{Vuilleumier2014}) & 1623 	& & 2.30 	& 5 $\pm$ 5	&  171 $\pm$ 15	&	5.3 $\pm$ 0.5			& 2.4 $\pm$ 0.2 & 1.6 $\pm$ 0.2    \\	
	& 1773 & 1772 & 2.28	& 5.0 	&	235 $\pm$ 20 &	4.1 $\pm$ 0.3 & 2.75 & 2.45			\\ 
	& 2073 & 2057 & 2.20	& 4.9 	&	335 $\pm$ 10	&	2.9 $\pm$ 0.3 & 4.75	 	&		4.05	\\ 
\\
	& 1653 & 1640 & 2.37	& 8.7 	& /				& 6.0 $\pm$ 0.3	& /	 	&	/		\\ 	
	& 1653	& 1648	& 2.37	& 8.8	& 195 $\pm$ 10	& 5.8 $\pm$ 0.2 & 1.84	& 1.74 \\
	& 1773 & 1773 & 2.35	& 10 	& 220 $\pm$ 20 & 4.9 $\pm$ 0.3	& 	2.38	& 2.18			\\
	& 2073 & 2046  & 2.27	& 9.5 	& 330 $\pm$ 10 & 3.4 $\pm$ 0.3	& 	4.00 	&	3.65		\\ 
\\ 
	& 1923 & 1906  & 2.42	& 20    & 236 $\pm$ 10 & 5.9 $\pm$ 0.4 	&	 2.50 & 2.37	\\
\\			
	& 1813 & 1796 & 2.52	& 28  	& 170 $\pm$ 10	& 7.2 $\pm$ 0.5	& 1.63	& 1.60	 		\\ 	
	& 1923 & 1919 & 2.51	& 30  	& 210 $\pm$ 10	& 5.6 $\pm$ 0.5	& 2.03 	& 2.02 		\\ 
	& 2073 & 2058 & 2.48	& 30	& 230 $\pm$ 10	& 5.0 $\pm$ 0.5	& 2.70 	& 2.56 		\\ 
\\ 
	& 1773  &  1771 & 2.66	&  45	& 120 $\pm$ 10   & 10.5 $\pm$ 0.5	& 1.05	& 1.11 	 \\ 
\texttt{AIMD} (Ref. \onlinecite{Vuilleumier2014}) & 1773 & & 2.64 & 45 $\pm$ 5	& 147 $\pm$ 15	&	10.5 	$\pm$ 2.0		& 1.2 $\pm$ 0.2 & 1.0 $\pm$ 0.2    \\	
	& 1923	& 1913 & 2.63 & 45	 	& 156 $\pm$ 10	& 8.2 $\pm$ 0.3	 & 1.57 & 1.57 	\\
\\
	& 2073 & 2063  & 2.63 & 50 & 180 $\pm$ 10 & 6.8 $\pm$ 0.3 & 2.00 & 2.00 \\
\\	 
	& 2063 & 2041 & 2.70	& 61	& 155 $\pm$ 5	& 8.1 $\pm$ 0.2 &	1.57 &	1.59 \\
	& 2063 & 2060 & 2.70	& 62  	&		/		& 7.9 	$\pm$ 0.3		&  	/	&	/ 				\\
	& 2073  & 2056 & 2.80 	& 79 	& 117 $\pm$ 5  & 9.7 $\pm$ 0.9 & 1.29 & 1.24	\\ 
\\
	& 2073  & 2040  & 2.90	& 99	&  90 $\pm$ 5  & 13.8  $\pm$ 0.4	& 0.93	& 	0.96 \\ 
\\	
	& 2073  & 2064  & 2.98	& 122	& 71 $\pm$ 5   & 16.3$\pm$ 0.9	& 0.79	& 	0.85 \\  
\texttt{AIMD} (Ref. \onlinecite{Vuilleumier2014}) & 2073 & & 2.94 & 122 $\pm$ 10 & 102 $\pm$ 20	& 12.4$\pm$ 2.5		& 1.2 $\pm$ 0.2 & 1.1 $\pm$ 0.2    \\	
\\ \hline 
\end{tabular}
\end{center}
\caption[Summary of the calculated properties for \ce{CaCO3}]{Thermodynamic averages and transport coefficients issued from all  MD simulations of $\mathbf{CaCO_3}$. Diffusion coefficients estimated from the AIMD simulations are also mentioned (\texttt{AIMD} Ref. \onlinecite{Vuilleumier2014}).}
\label{tantrans2}
\end{table}
\endgroup

\begingroup
\squeezetable
\begin{table}
\begin{center}
\begin{tabular}{l c c c c c c c c c} 
		&$T$ 	& $\langle T \rangle$ & 	$n$ 		& $\langle P\rangle$ 	&	$\sigma$	&	$\eta$		&  ${D}_{\text{Mg}}$ 				   &  ${D}_{\text{Ca}}$ 	&${D}_{\text{CO}_3}$	\\ 
		&(K) 	&			(K)	      &(g/cm$^{3}$)	& (kbar)	    &  (S/m) 			& (mPa$\cdot$s) & ($10^{-9}$m$^2$/s) & ($10^{-9}$m$^2$/s) & ($10^{-9}$m$^2$/s)\\ \hline
\\
\ce{CaMg(CO3)2}\\		
 		& 1773 & 1755 & 2.25 & 	1		& 272 $\pm$ 5		 & 4.1 $\pm$ 0.3 & 2.93 & 3.0 & 2.39	\\
\texttt{AIMD}  & 1773 	& 	& 2.25 	& 18 $\pm$ 10 	&  	140 $\pm$ 50	&					& 5.5 $\pm$ 0.3  & 5.0 $\pm$ 0.3 & 3.9 $\pm$ 0.5  \\	
\\
 	& 1773  & 1775 & 2.49	& 20	& 201 $\pm$ 10 & 6.3 $\pm$ 0.3		& 2.08	& 1.90 & 1.66	 	  	\\	
\\
   	& 1573  & 1571  & 2.62 	& 30 	& 112 $\pm$ 10 & 	12 $\pm$ 0.5		& 0.91	& 0.80	& 	0.77 	  	\\	
		& 1683  & 1700  & 2.59 	& 30	& 158 $\pm$ 10 	& & 1.37 & 1.20 & 1.12 \\
		& 1683  & 1679  & 2.59 	& 30	&  	/		  & 9.6 $\pm$ 1	& 	/	& 	/	& 	/		\\ 		
		& 1873  & 1859	& 2.55 	& 31	& 195 $\pm$ 15 & 6.9 $\pm$ 1	& 2.05	&  1.80	&  1.63 		\\	
		& 2073  & 2065 	& 2.51 	& 30	& 275$\pm$ 10	& 4.8$\pm$ 0.5	& 3.15 &  2.80	& 2.47 		\\ 
\\
		& 1653 & 1650 & 2.64 & 35 & 123 $\pm$ 10 & 11.7 $\pm$ 0.6 &  1.08 & 0.92 & 0.89 \\
\\
		& 1653 & 1685   & 2.67	   & 41	 &	118 $\pm$ 5 & 11.4 $\pm$ 0.2	& 1.06 & 0.92 & 0.91		 \\
		& 1873 & 1850	&  2.63    & 39	 & 173 $\pm$ 10 & 7.9 $\pm$ 0.2 & 1.65 & 1.47 & 1.40  \\
\\
		& 1653 & 1659	& 2.71		& 45 & 	103 $\pm$ 5			& 13.2 $\pm$ 0.3 & 0.90 & 0.78 & 0.76 \\		
\\
		& 1653 & 1627	& 2.74 		& 49 	& 92 $\pm$ 3 & 16.5 $\pm$ 0.5 &  0.73 & 0.62 & 0.62	\\
        & 1783 & 1782 	& 2.73 	    & 53	& 125 $\pm$ 5	& 11.1$\pm$ 0.5	& 1.13 & 0.98	&  0.96    		\\ 
\\
		& 1623  & 1604 	& 2.80 	& 59	& 72 $\pm$ 7  & 22.5$\pm$ 1.5	& 0.55 & 0.47	&  0.48 		\\ 
		& 1873  & 1878 	& 2.75 	& 60	& 130 $\pm$ 5 & 10.7 $\pm$ 0.5  & 1.33 &  1.15	& 1.11 		\\
		& 2073  & 2078  & 2.71 	& 60	& 188$\pm$ 10	& 7.5 $\pm$ 0.5	& 2.10 & 1.85	& 1.73 		\\ 
\\
		& 2073  & 2060	& 2.85 & 79		& 150 $\pm$ 10	& 9.3 $\pm$ 0.2  & 1.60 & 1.34 & 1.30	 \\
\\
		& 2073  & 2054 & 2.91 & 99 		& 118$\pm$ 5 & 12.1 $\pm$ 0.5 & 1.25 & 1.05 & 1.05 \\
\\
		& 2073  & 2067	& 2.99 	& 120	& 83$\pm$ 10	& 16.9 $\pm$ 0.5	& 1.02 & 0.85	& 0.85 		\\ 
\\
\hline
\end{tabular}
\end{center}
\caption[Summary of the calculated properties for \ce{CaMg(CO3)2}]{Thermodynamic averages and transport coefficients issued from all  MD simulations of $\mathbf{CaMg(CO_3)_2}$. Diffusion coefficients estimated from the AIMD simulations are also mentioned (\texttt{AIMD}).}
\label{tantrans3}
\end{table}
\endgroup

\begingroup
\squeezetable
\begin{table}
\begin{center}
\begin{tabular}{l c c c c c c c c c c} 
		&$T$ 	& $\langle T \rangle$ & 	$n$ 		& $\langle P\rangle$ 	&	$\sigma$	&	$\eta$		&  ${D}_{\text{X}_1}$ 					   &  ${D}_{\text{X}_2}$ 	&${D}_{\text{CO}_3}$	\\ 
		&(K) 	&			(K)	      &(g/cm$^{3}$)	& (kbar)	    &  (S/m) 			& (mPa$\cdot$s) & ($10^{-9}$m$^2$/s) & ($10^{-9}$m$^2$/s) & ($10^{-9}$m$^2$/s)\\ \hline
\\
\ce{Na2CO3}--\ce{MgCO3}\\
0.50:0.50 (mol)
	& 1323	& 1321	&  2.45 0.3-0.7& 30 & 98 $\pm$ 5 & 15.3 $\pm$ 1 & 1.52 & 0.56 & 0.48 \\ 
	& 1573  & 1576  & 2.40 	& 29 & 186 $\pm$10 & 	6.9	$\pm$ 0.2	& 3.80	& 1.48	&  1.30	 	\\
\\ \hline \\
\ce{K2CO3}--\ce{MgCO3}\\
0.75:0.25 (mol)	& 1573 & 1580 & 2.33 & 30 & 105 $\pm$ 5&  6.7 $\pm$ 0.5 & 3.06 & 1.10	& 1.22		  		\\	
\\ 
0.50:0.50 (mol)	
	& 1273  & 1247  & 2.48 	& 29 	&  20 $\pm$ 3 & 60 $\pm$ 7	& 0.42	& 0.13	&  0.125			\\
		    & 1373  & 1367  & 2.45 	& 30	&  38 $\pm$ 5 & 26.3	$\pm$ 1		& 0.87 & 0.32	&  0.30	 		\\
		& 1473  & 1468  & 2.42 	& 30	&  58 $\pm$ 5 & 16.5 $\pm$ 1	& 1.39 & 0.53		& 0.53	 	\\ 
		& 1573  & 1561  & 2.40 	& 30	& 75 $\pm$ 10 & 11.7	$\pm$ 0.5		& 1.98 & 0.81		& 0.79	 		\\ 
		& 1673 & 1687 & 2.37 & 30 & 102 $\pm$ 5	 & 7.7 $\pm$ 0.6  & 2.80  & 1.24 &  1.20\\	
\\
0.56:0.44 (mol) & 1673	& 1645	& 2.39 & 29 & 94 $\pm$ 5 & 9.6 $\pm$ 0.3 & 2.28 & 1.06 & 0.9 \\
\\
0.25: 0.75 (mol) 
	& 1673 	& 1654 	& 2.47	& 29 	& 102 $\pm$ 5 & 11.2 $\pm$ 0.2 & 1.70 & 1.06 & 0.88 \\	
\\ \hline \\
\ce{K2CO3}--\ce{CaCO3}\\
0.50:0.50 (mol)	& 1100  & 1094 	& 2.14 & 0  & 70 $\pm$ 5	&	15.1 $\pm$ 1.5	&  1.38	& 0.45	& 0.43	\\
	& 1200	& 1189	& 2.09	& 0 & 100 $\pm$ 10& 9.0 $\pm$ 0.5		& 2.15		& 0.74  & 	0.72	  \\
\\
    & 1223  & 1211  & 2.44 	& 25 	& 30 $\pm$ 1  & 35 $\pm$5 	& 0.61	& 0.24 	&  0.21	              	\\ 	 	 		
	& 1423  & 1403  & 2.38 	& 24    & 75 $\pm$ 10 & 12			&  1.64	& 0.68	& 0.65 	  \\ 	 		
\\
	& 1300  & 1288	& 2.46	& 30 	& 37 $\pm$ 5  & 28 $\pm$ 5 & 0.77 & 0.31 &  0.29 	\\
	& 1573	& 1584  & 2.39  & 30    & 95 $\pm$ 10 &	9	$\pm$ 0.5		& 2.45 & 1.12 	 	& 1.05 		\\
\\
 \hline
\end{tabular}
\end{center}
\caption[Summary of the calculated properties for binary mixtures]{Thermodynamic averages and transport coefficients issued from all  MD simulations of binary mixtures $\mathbf{Na_2CO_3}$--$\mathbf{MgCO_3}$, $\mathbf{K_2CO_3}$-$\mathbf{MgCO_3}$ and $\mathbf{K_2CO_3}$-$\mathbf{CaCO_3}$. The notation X$_1$ (= Na or K) and X$_2$ (= Mg or Ca) are assigned in the same order as appearing in the composition name (first column).}
\label{tantrans4}
\end{table}
\endgroup

\begingroup
\squeezetable
\begin{table}
\begin{center}
\begin{tabular}{l c c c c c c c c c c c} 
		&$T$ 	& $\langle T \rangle$ & 	$n$ 		& $\langle P\rangle$ 	&	$\sigma$	&	$\eta$		&  ${D}_{\text{X}_1}$ 					   &  ${D}_{\text{X}_2}$ 		   &  ${D}_{\text{Ca}}$ 	&${D}_{\text{CO}_3}$	\\ 
		&(K) 	&			(K)	      &(g/cm$^{3}$)	& (kbar)	    &  (S/m) 			& (mPa$\cdot$s) & ($10^{-9}$m$^2$/s) & ($10^{-9}$m$^2$/s) & ($10^{-9}$m$^2$/s)\\ \hline
\\
\ce{LiCO3}-\ce{NaCO3}-\ce{CaCO3}\\		
 0.21:0.49:0.30 (mol)	
	& 900 & 904 & 2.18 & 0 & 70 $\pm$ 5 & 25$\pm$ 3 & 0.98 & 0.78	& 0.28 & 0.24\\	
	& 1000	& 1000 & 2.13 & 0 & 120 $\pm$ 5 & 11.8 $\pm$ 1  & 1.80 & 1.42 & 0.54 & 0.48	\\	
\\ \hline \\ 
\ce{NaCO3}-\ce{K2CO3}-\ce{CaCO3}\\		
0.55:0.09:0.36 (mol)
	& 823 & 832 & 2.27 & 0	& 24.5 $\pm$ 1	& 68 $\pm$ 10 & 0.27 & 0.20 & 0.07 & 0.06  \\		
(natrocarbonatite)	& 1073 & 1070 & 2.14 & 0 & 105 $\pm$ 5 & 10.5 $\pm$ 0.5 & 1.74 & 1.43 & 0.61 & 0.54 \\
\\
0.25:0.25:0.50 (mol) \\
	& 1000 & 1005 & 2.21 & 0 	& 52 $\pm$ 5 & 35 $\pm$ 10	& 0.90 & 0.73 & 0.27 & 0.24\\			
	& 1100 & 1105 & 2.16 & 0 	& 82 $\pm$ 10 & 15.2 $\pm$ 1 & 1.27 & 1.50 & 0.50 & 0.45  	\\	
\\	
\hline 
 
\end{tabular}
\end{center}
\caption[Summary of the calculated properties for ternary mixtures]{Thermodynamic averages and transport coefficients issued from all  MD simulations of ternary mixtures $\mathbf{Li_2CO_3}$--$\mathbf{Na_2CO_3}$--$\mathbf{CaCO_3}$ and $\mathbf{Na_2CO_3}$--$\mathbf{K_2CO_3}$--$\mathbf{CaCO_3}$. The notation X$_1$ (= Li or Na) and X$_2$ (= Na or K) are assigned in the same order as appearing in the composition name (first column).}
\label{tantrans5}
\end{table}
\endgroup

\clearpage
\bibliography{main}